\documentclass[12pt,letterpaper]{article}
\pdfoutput=1

\usepackage[margin=1.7cm]{geometry}

\usepackage{amsfonts}
\usepackage{amsmath}
\usepackage{accents}
\usepackage{amssymb}
\usepackage{multirow}
\usepackage[dvipsnames]{xcolor}
\usepackage{graphicx}
\usepackage{authblk}

\usepackage{titlesec}

\titleformat{\paragraph}
{\normalfont\normalsize\bfseries}{\theparagraph}{1em}{}
\titlespacing*{\paragraph}
{0pt}{3.25ex plus 1ex minus .2ex}{1.5ex plus .2ex}

\usepackage{fancyhdr}
\usepackage[pdftex,bookmarks,hidelinks]{hyperref}
\usepackage[capitalize]{cleveref}
\usepackage[utf8]{inputenc}
\usepackage{tabularx}
\usepackage{enumitem}
\usepackage{cite}
\newcolumntype{C}[1]{>{\centering\arraybackslash}m{#1}}

\newcommand{\titlepagecornerblock}{}  
\fancypagestyle{empty}{
  \fancyhf{}

}
\fancypagestyle{titlepage}{
  \fancyhf{}

  \fancyhead[RO]{\titlepagecornerblock}
}
\fancypagestyle{plain}{
  \fancyhf{}
  \fancyfoot[C]{\bfseries \thepage} 

}
\fancypagestyle{simple}{
  \fancyhf{}
  \fancyhead[RO,LE]{\textsf{\footnotesize \thepage}}
  \fancyhead[LO,RE]{\textsf{\footnotesize \rightmark}}
  \fancyfoot[CO,CE]{\textsf{\footnotesize \explong}}
}
\let\OLDthebibliography\thebibliography
\renewcommand\thebibliography[1]{
  \OLDthebibliography{#1}
  \setlength{\parskip}{0pt}
  \setlength{\itemsep}{0pt plus 0.3ex}
}

\newif\ifdp
\newif\ifsp

\newcommand\snowmasstitle{\begin{center}\rule[-0.2in]{\hsize}{0.01in}\\\rule{\hsize}{0.01in}\\
\vskip 0.1in Submitted to the  Proceedings of the US Community Study\\ 
on the Future of Particle Physics (Snowmass 2021)\\ 
\rule{\hsize}{0.01in}\\\rule[+0.2in]{\hsize}{0.01in} \end{center}}

\hypersetup{
    pdftitle={BSM and nu flavor},
    pdfauthor={Snowmass white paper},
    final=true,
    colorlinks=true,
    linktocpage=true,
    urlcolor=Orchid,
    linkcolor=RoyalBlue,
    citecolor=OliveGreen,
    filecolor=black,
    pdfpagemode=UseOutlines,
    pdfborderstyle={/S/U},  
}

\begin{document}

\pagestyle{titlepage}
\cleardoublepage



\date{}

\title{\scshape\Large Snowmass White Paper: \\
Beyond the Standard Model effects on Neutrino Flavor\\
\normalsize 
\vskip -10pt
\snowmasstitle
}


\renewcommand\Authfont{\scshape\small}
\renewcommand\Affilfont{\itshape\footnotesize}


\author[1]{C.~A.~Arg{\"u}elles$^{\thanks{Main authors}}$}
\author[* 2]{G.~Barenboim}
\author[* 3]{M.~Bustamante}
\author[* 4]{P.~Coloma$^{\thanks{Editors (\href{mailto:pilar.coloma@ift.csic.es}{pilar.coloma@ift.csic.es},
\href{mailto:dvanegas@udemedellin.edu.co}{dvanegas@udemedellin.edu.co},
\href{mailto:teppei.katori@kcl.ac.uk}{teppei.katori@kcl.ac.uk}
)
}}$}
\author[* 5]{P.~B.~Denton}
\author[* 6,7]{I.~Esteban}
\author[* 8]{Y.~Farzan}
\author[* 4,9]{E.~Fern\'andez Mart\'inez}
\author[$\dagger$* 10]{D.~V. Forero}
\author[* 11]{A.~M. Gago}
\author[$\dagger$* 12]{T.~Katori}
\author[* 13,14]{R.~Lehnert}
\author[* 15]{M.~Ross-Lonergan}
\author[* 16,17]{A.~M.~Suliga}
\author[* 18]{Z.~Tabrizi}

\author[19]{L.~Anchordoqui}
\author[20]{K.~Chakraborty}
\author[21]{J.~Conrad}
\author[22]{A.~Das}
\author[23]{C.~S.~Fong}
\author[24]{B.~R.~Littlejohn}
\author[4]{M.~Maltoni}
\author[25]{D.~Parno}
\author[26]{J.~Spitz}
\author[27]{J.~Tang}
\author[28]{S.~Wissel}
\vspace{-1.5cm}

\affil[ 1]{Department of Physics \& Laboratory for Particle Physics and Cosmology, Harvard University, Cambridge, MA 02138, USA}
\affil[ 2]{Departament de F\'isica Te\`orica and IFIC, Universitat de Val\`encia-CSIC, E-46100 Burjassot, Spain}
\affil[ 3]{Niels Bohr Institute, University of Copenhagen, DK-2100 Copenhagen, Denmark}
\affil[ 4]{Instituto de Fisica Te\'orica UAM-CSIC, Universidad Aut\'onoma de Madrid, 28049 Madrid, Spain}
\affil[ 5]{High Energy Theory Group, Physics Department, Brookhaven National Laboratory, Upton, NY 11973, USA}
\affil[6]{Center for Cosmology and AstroParticle Physics (CCAPP), Ohio State University, Columbus, OH 43210, USA}
\affil[7]{Department of Physics, Ohio State University, Columbus, OH 43210, USA}
\affil[8]{School of Physics, Institute for Research in Fundamental Sciences (IPM), P.O. Box 19395-5531, Tehran, Iran}
\affil[9]{Departamento de Fisica Te\'orica, Universidad Aut\'onoma de Madrid, 28049 Madrid, Spain}
\affil[10]{Universidad de Medell\'{i}n, Carrera 87 No 30 - 65 Medell\'{i}n, Colombia}
\affil[11]{Secci\'on F\'isica, Departamento de Ciencias, Pontificia Universidad Cat\'olica del Per\'u, Apartado 1761, Lima, Per\'u}
\affil[12]{Department of Physics, King's College London, WC2R 2LS London, UK}
\affil[13]{Department of Physics, Indiana University, Bloomington, IN 47405, USA}
\affil[14]{Indiana University Center for Spacetime Symmetries, Bloomington, IN 47405, USA}
\affil[15]{Department of Physics, Columbia University, New York, NY 10027, USA}
\affil[16]{Department of Physics, University of California Berkeley, Berkeley, CA 94720, USA} 
\affil[17]{Department of Physics, University of Wisconsin-Madison, Madison, WI 53706, USA}
\affil[18]{Department of Physics, Northwestern University, Evanston, IL 60208, USA}
%
\affil[19]{Lehman College, City University of New York, Bronx, NY 10468, USA}
\affil[20]{Physical Research Laboratory, University Area, Ahmedabad, Gujarat 380009, India}
\affil[21]{Massachusetts Institute of Technology, Cambridge, MA 02139, USA}
\affil[22]{Hokkaido University, Sapporo, Hokkaido 060-0808, Japan}
\affil[23]{Universidade Federal do ABC, Santo Andr\'e - SP, 09210-580, Brazil}
\affil[24]{Illinois institute of Technology, Chicago, IL 60616, USA}
\affil[25]{Carnegie Mellon University, Pittsburgh, PA 15213, USA}
\affil[26]{University of Michigan, Ann Arbor, MI 48109, USA}
\affil[27]{Sun Yat-sen University, Guangzhou, 510275, P. R. China}
\affil[28]{Pennsylvania State University, University Park, PA 16802, USA}
\maketitle
\clearpage 
Endorsed by
\begin{itemize}[noitemsep]
\item A.~Ariga, Chiba University, Japan
\item M.~Bergevin, LLNL, USA
\item M.~Bishai, Brookhaven National Laboratory, NY, USA
\item J.~Boyd, CERN, Switzerland
\item G.~S.~Davies, University of Mississippi, MI, USA
\item A.~De Roeck, CERN, Switzerland
\item G.~Elor, Johannes Gutenberg University of Mainz, Germany
\item J.~Feng, University of California, Irvine, CA, USA
\item D.~F.~G.~Fiorillo, Niels Bohr Institute, Copenhagen, \item S. Gariazzo, INFN Turin, Italy
\item J.~Gehrlein, Brookhaven National Laboratory, NY, USA
\item C.~Giunti, INFN Turin, Italy
\item S.~Goswami, Physical Research Laboratory, India
\item E.~Grohs, North Carolina State University, Raleigh, USA
\item R.~Gupta, Los Alamos national Laboratory, USA
\item F.~Halzen, University of Wisconsin-Madison, WI, USA
\item L.~Johns, University of California, Berkeley, CA, USA
\item B.~Jones, University of Texas at Arlington, USA
\item E.~Kemp, Universidade Estadual de Campinas (UNICAMP),Campinas, Brasil
\item A.~N.~Khan, Max Planck Institute, Heidelberg, Germany. 
\item D.~Kim, Texas A\&M University, College Station, USA
\item I.~T.~Lim, Chonnam National University, South Korea
\item A.~Lopez Moreno, King's College London, UK
\item K.~Mahn, Michigan State University, East Lansing, MI, USA
\item X.~Marcano, Universidad Aut\'onoma de Madrid, Spain
\item T.~Maruyama, KEK, Japan
\item P.~Mehta, Jawaharlal Nehru University, New Delhi, India
\item H.~Minakata, Tokyo Metropolitan University, Tokyo, Japan
\item R.~Neilson, Drexel University, Philadelphia, USA
\item N.~Otte, Georgia Institute of Technology, Atlanta, USA
\item V.~Pandey, University of Florida, Gainesville, FL, USA
\item S.~J. Parke, Fermilab, IL, USA
\item S.~Roy, INFN Napoli, Italy
\item K.~Scholberg, Duke University, USA
\item P.~Shanahan, Fermilab, USA
\item A.~Sousa, University of Cincinnati, OH, USA
\item T.~Stuttard, Niels Bohr Institute, Copenhagen, Denmark 
\item I.~Taboada, Georgia Institute of Technology, Atlanta, USA
\item C.~A.~Ternes, INFN, Sezione di Torino, Italy
\item T.~Thakore, University of Cincinnati, OH, USA
\item A.~Tonero, Carleton University, Ottawa, Canada
\item Y.-D. Tsai, University of California, Irvine, CA, USA
\item Y.-T.~Tsai, SLAC National Accelerator Laboratory, USA
\item E.~Vitagliano, University of California (UCLA), Los Angeles, CA, USA
\item M.~Wurm, Johannes Gutenberg University of Mainz, Germany
\item J.~Yu, University of Texas at Arlington, USA
\end{itemize}
 \clearpage
 
\renewcommand{\familydefault}{\sfdefault}
\renewcommand{\thepage}{\roman{page}}
\setcounter{page}{0}

\pagestyle{plain} 
\clearpage
\setcounter{tocdepth}{3}
\textsf{\tableofcontents}





\renewcommand{\thepage}{\arabic{page}}
\setcounter{page}{1}

\pagestyle{fancy}

\fancyhead{}
\fancyhead[RO]{\textsf{\footnotesize \thepage}}
\fancyhead[LO]{\textsf{\footnotesize \rightmark}}

\fancyfoot{}
\fancyfoot[RO]{\textsf{\footnotesize Snowmass 2021}}
\fancyfoot[LO]{\textsf{\footnotesize BSM Effects on Neutrino Flavor}}
\fancypagestyle{plain}{}

\renewcommand{\headrule}{\vspace{-4mm}\color[gray]{0.5}{\rule{\headwidth}{0.5pt}}}



\clearpage

\section*{Executive Summary}
\label{sec:ExecSummary}

Neutrinos are one of the most promising messengers for signals of new physics Beyond the Standard Model (BSM). On the theoretical side, their elusive nature, combined with their unknown mass mechanism, seems to indicate that the neutrino sector is indeed opening a window to new physics. On the experimental side, several long-standing anomalies have been reported in the past decades, providing a strong motivation to thoroughly test the standard three-neutrino oscillation paradigm. This can be done in three main ways. First, neutrino oscillation experiments are very precise interferometers, sensitive to subleading effects from new physics affecting neutrino flavor transitions. On a separate front, neutrino telescopes provide a unique avenue to probe BSM effects, given the very long distances traveled by the detected neutrinos (which range from the Earth radius to several gigaparsecs), as well as their ultra-high energies. Finally, astrophysical observations (such as a nearby core-collapse supernovae) in the upcoming decade will provide invaluable information and allow us to test neutrino propagation in extremely dense environments.

In the past, neutrino physics has been driven by data, from the postulation of the neutrino by Pauli to the discovery of neutrino oscillations, which were awarded the Nobel Prize in 2015. While the upcoming generation of oscillation experiments aims to measure the leptonic CP phase and the neutrino mass ordering, it will also test the standard picture with an unprecedented level of precision. For the community to succeed in this goal, a rich experimental program is required, and should be complemented with a similar effort by the community working in phenomenology and theory. We highlight the following main issues, which can largely impact our ability to constrain BSM scenarios in the next two decades:
\begin{description}
\item[Neutrino oscillations should be tested in different environments.] This is extremely useful to break degeneracies between standard and BSM parameters, and to make sure that the three-neutrino paradigm is robust. This entails contrasting the oscillation pattern in matter and vacuum,  as well as on different oscillations channels, and/or for experiments relying on different detection mechanisms, among others. In fact, since the last Snowmass process new data from T2K and NOvA have become available. This is already enabling us to test the consistency of the standard neutrino oscillation picture. In the future, DUNE, JUNO, and T2HK will continue this endeavor, with a much higher level of precision. This also extends to non-neutrino experiments: as an example, dark matter direct detection experiments are sensitive to BSM effects on the neutrino floor.
\item[Global fits between different data sets should be encouraged.] Global fits to neutrino data are very powerful to unveil subleading effects in the oscillation probabilities, and to test the robustness of the three-neutrino framework. In fact, these provided evidence for a non-zero $\theta_{13}$ before it was determined experimentally. Similarly, global fits will be critical to unveil new physics effects on neutrino flavor. Experimental collaborations are already working on joint fits for the determination of the neutrino mixing parameters (between T2K and NOvA, ongoing), and analogous efforts have been carried out by reactor experiments searching for eV-scale sterile neutrinos. These should be extended to constrain other new physics models as well. 
\item[Accurate simulations, and control of systematic uncertainties, is required.] New-generation neutrino oscillation facilities feature beams of unprecedented luminosity which, when observed by their near detectors, will provide a very powerful tool to explore new physics effects (as long as these do not require long baselines to develop). Given the very high statistics, the bottleneck for the sensitivity to these searches  is always the level of understanding of the signal sample. That is, systematic uncertainties are critical and must be evaluated and modelled thoroughly in order to derive reliable constraints. Similar restrictions apply to BSM searches using neutrino telescopes, which are subject to large uncertainties stemming from our poor knowledge of neutrino fluxes and cross sections at such high energies. At neutrino telescopes, improving neutrino flavor identification will be key in order to improve their capabilities to test for BSM effects.
\item[The use of general theoretical frameworks and parametrizations is desirable.] The use of general parametrizations allows to easily recast the obtained bounds into specific BSM models. An example is the use of effective operators to parametrize possible effects from new interactions at low energies (a similar approach is also employed in other areas of particle physics). Other examples include the parametrizations used to describe deviations from unitarity, or SME coefficients in scenarios where Lorentz symmetry is violated. This has to be complemented with an effort on the theory side, to make sure that the bounds on effective parameters are correctly interpreted, and matched onto viable models. 
\item[Experiments should try to unlock their full potential, should try everything within reach. ]
So far, the new physics seems to evade our searches in the energy frontier, and thus it could yield unexpected signals. Neutrino experiments covered in this white paper span over 20 orders of magnitude in energy, from meV to EeV, and all of them should be encouraged to look for new signals from BSM. 
Because of this, experiments should be flexible and not single-purpose, 
and those experiments still in their designing stages should take into account all possibilities. 
For example, at long-baseline oscillation experiments, strong emphasis is placed on the charged-current measurements for the standard oscillation program; 
however, for BSM searches neutral-current measurements are equally important. 
\item[Data should be handled with great care not to miss any signals.] Experimental data may hide the next discovery in unknown ways. 
For example, at long-baseline oscillation experiments the neutrino flux measurements at the near detectors are assumed to be unoscillated so they can be used to extract the oscillation parameters at the far detector. However, near detectors themselves are sensitive to certain BSM effects; thus, unexpected anomalies (e.g., on the event rate normalization at near detectors) should not be completely dismissed.  
Collaborations should also facilitate information to make it possible for the community to analyze their data to look for BSM effects. Two notable examples in this context are the Icecube and COHERENT collaborations: their data release efforts have empowered the theoretical community to reinterpret their results and to derive new bounds on BSM scenarios, or to propose new experimental searches to extract their full potential.
\end{description}


\cleardoublepage

\section{Introduction}

The discovery of neutrino masses stands as one of the strongest evidence pointing to the existence of physics beyond the Standard Model (BSM). Our progress in the field of neutrino physics has been mainly driven by data, which lead to the discovery of neutrino oscillations and the award of the 2015 Nobel Prize. This trend seems to continue to the future, and the momentum of both theoretical and experimental studies of neutrino is unstoppable~\cite{Athar:2021xsd}. 

In the near future, we aim to significantly improve our understanding of neutrino flavor transitions, since these are uniquely positioned to discover BSM effects. The basis for this assessment is three-fold: %
\begin{description}
\item[Neutrino physics is one of the most promising places to discover BSM effects.] The violation of individual lepton flavor symmetries already supports the prevailing view that global symmetries are accidental. On top of this, the smallness of neutrino masses seems anomalous compared with other Standard Model (SM) fermions, and a number of BSM mechanisms have been proposed to explain it including models allowing for the violation of total lepton-number conservation. 
The observed mixing pattern in the neutrino sector (with large mixing angles) is also unexpected \emph{a priori}, as it is drastically different from the small mixing observed in the quark sector. 
These peculiar features suggest neutrinos may be the key to discovering BSM physics which can explain this so-called flavor puzzle, and boost our understanding of particle physics. Persistent data anomalies in neutrino physics also imply a big discovery is just around the corner.
\item[Neutrino oscillations are extremely precise interferometers.] The tiny squared mass-splittings in the neutrino sector allows us to observe quantum interference at macroscopic scales. Similarly, any new additional couplings of neutrinos to new degrees of freedom, albeit tiny, may lead to measurable consequences and an anomalous pattern in flavor transitions.
This can be seen as a correction on the dispersion relation due to new BSM effects on neutrino masses. Or, quantum states of neutrinos themselves may exhibit non-standard decoherence due to the BSM effects. New physics may also modify neutrino flavor at production, propagation, or detection processes. 
\item[Many corresponding test facilities exist or are being commissioned.] The main goal of current and next-generation long-baseline neutrino oscillation experiments is to provide an unambiguous test for CP violation in the leptonic sector, and to measure the neutrino mass ordering. The current long-baseline experiments (T2K and NOvA) will continue taking data in the upcoming years, and new long-baseline facilities (T2HK, DUNE, JUNO) will start running in the next decade. These experiments will also serve as solar and atmospheric neutrino observatories, and will be prepared to detect neutrinos from core-collapse supernovae. In parallel, the observed short-baseline anomalies have motivated a strong experimental effort with many new projects running, being built, or conceived. These will not only study the properties of neutrinos, but also use neutrinos as unique probes for BSM effects. New proposals for neutrino telescopes, such as IceCube-Gen2, have the potential to test for BSM physics over a wide range of observables, from neutrino properties to multi-messenger astronomy. Overall, measurements performed in neutrino oscillations will be complemented by new, precise tests of the impact of neutrinos in cosmology and upper bounds on the effective electron neutrino mass. A strong effort dedicated to neutrinoless double-beta decay searches will also provide complementary information.
\end{description}

\newpage

Stimulated with these motivations, this white paper explores the phenomenological implications on neutrino flavor observables which stem from four main categories of BSM effects. 

{\it The existence of additional neutrino states.}  Additional neutrino states, usually referred to as sterile neutrinos, have attracted great attention in the community in the past decades. Two dedicated white papers will be submitted to the Snowmass process (on eV-scale sterile neutrinos~\cite{LightSteriles-Snowmass} and Heavy Neutral Lepton searches~\cite{HNL-Snowmass}).
Here, we will address the indirect phenomenological consequences derived from these additional neutrino states, through deviations from the unitarity of the leptonic mixing matrix. 

{\it New neutrino interactions.} The impact of new neutrino interactions, 
or non-standard interactions  (NSI) is covered in this white paper in some detail since in the past few years it is a topic that has gathered a lot of attention in the community. 
This is motivated by a renewed interest in NSI from the theoretical point of view: on one side, on viable NSI models induced by light mediators (well below the EW scale), which could lead to potentially large effects in neutrino oscillations; and on the other, on the use of the SMEFT formalism to study low-energy NSI induced by heavy mediators.

{\it New neutrino properties such as neutrino decay.} As we are unsure about the mechanism that is behind neutrino masses, it is important to test the neutrino lifetime experimentally. Neutrino decay may also cause anomalous neutrino flavor transitions, affecting the flavor pattern for ultra-high energy neutrinos and neutrino oscillations.  

{\it Small violations of fundamental laws of physics.}  Finally, the exquisite sensitivity of neutrino flavor transitions allows to search for the violation of fundamental space-time symmetries, the very first principles of quantum field theory and general relativity: Lorentz and CPT symmetries, and quantum  decoherence. 

The phenomenological consequences and experimental sensitivities to these four types of effects depend largely on the neutrino energy. Thus, experimental technologies and facilities covered in this white paper have been categorized according to three main energy regions, outlined below.

{\it Low-energy ($\lesssim 100$~MeV) neutrino experiments} include 
solar, reactor, and supernova neutrinos, as well as new efforts to measure coherent elastic neutrino-nucleus scattering (CE$\nu$NS). We will comment on the interplay and complementarity between different data sets, and highlight potential new avenues for discovery.
Large scale direct dark matter detectors may also be sensitive to new effects in the neutrino sector and provide additional sensitivity to BSM effects.     

{\it Medium-energy ($100$~MeV$\sim 100$~GeV) neutrino experiments} include both long-baseline experiments and atmospheric neutrino observatories. 
The highest sensitivity on neutrino oscillation parameter measurements
makes them a natural place to look for effects coming from new interactions or non-unitarity of the leptonic mixing matrix. Detectors range 
from few meter scale near detectors of long-baseline neutrino experiments, to km-scale low-energy arrays for neutrino telescopes. 

{\it High-energy ($\gtrsim 100$~GeV) neutrino experiments} are neutrino telescopes mainly designed for multi-messenger astronomy, but with great potential to discover BSM effects in the neutrino sector. The downside is that flavor identification is extremely challenging. In the next decade, we may see the first cosmogenic neutrinos with unseen information of particle physics and astrophysics.

This white paper is structured in two separate sections, covering the main phenomenological (Sec.~\ref{sec:th}) and experimental (Sec.~\ref{sec:exp}) aspects of neutrino flavor in current and future facilities. Each section has been further divided into the main categories outlined above. Finally, Sec.~\ref{sec:summary} presents a summary of the white paper and outlook for the future.

\clearpage

\section{Theoretical aspects}
\label{sec:th}

A possible way to classify the plethora of new physics models can be as follows. New physics effects can stem from the inclusion of: $(i)$ additional neutrinos, $(ii)$ new neutrino interactions, $(iii)$ direct consequences from the fact that neutrinos are massive, such as neutrino decay, and $(iv)$ small violations of fundamental symmetries, such as CPT or Lorentz violation. In this section we review the main phenomenological implications for these four categories separately.
\subsection{New physics from the existence of additional neutrino states}
\label{sec:steriles}

\textbf{Main authors:} E. Fernandez-Mart\'inez and D. V. Forero
\vspace{2mm}

Among all possible extensions of the Standard Model (SM) of particle physics to accommodate neutrino masses and mixings to address the evidence for neutrino flavor change from the observed oscillation phenomenon, the addition of extra, heavy neutrino states is arguably the simplest~\cite{Minkowski:1977sc,Mohapatra:1979ia,Yanagida:1979as,GellMann:1980vs}. Indeed, right-handed neutrinos, in analogy to all other fermions, allow for neutrino masses but their gauge singlet \emph{sterile} nature implies their mass scale could be very different and of a Majorana, lepton-number-violating, nature. Depending on the value of this new mass scale, the phenomenology of these new states can be very different. 

Given their contribution to the active neutrino masses, typically through regular Yukawa couplings, a very general consequence of these models is that the new states will mix with the active neutrino flavors so that the full neutrino mixing matrix is larger than $3\times3$. In particular the flavor states $\nu_{\alpha} = \mathcal{U}_{\alpha i} \nu_i$ with $\alpha =e, \mu , \tau$ and $i$ running not only over the known 3 light neutrino mass eigenstates but also through the extra heavy ones. As a consequence, the PMNS matrix that describes the $W$ interactions with the charged leptons and the 3 light neutrinos, that is the $3\times3$ upper-left submatrix of $\mathcal{U}$, is not
unitary. We will dub this matrix $N$ to stress its non-unitary nature. One of the possible general ways to parameterize these unitarity deviations in $N$ is through a triangular matrix~\cite{Escrihuela:2015wra}~\footnote{For a similar parameterization corresponding to a $(3+1)$ and a $(3+3)$-dimensional mixing matrix,  see Refs.~\cite{Xing:2007zj,Xing:2011ur}}
 \begin{equation}
  N = 
 \left\lgroup
 \begin{array}{ccc} 
 1-\alpha_{ee} & 0 & 0 \\
 \alpha_{\mu e} & 1-\alpha_{\mu \mu} & 0 \\
  \alpha_{\tau e} & \alpha_{\tau \mu} & 1-\alpha_{\tau \tau}
 \end{array}
 \right \rgroup U \,,
 \label{eq:triangular}
 \end{equation}
with $U$ a unitary matrix that tends to the usual PMNS matrix when the non-unitary parameters $\alpha_{\beta \gamma} \rightarrow 0$~\footnote{The original parameterization in ref.~\cite{Escrihuela:2015wra} uses $\alpha_{ij}$ instead of $\alpha_{\beta\gamma}$. The equivalence between the two notations is as follows: $\alpha_{ii} = 1-\alpha_{\beta\beta}$ and $\alpha_{ij} = \alpha_{\beta\gamma}$.} .
 The triangular
matrix in this equation accounts for the non-unitarity of the $3\times 3$ matrix for any number of extra neutrino species. This parameterization has been shown to be particularly well-suited for oscillation searches~\cite{Escrihuela:2015wra,Blennow:2016jkn} since, compared to other alternatives, it minimizes the departures of its unitary component $U$ from the mixing angles that are directly measured in neutrino oscillation experiments when unitarity is assumed.

\subsubsection{Neutrino oscillations in presence of heavy sterile neutrinos}

If the new states are too heavy to be produced in the neutrino beam, the flavor states in which the neutrinos are produced and detected are given by $|\nu_\alpha \rangle = N^*_{\alpha i} |\nu_i \rangle $. That is, the truncated sum over the light accessible eigenstates of the full matrix $\mathcal{U}$. Hence, the flavor basis is no longer orthonormal given that $N$ is non-unitary. It is therefore convenient to study the evolution of the states and the oscillation phenomenon on the mass basis so that the Hamiltonian is simply given by:
\begin{equation}
\label{eq:Hnonunitarity}
H = \frac{1}{2E}\begin{pmatrix}
0 & 0 & 0 \\
0 & \Delta m_{21}^2 & 0 \\
0 & 0 & \Delta m_{31}^2
\end{pmatrix} + N^\dagger  \begin{pmatrix}
V_{\rm CC}+V_{\rm NC} & 0 & 0 \\
0 & V_{\rm NC} & 0 \\
0 & 0 & V_{\rm NC}
\end{pmatrix}  N,
\end{equation}
where $V_{\rm CC}=\sqrt{2}G_Fn_e$ and $V_{\rm NC}=-G_F n_n/\sqrt{2}$ are the charged-current (CC) and neutral-current (NC) matter potentials, respectively. The amplitude for a neutrino in the mass eigenstate $j$ to interact as a neutrino of flavor $\beta$ is given by the mixing 
matrix element $N_{\beta j}$, which means that the oscillation probability will be given by
\begin{equation}
\label{eq:nonunitarityoscillationformula}
P_{\alpha\beta} = |(N \exp(- i H L) N^\dagger)_{\beta\alpha}|^2.
\end{equation}
We can remark two important differences with the standard case when $N$ is non-unitary. Firstly, in the limit $L \rightarrow 0$, $P_{\alpha\beta} \neq \delta_{\alpha \beta}$. This \emph{zero distance effect} is a direct consequence of the flavor eigenstates not being orthonormal. Another interesting difference is that the NC matter potential now plays a relevant role, since $N^\dagger N $ is no longer the identity in eq.~(\ref{eq:Hnonunitarity}). Non-standard matter effects as well as \emph{zero distance effects} can thus be a way to probe for unitarity deviations in neutrino oscillations.

\subsubsection{Neutrino oscillations in presence of light sterile neutrinos}

Conversely, if the extra neutrinos are light enough to be produced in the beam, these new mass eigenstates will introduce new oscillation frequencies and will participate in the neutrino oscillation phenomenon. Furthermore, oscillations will be sensitive to the whole mixing matrix $\mathcal{U}$ and not just to the $3 \times 3$ truncated matrix $N$. This scenario has been studied in great detail in connection to the standing \emph{short-baseline neutrino oscillation anomalies} and is covered in a dedicated white paper. Here we will instead consider the simplified scenario in which all the new oscillation frequencies introduced by the additional states are large such that $\Delta m^2_{iJ}/2E \gg L^{-1}, V_{\rm CC}, V_{\rm NC}$. In this limit, the new oscillation frequencies are \emph{averaged out} and the oscillation probability becomes the same as eq.~(\ref{eq:nonunitarityoscillationformula}) at leading order in the mixing between the new mass eigenstates and the active flavors~\cite{Blennow:2016jkn}. Thus, ``high-energy non-unitarity'' and ``averaged-out light steriles'' have exactly the same impact in neutrino flavor oscillations, up to quartic terms in the heavy-active mixing\footnote{Refs.~\cite{Fong:2016yyh,Fong:2017gke} concentrate on this subleading difference between the two regimes to try to distinguish between them.} or, equivalently, $\mathcal{O}(\alpha^2)$. When the propagation is through matter, differences quadratic in the heavy-active mixing are also present, although further suppressed by the matter potential over the new mass splittings. See Ref.~ \cite{Fong:2017gke} for a thorough discussion.

\subsubsection{Normalization of the oscillation probability}

Notice that, even though the evolution operator in Eq.~(\ref{eq:nonunitarityoscillationformula}) is unitary, $N$ is not and as such the ``probabilities'' $P_{\alpha \beta}$ do not add up to 1 and at $L=0$ $P_{\alpha \alpha} \neq 1$. Therefore, this amplitude cannot be directly interpreted as an oscillation probability of $\nu_\alpha$ transitioning to a $\nu_\beta$, at least not in the usual sense. In order to make the connection with a given experimental result, the relevant question is how the expected number of events is estimated. If the measured number of events associated with a charged lepton of flavor $\beta$ from a neutrino source produced with a charged lepton of flavor $\alpha$ is compared with the SM expectations, then Eq.~(\ref{eq:nonunitarityoscillationformula}) does indeed correspond to the ``oscillation probability'' inferred from this data~\cite{Antusch:2006vwa}. 

However, in most situations this is not the case and the neutrino flux and/or the detection cross section are rather estimated in a data-driven way. For instance, a common scenario is to have a near detector to calibrate the neutrino flux and cross section for the far detector measurement. In this situation, the unitarity deviations might also affect the measurement at the near detector. This will always be the case when the sterile neutrinos are too heavy to be produced or when they are heavy enough to be in the \emph{averaged-out} regime also for the shorter baseline of the near detector $L_{ND}$ such that $\Delta m^2_{iJ}/2E \gg L_{ND}^{-1}$. Thus, the oscillation probability inferred in this scenario would rather correspond to~\cite{Blennow:2016jkn,Forero:2021azc}:
\begin{equation}
\label{eq:experimentalprobability}
\mathcal P_{\alpha\beta} = \left| (N \exp(- i H L) N^\dagger)_{\beta\alpha} \right|^2/((NN^\dagger)_{\alpha\alpha})^2.
\end{equation}
And some of the sensitivity to the non-unitarity parameters would cancel in the ratio. Thus, it is very important to consider the impact of the unitarity deviations not only at the detector probing the neutrino oscillations, but also at the measurements used to calibrate that expectation, since they might also be affected depending on the mass of the new sterile states as well as  the corresponding baseline.

\subsubsection{Present constraints on non-unitarity parameters}

Present bounds on the unitarity deviations encoded in the $\alpha$ parameters strongly depend on the mass scale of the new sterile states. Indeed, if the new sterile neutrinos are sufficiently heavy, they wont be kinematially accessible in weak processes involving neutrinos such as $W$, $Z$, meson and charged lepton decays. As such, when summing over the available mass eigenstates, the non-unitary correction $|(TT^\dagger)_{\beta \beta} |^2\sim |1 -2 \alpha_{\beta \beta}|^2$ appears in all these processes, where $T$ is the triangular matrix of Eq.~(\ref{eq:triangular}) and $\beta$ is the flavor of the charged lepton in the neutrino vertex. Thus, lepton universality, flavor and electroweak precision tests are excellent probes of unitarity deviations of the PMNS matrix~\cite{Shrock:1980vy,Schechter:1980gr,Shrock:1980ct,Shrock:1981wq,Langacker:1988ur,Bilenky:1992wv,Nardi:1994iv,Tommasini:1995ii,Antusch:2006vwa,FernandezMartinez:2007ms,Antusch:2008tz,Biggio:2008in,Antusch:2009pm,Forero:2011pc,Alonso:2012ji,Antusch:2014woa,Abada:2015trh,Fernandez-Martinez:2015hxa,Escrihuela:2015wra,Parke:2015goa,Miranda:2016wdr,Fong:2016yyh,Escrihuela:2016ube}. For recent global fits to all flavor and electroweak precision data summarizing present bounds on non-unitarity see Refs.~\cite{Antusch:2014woa,Fernandez-Martinez:2016lgt,Chrzaszcz:2019inj,Coutinho:2019aiy}. 

These bounds only apply if the extra sterile neutrinos are too heavy to be produced in the corresponding observable. At lower scales, between the MeV and tens of GeVs, these \emph{heavy neutral leptons} may be directly produced and searched for through their decays at collider and beam dump experiments and looking for kinks and peaks in meson or even beta decays down to the keV scale. In this intermediate energy range, much more stringent constraints apply depending on their actual mass~\cite{Atre:2009rg,Abada:2007ux,Gorbunov:2007ak,Abada:2016plb,Abada:2018nio,Abada:2018sfh,Ballett:2019bgd,Berryman:2019dme,Abada:2019bac,Krasnov:2019kdc,Bryman:2019ssi,Bryman:2019bjg,Bondarenko:2019yob, Drewes:2015iva, Chrzaszcz:2019inj,Gorbunov:2020rjx,Coloma:2020lgy} (see Ref.~\cite{HNL-Snowmass} for details).

Finally, below the MeV scale, the sterile neutrinos may be produced in neutrino sources and participate, together with their SM active neutrino counterparts in the neutrino oscillation phenomenon. Depending on their mass, the beam energy and baseline, they may introduce new oscillation frequencies that can be searched for (or decohere, or average out their oscillations), leading to the effects discussed in Eq.~(\ref{eq:nonunitarityoscillationformula}). 

\begin{table}[t!]
\label{tab:unitaritybounds}
\begin{center}
\begin{tabular}{|c| c| c|}
\hline
    & ``flavor+electroweak'' & ``Averaged-out oscillations'' \\
     & $m>$ EW ($2\sigma$ limit) &  $\Delta m^2 \gtrsim 0.1$~eV$^2$ (90\% CL) \\ \hline\hline
$\alpha_{ee} $ & $1.3 \cdot 10^{-3}$~\cite{Fernandez-Martinez:2016lgt} & $8.4 \cdot 10^{-3} $~\cite{Goldhagen:2021kxe}  \\
$\alpha_{\mu\mu}$ & $2.2 \cdot 10^{-4}$~\cite{Fernandez-Martinez:2016lgt} & $5.0 \cdot 10^{-3}$~\cite{Forero:2021azc} \\
$\alpha_{\tau\tau}$ & $2.8 \cdot 10^{-3}$~\cite{Fernandez-Martinez:2016lgt} & $6.5 \cdot 10^{-2}$~\cite{Dentler:2018sju} \\
$\lvert\alpha_{\mu e}\rvert$ & $6.8 \cdot 10^{-4} \; (2.4 \cdot 10^{-5})$~\cite{Fernandez-Martinez:2016lgt} & $9.2 \cdot 10^{-3} $\\
$\lvert\alpha_{\tau e}\rvert$ & $2.7 \cdot 10^{-3}$~\cite{Fernandez-Martinez:2016lgt} & $1.4 \cdot 10^{-2}$ \\
$\lvert\alpha_{\tau\mu}\rvert$ & $1.2 \cdot 10^{-3}$~\cite{Fernandez-Martinez:2016lgt} & $1.1 \cdot 10^{-2}$ \\ \hline
\end{tabular}
\caption{\label{tab:NUbounds} \textbf{Current upper bounds on the $\alpha$ parameters.} 
The bounds in the middle column (shown at $2\sigma$, 1 d.o.f.) apply for $\Delta m^2 \gtrsim 100$~eV$^2$ and are taken to a global fit to flavor and electroweak precision observables from Ref.~\cite{Fernandez-Martinez:2016lgt}. The bounds in the right column (shown at 90\% CL, for 1 d.o.f.) apply for $\Delta m^2 \gtrsim 0.1$~eV$^2$ and come from neutrino oscillation searches, when the sterile neutrinos would be in the averaged-out regime for such a mass scale. The second number quoted in parenthesis for the $\alpha_{\mu e}$ element includes the $\mu \to e \gamma$ observable, which can in principle be evaded~\cite{Forero:2011pc}. The bounds without a reference are indirectly obtained from constraints on the diagonal parameters via $\alpha_{\alpha \beta} \leq 2 \sqrt{\alpha_{\alpha \alpha} \alpha_{\beta \beta}}$, which follows from the unitarity of the full $\mathcal{U}$ mixing matrix.}
\end{center}
\end{table}

In the intermediate mass regime, between the MeV\footnote{Or down to the keV scale for the electron flavor from $\beta$ decays} and the electroweak scales, the constraints from direct production are very strong and preclude their observation in the neutrino oscillation phenomenon. Above the electroweak scale, the constraints from the PMNS unitarity are stringent but may still allow to probe for them at very precise future facilities. These constraints have been summarized in the middle column of Table~\ref{tab:NUbounds}. For much lighter sterile neutrino masses, bounds can be derived through their impact in the neutrino oscillation phenomenon~\cite{FernandezMartinez:2007ms,Antusch:2009pm,Parke:2015goa,Miranda:2016wdr,Ge:2016xya,Blennow:2016jkn,Escrihuela:2016ube,Kosmas:2017zbh,Ellis:2020ehi,Ellis:2020hus,Miranda:2020syh,Denton:2021mso,Denton:2021rsa}. We list in the last column of Tab.~\ref{tab:NUbounds} the most stringent present constraints we found on the different elements for the new mass splittings $\Delta m^2  \gtrsim 0.1$~eV$^2$ would be in the \emph{averaged-out regime} for the corresponding observable. This way the bounds become independent from the new mass scale as long as $\Delta m^2  \gtrsim 0.1$~eV$^2$. Nevertheless, for particular values of the new mass scale, slightly stronger constraints might apply if they correspond to an oscillation maximum of some facility. Thus, the bounds in Tab.~\ref{tab:NUbounds} should be regarded as conservative but applicable over a long range of masses. The bound on $\alpha_{ee}$ comes from solar neutrino data from Ref.~\cite{Goldhagen:2021kxe}. The bound quoted is, conservatively, for the GS98 solar model while the AGSS09 model would lead to constraints about a factor 2 stronger. The new estimation of fluxes for reactor antineutrinos point to a solution of the reactor anomaly and the possibility to use reactor data also as constraints for $\alpha_{ee}$. These constraints would be similar to those from solar data but more dependent on the actual flux assumed~\cite{Giunti:2021kab}. The most stringent bound on $\alpha_{\mu \mu}$ comes from MINOS/MINOS+ data~\cite{MINOS:2017cae} and we quote a result from a global fit from Ref.~\cite{Forero:2021azc}. For $\alpha_{\tau \tau}$ the bound comes from a combination of different atmospheric neutrino data from Ref.~\cite{Dentler:2018sju}. Finally, the off-diagonal bounds are obtained indirectly from 
constraints on the diagonal parameters via $\alpha_{\alpha \beta} \leq 2 \sqrt{\alpha_{\alpha \alpha} \alpha_{\beta \beta}}$, which is a consequence of the unitarity of the full $\mathcal{U}$ mixing matrix (these bounds turn out to be stronger than direct constraints of the $\alpha \rightarrow \beta$).

Apart from constraining the unitarity-violating parameters $\alpha$, an interesting question is what are the present bounds on the different elements of $N$ when unitarity is not assumed. This has been explored in Refs.~\cite{Qian:2013ora,Parke:2015goa,Denton:2021mso,Denton:2021rsa}. Notice however that the present bounds on the possible unitarity deviations encoded by $\alpha$ from Tab.~\ref{tab:NUbounds} are of the same order or stronger than the present accuracy on the mixing angles of the unitary parametrization (see for instance Ref.~\cite{Esteban:2020cvm}). Thus, when these are taken into account, the constraints on the individual matrix entries are not significantly affected by not assuming unitarity. 


\subsubsection{Neutrino oscillations in the presence of Large Extra Dimensions (LED)}

The main motivation for introducing extra space-time dimensions was to alleviate the so-called hierarchy problem, i.e. the large difference between the electroweak and the GUT~\cite{Dienes:1998vh,Dienes:1998vg} or the Planck energy scales~\cite{ArkaniHamed:1998rs,ArkaniHamed:1998nn,Antoniadis:1998ig}. More interestingly, models with large extra dimensions (LED) can also accommodate non-zero neutrino masses~\cite{Dienes:1998sb,Arkani-Hamed:1998wuz, Mohapatra:1999zd, Barbieri:2000mg, Mohapatra:2000wn}.  Since right-handed  neutrinos  are  singlets  under  the  Standard Model  (SM)  gauge  group,  it is often assumed that they can propagate in the bulk while the SM particles are confined in the four-dimensional brane. In this context, neutrinos acquire a Dirac mass that is naturally small, since the neutrino four-dimensional neutrino Yukawa couplings are suppressed relative to charged-fermion
Yukawa couplings by a factor proportional to the volume of the extra dimensions. This provides an alternative to the usual seesaw mechanism~\cite{Dienes:1998sb} for the generation of the neutrino masses, while avoiding the need for large energy scales.

\begin{figure}[!htb]
\center
\includegraphics[width=0.35\textwidth]{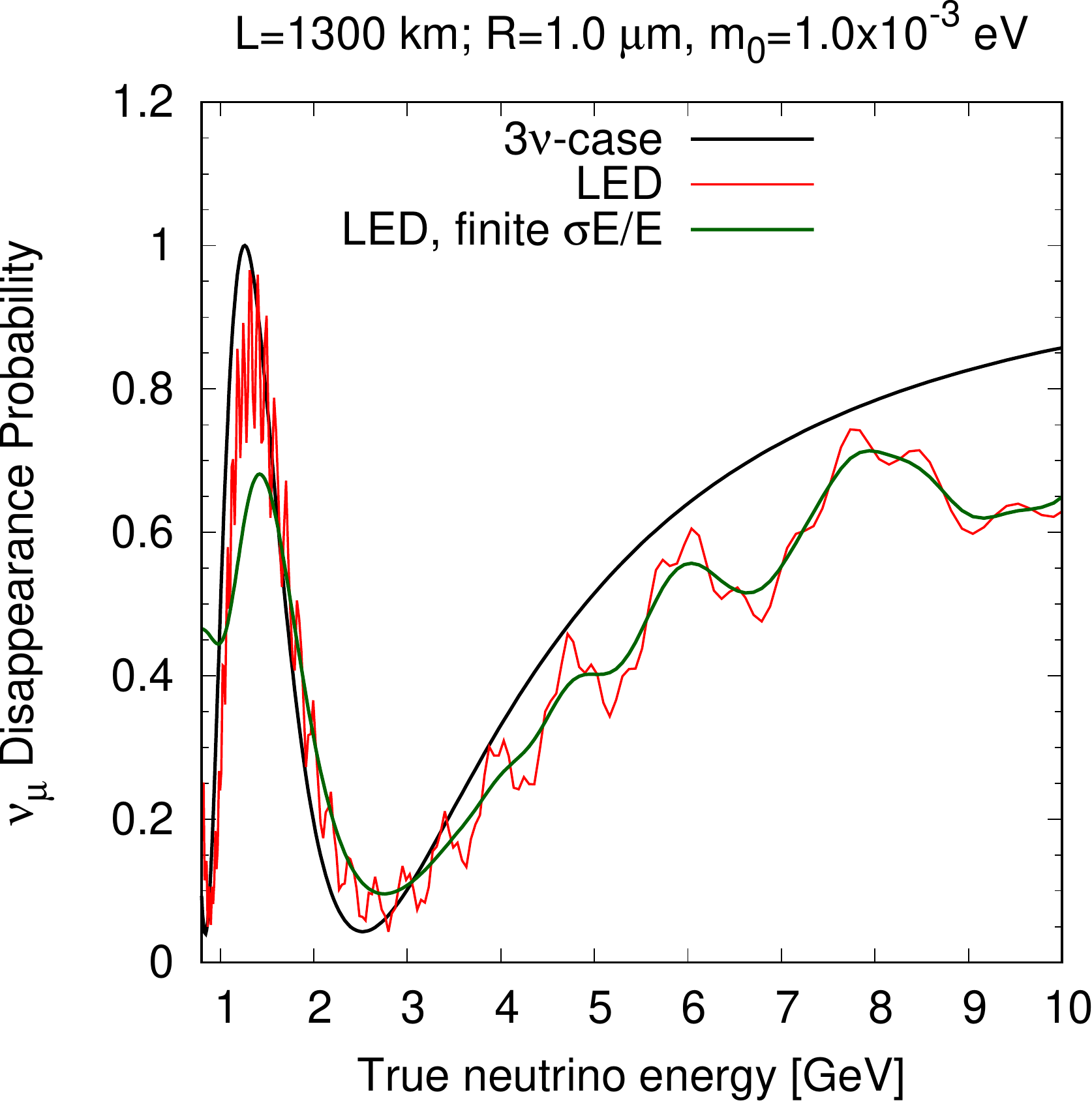}
\includegraphics[width=0.55\textwidth]{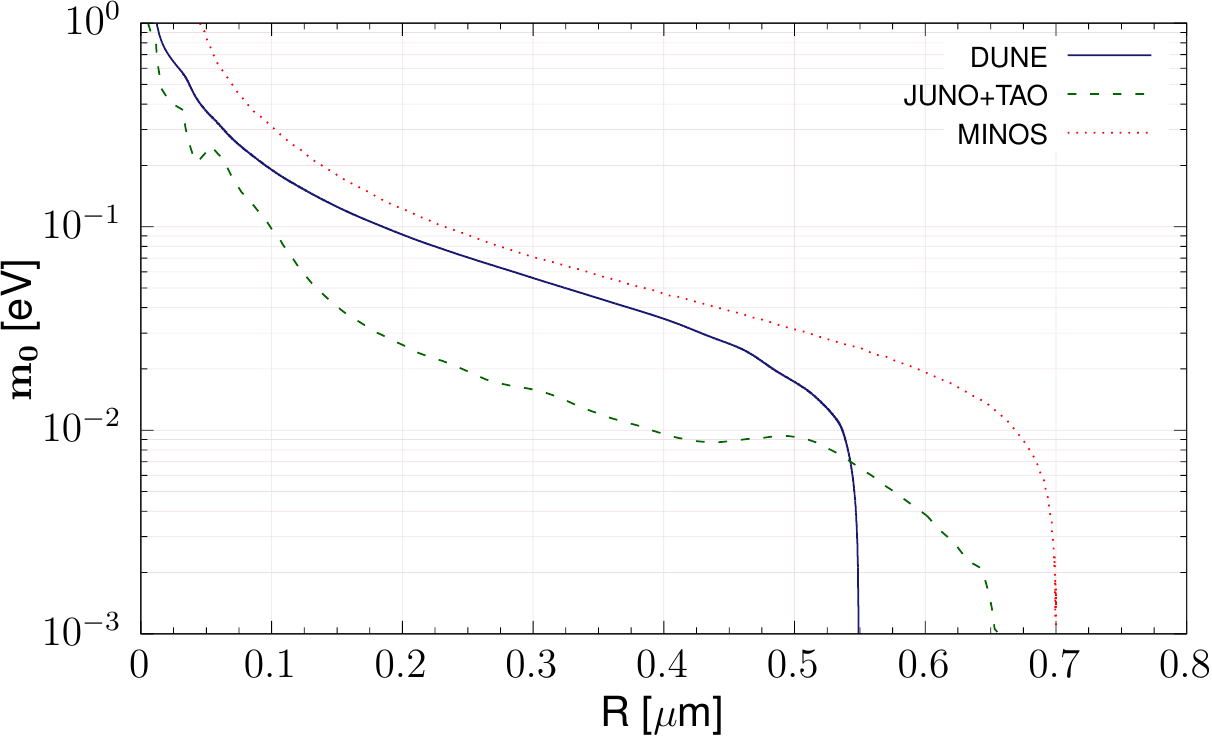}
\caption{\textbf{Left: LED impact on the oscillation probabilities. } LED-induced muon neutrino disappearance probability (red and green lines) in comparison to the three-neutrino flavor case ($3\nu$-case, black line), as the function of neutrino energy with $1300$~km baseline. The LED-induced oscillations without (with) the detector energy resolution are shown by the red (green) line. Particular values of $m_0$ and $R$ were chosen to enhance the LED effect, and the standard oscillation parameters were fixed to their best fit values from Ref.~\cite{deSalas:2020pgw}.
\textbf{Right: Future sensitivities to LED in neutrino oscillation experiments. } DUNE sensitivity line taken from Ref.~\cite{DUNE:2020fgq} and JUNO+TAO sensitivity line taken from Ref.~\cite{Basto-Gonzalez:2021aus}, in comparison to the MINOS sensitivity from Ref.~\cite{MINOS:2016vvv}. All lines correspond to 90\% C.L. contours for normal mass ordering. The excluded region is towards the right of the lines. Finally, note that the actual bound derived from MINOS data is stronger than the expected sensitivity shown in this figure, see text for details.} \label{fig:LED}
\end{figure}

In the four-dimensional brane, the Kaluza-Klein (KK) modes of the right-handed neutrino behave as an infinite tower of sterile neutrinos. If these are light enough, they will induce modifications of the neutrino oscillation probabilities. The LED-induced oscillation probability is shown in Fig.~\ref{fig:LED} left, for the case of muon neutrino disappearance.

From the probability plots in Fig.~\ref{fig:LED} left, the main features present in LED-induced oscillations are:
\begin{itemize}[nosep]
\item A global reduction of survival probabilities respect to the three-flavor case, which are typically noticeable at high energies. 
\item Appearance of modulations and fast oscillations to Kaluza-Klein states on top of the main oscillation.
\end{itemize}
So far, the only experimental collaboration that has constrained the LED parameter space is MINOS~\cite{MINOS:2016vvv}. 
Since evidence for deviations from the standard three-flavor oscillations was not found, a limit on the compactification radius $R<0.45\,\mu\text{m}$ at 90\% of C.L. (for the conservative case $m_0\to0$) was reported. An analysis including MINOS/MINOS+ data~\cite{DeRijck:2017ynh} could set even stronger limits. In a recent analysis of the KATRIN, Daya Bay and MINOS data~\cite{Forero:2022skg}, the bound $R<0.20~\mu\text{m}$ ($R<0.10~\mu\text{m}$) at 90\% C.L. for normal (inverted) neutrino mass ordering was reported. Before the MINOS analysis, the same LED-induced oscillations were probed at accelerator~\cite{Machado:2011jt}, reactor~\cite{Machado:2011jt,Machado:2011kt,Girardi:2014gna,DiIura:2014csa}, and atmospheric~\cite{Esmaili:2014esa} neutrino experiments, producing a similar bound. 
 
As shown in Fig.~\ref{fig:LED} left, LED-induced oscillations imprint particular features in the event energy spectrum which can be exploited at the analysis level. With the expected energy resolution, future neutrino experiments can be sensitive to the main modulation of the fast oscillations to Kaluza-Klein states (green line in Fig.~\ref{fig:LED} left). Sensitivity to LED-induced oscillations is expected to be limited by systematical uncertainties that impact the shape of the spectrum, especially at the near detectors. This has been implemented in forecast studies, through simulations of data at different experiments like DUNE~\cite{Berryman:2016szd,DUNE:2020fgq}, SBND~\cite{Stenico:2018jpl}, and JUNO~\cite{Basto-Gonzalez:2021aus}. Since LED oscillations occur at short and long baselines simultaneously, a two-detector analysis has better prospects for testing this hypothesis.

In Fig.~\ref{fig:LED} (right panel), a comparison of the DUNE~\cite{DUNE:2020fgq}, JUNO~\cite{Basto-Gonzalez:2021aus}, and MINOS~\cite{MINOS:2016vvv} sensitivity to LED-induced oscillations is shown. For the DUNE analysis only the far detector spectral information was used to test the LED hypothesis, and 5\% systematic errors were assumed in the simulation. As can be seen from the figure, DUNE will be sensitive to $R\gtrsim 0.55\mu\text{m}$ at 90\% of C.L. The same figure shows the sensitivity for a reactor neutrino experiment which uses a combination of near (TAO) and far (JUNO) detectors. Under the assumption of 1\% systematic errors, the JUNO+TAO combination would be sensitive to $R\gtrsim 0.65\mu\text{m}$ at 90\% of C.L. This figure shows a synergy of future reactor and accelerator neutrino experiments, which will use different detector technologies and complementary oscillation channels. Based on these results, there are good prospects for probing LED-induced oscillations at DUNE and JUNO+TAO.
 
Besides neutrino oscillations~\cite{Davoudiasl:2002fq}, models with LED can be tested in beta-decay experiments~\cite{Basto-Gonzalez:2012nel} and tabletop experiments~\cite{Hoyle:2000cv,Adelberger:2003zx,ParticleDataGroup:2020ssz}. However, neutrino oscillation experiments provide a stronger constraint on $R$, which is at the sub-micrometer scale. 

In the forthcoming precision Era, new physics signals might emerge as subleading effects of the three-neutrino paradigm or as a new oscillation phase(s). Since the LED model predicts a large number of sterile states, it is worth testing this model in the same footing as it is being done for the $3+1$ scenario (for a discussion on the equivalence between a $3+N$ sterile neutrino framework and neutrino oscillations in presence of LED see, e.g. Ref.~\cite{Esmaili:2014esa}).

\subsection{New interactions in the neutrino sector }
\label{sec:inter}

\textbf{Main authors:} P. Coloma, Y. Farzan, Z. Tabrizi
\vspace{2mm}

This section discusses the impact of new interactions in neutrino production, detection and propagation. We start reviewing the introduction of such effects within the Effective Field Theory (EFT) formalism, for operators up to dimension 6. Section~\ref{subsec:inter-eft} discusses the inclusion of operators leading to charged-current (CC) interactions, while Sec.~\ref{subsec:inter-NSI} focuses on the formalism of neutral-current (NC) Non-Standard Interactions (NSI) of neutrinos with matter fields. As we will see in Sec.~\ref{subsec:inter-models} it is theoretically challenging to build a model that can lead to large NSI effects in the neutrino sector, while respecting bounds from charged lepton experiments. A common way out of this is to invoke the existence of light weakly coupled mediators. This possibility is entertained in Sec.~\ref{subsec:inter-LIGHTnew}, which includes a thorough discussion of current constraints on such scenarios. Finally, the last part of this section focuses on operators linking the active neutrinos to the dark sector: Sec.~\ref{subsec:inter-newops} discusses the phenomenological implications of new operators involving heavy neutral leptons, while Sec.~\ref{subsec:inter-DMbg} presents the main implications of neutrino interactions with dark matter.

\subsubsection{General effective interactions between neutrinos: the charged-current case} 
\label{subsec:inter-eft}

From an EFT point of view, the most general Lagrangian which describes the new physics above the electroweak scale is the Standard Model Effective Field Theory (SMEFT). It has the exact same particle content and local symmetry as the SM, the difference is that higher dimensional non-renormalizable operators are added to the SM Lagrangian. In this way we can provide an effective description of the physical effects caused by heavy BSM particles, by e.g. investigating how the vertices of the SM gauge bosons and fermions are modified \cite{Falkowski:2017pss}. At lower energies, e.g. below the $Z$ boson mass $m_Z$, the relevant Lagrangian is the one of the weak effective field theory (WEFT), for which the heavy SM particles ($Z$ and $W$ bosons, Higgs and top quark) are integrated out from the theory. The parameters of WEFT can be matched to the ones of SMEFT at a scale $m_Z$~\cite{Falkowski:2019xoe}.  

At the SM level, the interaction between neutrinos and charged leptons or quarks is described by the Fermi interaction. Focusing only on the left handed neutrinos, the Charged-Current (CC) part of the WEFT Lagrangian is:
\begin{align} 
    \mathcal{L}_{\rm WEFT} 
    &\supset
    - \,2\sqrt{2} G_F V_{jk}  \Big\{
      [ {\bf 1} + \epsilon_L^{jk}]_{\alpha\beta}
             (\bar{u}^j \gamma^\mu P_L d^k) (\bar\ell_\alpha \gamma_\mu P_L \nu_\beta)
    \,+\, [\epsilon_R^{jk}]_{\alpha\beta} (\bar{u}^j \gamma^\mu P_R d^k)
                                     (\bar\ell_\alpha \gamma_\mu P_L \nu_\beta) \nonumber\\
    &\quad +\,
      \frac{1}{2} [\epsilon_S^{jk}]_{\alpha\beta}
             (\bar{u}^j d^k) (\bar\ell_\alpha P_L \nu_\beta)
    - \frac{1}{2} [\epsilon_P^{jk}]_{\alpha\beta}
             (\bar{u}^j \gamma_5 d^k) (\bar\ell_\alpha P_L \nu_\beta) \nonumber\\
    &\quad +\,
      \frac{1}{4} [\epsilon_T^{jk}]_{\alpha\beta} (\bar{u}^j \sigma^{\mu\nu} P_L d^k)
                                      (\bar\ell_\alpha \sigma_{\mu\nu} P_L \nu_\beta)
    + {\rm{h.c.}} \Big \} \,.
  \label{eq:EFT_lweft}
\end{align}
where $G_F$ is the Fermi constant, $V$ is the CKM matrix, $P_{L,R} = \frac{1}{2} (1 \mp \gamma^5)$ are the chirality projection operators, $\sigma^{\mu\nu} = \tfrac{i}{2} [\gamma^\mu,\gamma^\nu]$, and $\ell_\alpha$ are the charged leptons, while $u^j$ ($d^k$) stand for the up-type (down-type) quarks ($j$ and $k$ are quark mass eigenstates). The new interactions between neutrinos, charged leptons and quarks are parameterized in terms of the dimensionless Wilson coefficients $[\epsilon_X^{jk}]_{\alpha\beta}$, where $X=L,R,S,P,T$ stands for the left- and right-handed, scalar, pseudo-scalar and tensor interactions, respectively. 

\begin{figure}[ht!]
  \centering
  \includegraphics[width=0.85\textwidth]{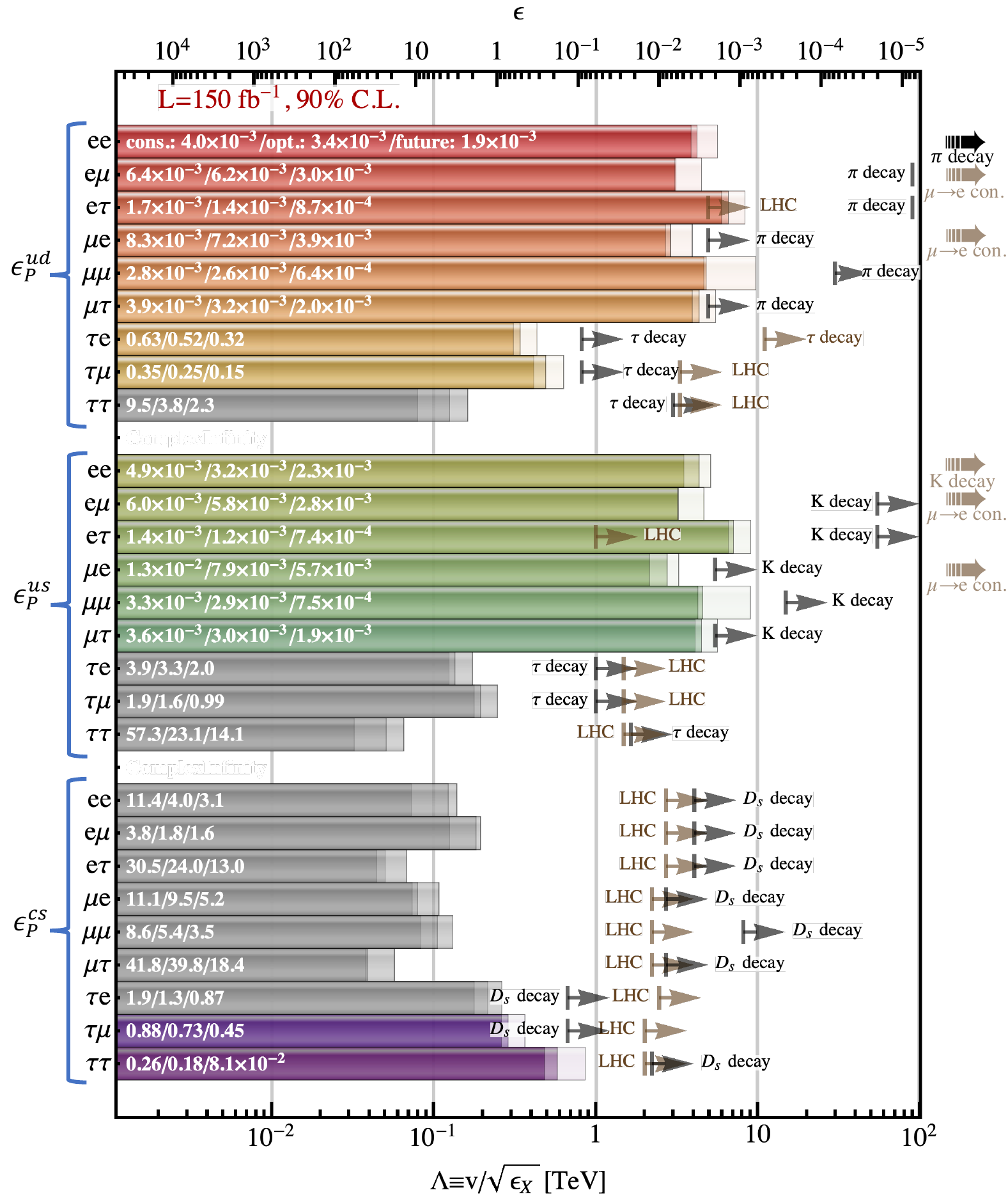}
  \caption{\textbf{Current and future projections for the sensitivity to pseudo-scalar CC operators.} The projected constraints from FASER$\nu$ experiment are shown in colored/grey bars. The black and brown arrows show the corresponding constraints from low energy meson decay experiments and high energy SMEFT constraints from LHC, respectively. Figure taken from Ref.~\cite{Falkowski:2021bkq}, see \cite{Falkowski:2021bkq} for more details.
  }
  \label{fig:EFTFASERnu}
\end{figure}

We can systematically study these new interactions at the neutrino oscillation experiments using the approach introduced in Refs.~\cite{Falkowski:2019xoe,Falkowski:2019kfn}. From the reactor neutrino experiments Daya Bay and RENO and taking into account the EFT effect on nuclear beta and inverse beta decay interactions we can get constraints on the right handed, scalar and tensor interactions using the $\bar\nu_e\to\bar\nu_e$ transition: $[\epsilon_R^{ud}]_{e\mu,e\tau}\sim(0.05-0.1)$, where $X=R,S,T$~\cite{Falkowski:2019xoe}. Using the same approach the SMEFT coefficients were studied at the FASER$\nu$ detector at LHC, where new physics effects on the neutrino production through meson decays and neutrino detection through Deep Inelastic Scattering (DIS) were studied \cite{Falkowski:2021bkq}. It was shown that FASER$\nu$ detector has the capability to constraint new physics at the multi-TeV scale, for some 4-fermion interactions giving constraints that are much better than the existing constraints in the literature (see e.g. Fig.~\ref{fig:EFTFASERnu}). 

The resulting bounds can then be connected to the parameters within a UV-complete scenario. For example, left-right symmetric models can generate the effective operators proportional to $\epsilon_R$~\cite{Beg:1977ti,Langacker:1989xa}. On the other hand, the scalar and pseudo-scalar interactions are usually generated in models with leptoquarks~\cite{Buchmuller:1986zs,Davidson:1993qk,Dorsner:2016wpm}. More generally, these Wilson coefficients can also be matched to the SMEFT parameters and, in this way, the low-energy neutrino experiments can be sensitive to heavy states in a specific BSM theory.

\subsubsection{Neutral-Current Non-Standard neutrino Interactions}
\label{subsec:inter-NSI}
As discussed in the previous section, CC operators involving neutrino fields may be tested using a variety of experiments. However, effective operators including two neutrino fields (NC-like) are harder to constrain experimentally. In the case of purely vector operators, the best bounds come from oscillation experiments, which are sensitive to modifications in the matter potential that neutrinos feel as they propagate in a medium. The effect of NC-like vector interactions involving two neutrino fields is conveniently introduced using four-fermion operators of the form
\begin{equation} 
\label{eq:NSI}
-2\sqrt{2} G_F \epsilon^{f,P}_{\alpha \beta}( \bar{\nu}_\alpha \gamma^\mu P_L \nu_\beta)( \bar{f} \gamma_\mu P f),
\end{equation} 
where $f \in \{u,d,e\}$ and $P\equiv P_L,P_R$. Similarly to the standard weak interaction between neutrinos and matter fields, these Non-Standard Interaction (NSI) operators will induce a position-dependent effective potential in matter:
\begin{equation}
    V_{NSI}(x) = \sqrt{2} G_F n_e (x) \left( 
    \begin{array}{ccc}
         1 + \mathcal{E}_{ee} & \mathcal{E}_{e\mu} & \mathcal{E}_{e\tau}\\
         \mathcal{E}_{e\mu}^* &  \mathcal{E}_{\mu\mu} & \mathcal{E}_{\mu\tau} \\
         \mathcal{E}_{e\tau}^* & \mathcal{E}_{\mu\tau}^* & \mathcal{E}_{\tau\tau}
    \end{array}\right) \, , 
    \quad \mathrm{with} \quad  
    \mathcal{E}_{\alpha\beta} \equiv \sum_{f,P} \frac{n_f(x)}{n_e(x)}\epsilon_{\alpha\beta}^{f,P}, 
\end{equation}
where $n_f(x)$ stands for the $f$ fermion density in the medium at a given point $x$ along the neutrino path, and the ``1'' in the first entry of the matrix refers to the SM contribution to the potential. At this point it should be stressed that, since neutrino oscillations are only determined by the differences of the eigenvalues of the full Hamiltonian, they are only sensitive to two linear combinations of the diagonal NSI parameters. Global fits to oscillation data have been performed for NSI with quarks~\cite{Esteban:2018ppq,Gonzalez-Garcia:2011vlg, Gonzalez-Garcia:2013usa} and set relatively strong bounds on the size of the off-diagonal NSI coefficients. For NSI with electrons, NSI limits from single experiments are available (e.g. for Borexino~\cite{Borexino:2019mhy} or IceCube~\cite{IceCubeCollaboration:2021euf}) as well as from the combination of $\nu-e$ scattering experiments and collider measurements~\cite{Forero:2011zz}. The combination of solar and reactor neutrino data has also been studied in e.g. Ref.~\cite{Bolanos:2008km}, but so far no global oscillation analysis (including also atmospheric and long-baseline oscillation data) is available in the literature. 

In the presence of NSI new degeneracies typically arise between the standard and non-standard parameters. In many cases these are only approximate and can be lifted (at least partially) through combinations of different data sets. The problem of new degeneracies arising in presence of NSI has been extensively studied in the literature (for an incomplete list, see   Refs.~\cite{Coloma:2011rq,Chatterjee:2020kkm,Coloma:2015kiu,Masud:2015xva,deGouvea:2015ndi,Liao:2016hsa,Huitu:2016bmb,Masud:2016bvp,Masud:2016nuj,Blennow:2016etl,Ge:2016dlx,Forero:2016ghr,Yasuda:2020cff,Bakhti:2016prn,Agarwalla:2016fkh,Forero:2016cmb,Denton:2020uda,Liao:2016bgf,Soumya:2019kto,Masud:2018pig,Masud:2017bcf,Rout:2017udo}). Strategies to solve them typically involve a combination of data taken at low energies or short baselines (which are less sensitive to NSI in propagation), or from different oscillation channels (which are sensitive to a different combination of parameters). As an explicit example, it has been demonstrated that even if data are compatible with the SM, the sensitivity of future long-baseline experiments to leptonic CP violation and the neutrino mass ordering can be compromised by the existence of NSI within present bounds~\cite{Coloma:2015kiu,Masud:2015xva,Liao:2016orc}. It is thus pressing to understand if complementary channels can probe NSI and resolve them. Fortunately, these degeneracies are not very relevant to present data: the current determination of leptonic CP violation and the mass ordering is dominated by T2K and reactor experiments, less sensitive to matter effects; and atmospheric constraints on NSI are strong enough to avoid large distortions~\cite{Esteban:2019lfo,Esteban:2020cvm, Esteban:2020itz,Esteban:2020opq}.

A more profound degeneracy involves the neutrino mass ordering, the solar mixing angle, and the NSI parameters. This was first observed in Refs.~\cite{Miranda:2004nb,Escrihuela:2009up} where it was found that, for large enough NSI, solar data can be explained by a solar mixing angle in the second octant: the so-called ``LMA-dark'' solution. Such degeneracy stems from an invariance at the level of the Hamiltonian describing neutrino evolution in the presence of NSI~\cite{Gonzalez-Garcia:2011vlg,Gonzalez-Garcia:2013usa,Bakhti:2014pva,Coloma:2016gei}. Although a partial breaking of this so-called ``generalized mass ordering degeneracy'' is a priori feasible by combining data with sufficiently different matter composition~\cite{Coloma:2016gei} ({\it i.e.,} $n_n(x)/n_e(x)$), global fits to oscillation data have shown that this only leads to a mild lifting of the degenerate solution~\cite{Coloma:2019mbs,Esteban:2018ppq}. Consequently, additional input is required besides oscillation data.

Neutrino scattering data is also sensitive to  effective operators in Eq.~\eqref{eq:NSI}. Most importantly, while the matter potential affecting oscillations only depends on two linear combinations of the diagonal parameters, scattering experiments are sensitive to the diagonal NSI parameters individually and thus provide complementary information. Neutrino scattering experiments such as NuTeV, CHARM, and COHERENT have measured NC neutrino scattering on nuclei and are therefore sensitive to NC-NSI with up- and down-type quarks~\cite{Davidson:2003ha,Biggio:2009nt,Biggio:2009kv,Coloma:2017egw}. Of particular relevance in this context are measurements of coherent elastic neutrino-nucleus scattering (CE$\nu$NS) at very low momentum transfers (for the derived bounds see e.g. Refs.~\cite{Coloma:2017ncl,Coloma:2019mbs,Coloma:2022avw,Liao:2022hno,Liao:2017uzy,Giunti:2019xpr,Papoulias:2017qdn}) since these are sensitive to a wider set of NSI models, as will be explained in more detail in Sec.~\ref{subsec:inter-models}. Regarding NSI with electrons, in this case the strongest bounds come from $\nu - e$ elastic scattering measurements (see e.g. Ref.~\cite{Barranco:2005ps, Forero:2011zz} and references therein). In the future, the IsoDAR proposal~\cite{Alonso:2021kyu} may be able to improve over current constraints and test NSI with electrons at the few percent level. 

Besides improving the overall sensitivity to NSI parameters, the combination of oscillation and scattering data also disfavors the LMA-Dark solution. In fact, the analysis in Ref.~\cite{Coloma:2017ncl} shows that the global combination of oscillation and COHERENT CsI data rejects it at more than $3\sigma$, for NSI involving only up- or down-quarks. However, for more general NSI models allowing NSI with up- and down-quarks simultaneously the dependence of the CE$\nu$NS cross section allows for a cancellation of the effective couplings to neutrons and protons, leading to a vanishing effect for CE$\nu$NS on a given nucleus (see e.g. Refs.~\cite{Esteban:2018ppq, Giunti:2019xpr} for a discussion). This can be partially broken with the addition of oscillation data~\cite{Esteban:2018ppq,Coloma:2019mbs}, or from the combination of CE$\nu$NS data on different nuclei, see e.g. Refs.~\cite{Scholberg:2005qs, Barranco:2005yy, Coloma:2017egw, Miranda:2020tif}. For example, the combination of oscillations, present COHERENT data and future data obtained using a silicon target at European Spallation Source~\cite{Baxter:2019mcx} will be able to discriminate between the standard LMA and LMA-Dark solutions at more than 4 $\sigma$ C.L., regardless of the relative strength for the NSI couplings to up- and down-type quarks~\cite{Chaves:2021pey}. 

NSI effects can also be induced by new scalar mediator, leading to the effective four-Fermion interaction~\cite{Ge:2018uhz}
\begin{equation}
\sqrt{2} G_F \eta_{\alpha \beta}( \bar{\nu}_\alpha  \nu_\beta)( \bar{f}  f), \label{scalar-med}
\end{equation}
where $\eta_{\alpha\beta}$ are dimensionless parameters. In a medium such as the Earth or the Sun where its constituents are non-relativistic, this term will induce an effective mass $\delta m_{\alpha\beta}=\sqrt{2}G_F\eta_{\alpha \beta} (n_f+n_{\bar{f}})$, with the same Lorentz structure as a Dirac neutrino mass.\footnote{Of course, in the Earth, in Sun and in other stars, $n_{\bar{f}}=0$ but in the early universe, $n_f=n_{\bar{f}}$.} Thus, the correction to the dispersion relation $E\simeq p +m^2/(2E)$ will be through a shift as $m^2/(2E)\to (m+ \delta m)^2/(2E) $ which is suppressed by the energy. This contrasts with the vector-mediated NSI, where the contribution to the dispersion relation is of form $E\simeq p+m^2/(2E)+V$ where $V=\sqrt{2}G_F\epsilon_{\alpha \beta} (n_f-n_{\bar{f}})$ (see Eq.~\eqref{eq:NSI}) and therefore not suppressed with the  energy. Given the current constraints, the  effects induced by the operators in Eq.~(\ref{scalar-med}) will be negligible for solar, atmospheric and terrestrial  neutrinos~\cite{Smirnov:2019cae,Babu:2019iml}. However, since the effect is proportional to $n_f+n_{\bar{f}}$, in very dense environments such as the supernova core it can be important \cite{Babu:2019iml}. Also, unlike the case of NSI induced by a vector mediator, the scalar-mediated contributions from matter and antimatter add up in the effective mass, leading to sizable effects in the early Universe~\cite{Babu:2019iml}.

\subsubsection{Model building aspects: heavy vs light mediators}
\label{subsec:inter-models}

It is worth pointing out that the interactions included in Eq.~\eqref{eq:NSI} are not gauge invariant. If these operators are obtained from a new theory at high energies, gauge invariance generically implies the simultaneous generation of similar operators involving charged leptons, for which tight experimental constraints exist~\cite{Gavela:2008ra,Antusch:2008tz}. Unless fined-tuned cancellations are invoked~\cite{Gavela:2008ra}, this makes it hard to build a model that leads to sizable NSI effects in neutrino oscillation experiments. 

A possibility to generate large NSI effects arises in radiative neutrino mass models, which explain the smallness of neutrino masses by allowing the Majorana mass terms to be induced only at the loop level. In this case, if the new mediators inside the loop have masses below the TeV scale they can induce NSI effects as well, which can have observable effects at the production, detection and oscillation of neutrinos. Reference~\cite{Babu:2019mfe} focuses on ultraviolet (UV) completion of a series of models leading to large NSI in the neutrino sector. In certain leptoquark scenarios, for example, the NSI with SM fermions are generated at tree level which leads to $\epsilon$ values as large as $\mathcal{O}(0.1)$. At the same time these values are consistent with constraints from Charged Lepton Flavor Violation (CLFV), direct collider searches as well as electroweak precision measurements, see Ref.~\cite{Babu:2019mfe} for details. 

A second possibility to avoid the tight bounds from CLFV observables is to consider that the new physics is weakly coupled to the SM, via new mediators with masses well \emph{below} the EW scale. In this regard it is relevant to note that, as a neutrino propagates in a medium, if the matter constituents are charged under the new $U(1)'$ symmetry\footnote{Throughout this document we will focus on \emph{flavored} $U(1)'$ models (that is, models where the different neutrino flavors have different charges under the new interaction) since only these can have an impact on neutrino oscillation probabilities.} the amplitude of the new interaction will be proportional to the propagator $g'^2/(q^2 + M_{Z'}^2)\to g'^2/M_{Z'}^2$ since in forward coherent scattering the momentum transfer is very small. Thus, for light enough $Z'$ masses, the new interaction may still generate large NSI-like effects even for small enough couplings to evade the bounds from charged lepton experiments. At the same time, for weakly-coupled $Z'$ models, bounds from scattering experiments with momentum transfer above the mediator mass ($q^2 \gg M_{Z'}^2$) will be suppressed since the cross section in this case is proportional to $ \sim g'^2/q^2$. For example, references \cite{Farzan:2019xor,Farzan:2017xzy,Farzan:2016fmy,Farzan:2016wym,Farzan:2015hkd,Farzan:2015doa} show that it is possible to build viable NSI models with $\epsilon_{\alpha \beta}$ as large as 0.1$-$1
({\it i.e.,} NSI with a strength comparable to that of weak interactions)  by invoking a  new light $Z'$ with a mass of $\mathcal{O}(\mathrm{MeV})$ or below. 

Oscillation bounds apply to NSI induced by light mediators as long as the range of the new interaction is shorter than the scale over which the matter density extends. This condition is typically fulfilled for $Z'$ masses above $M_{Z'} \gtrsim 10^{-12}~\mathrm{eV}$. Below this mass range the contact interaction approximation is no longer valid, as the new force becomes \emph{long-range}. For such ultra-light mediators, the impact of the new interaction can still be described in terms of a modified matter potential, using the NSI formalism. However, in this case the matter potential should be computed taking the average of the matter density within a radius $r \sim 1 / M_{Z'}$ around the neutrino position as it propagates through a medium~\cite{Grifols:2003gy,Gonzalez-Garcia:2006vic,Davoudiasl:2011sz,Smirnov:2019cae}.

\subsubsection{New neutrino interactions with light mediators}
\label{subsec:inter-LIGHTnew}

Once we abandon the EFT approach, in order to write down the renormalizable interactions between neutrino fields and new light mediators, a choice has to be made regarding the nature of the new particle introduced. A possibility is to include a new light scalar $\phi$, which is coupled to the neutrino field via a Yukawa-like interaction. In the literature, $\phi$ is sometimes referred to as a Majoron~\cite{Chikashige:1980ui,Gelmini:1980re,Schechter:1981cv}. Its coupling to neutrinos can be either lepton-number conserving ($\phi \bar{\nu}\nu$) or lepton-number violating ($\phi \bar{\nu}^c \nu$). 
A second possibility is to couple the neutrinos to a light gauge boson in the form $Z'_\mu \bar{\nu} \gamma^\mu \nu$, which arises naturally in scenarios where the SM gauge group is enlarged with a local $U(1)$ symmetry. A usual choice in this case is to gauge a linear combination of lepton-flavor and baryon numbers since it would be anomaly-free. In this case, the coupling will be flavor-diagonal (although it is also possible to generate also off-diagonal interactions, as shown for example in Ref.~\cite{Farzan:2015hkd}) but not necessarily flavor-universal, which would lead to an effect in neutrino oscillations.

Besides neutrino oscillation experiments (see Secs.~\ref{subsec:inter-NSI} and~\ref{subsec:inter-models}), the existence of new light mediators coupled to neutrinos can be probed in multiple ways.

{\it Low-energy experiments: CE$\nu$NS and $0\nu\beta\beta$} --- From a purely phenomenological perspective, an interaction of the form $\phi \bar{\nu}_e^c\nu_e$ with $\phi$ with a mass of $O(\mathrm{MeV})$ or lighter can be tested by neutrinoless double beta-decay experiments \cite{KamLAND-Zen:2012uen,Agostini:2015nwa}.
If the mediator also couples to quarks, it can be tested by Coherent Elastic $\nu$ Nucleus Scattering (CE$\nu$NS) measurements at the COHERENT experiment~\cite{Farzan:2018gtr, Coloma:2020gfv,Coloma:2022avw, Cadeddu:2020nbr,Papoulias:2017qdn,Miranda:2020tif,Abdullah:2018ykz,Corona:2022wlb,Shoemaker:2017lzs}. Relevant limits on light mediators have also been obtained from CE$\nu$NS searches using reactor neutrinos at CONUS~\cite{CONUS:2021dwh}, CONNIE~\cite{CONNIE:2019xid}, and the Dresden-II reactor experiment~\cite{Coloma:2022avw,Liao:2022hno,Sierra:2022ryd}. Proposals for similar searches have been put forward at the European Spallation Source~\cite{Baxter:2019mcx,Bertuzzo:2021opb} and at Los Alamos National Laboratory~\cite{Shoemaker:2021hvm}.

{\it Fixed-target experiments} --- The new mediator may be produced in meson decays (e.g., $~K^+(\pi^+)\to  l_\alpha^+ \nu_\beta Z'$ or $~K^+(\pi^+)\to  l_\alpha^+ \nu_\beta \phi$, where $l_\alpha \in \{e,\mu\}.$), a scenario tightly constrained by KLOE, NA62 and E494~\cite{Bakhti:2017jhm,Laha:2013xua,NA62:2021bji}. 
Such processes can also take place in the source of neutrino beam experiments such as DUNE, resulting in a minor but detectable distortion of the flavor composition and energy spectrum of the neutrino flux~\cite{Bakhti:2018avv}. If the new mediator is produced on-shell, it may also propagate to the detector and decay back into SM particles. For a $Z'$ coupled to both quarks and leptons, this is strongly constrained by CHARM\cite{Gninenko:2012eq,Tsai:2019mtm}, $\nu$-CAL~\cite{Tsai:2019mtm,Blumlein:2011mv}, NOMAD~\cite{Gninenko:2011uv,NOMAD:1998pxi} and PS191~\cite{Gninenko:2011uv, Bernardi:1985ny}. If the $Z'$ is coupled to $L_e$ very tight constraints are obtained from the electron beam-dump experiments E137, E141 and E774~\cite{Bross:1989mp, Riordan:1987aw, Bjorken:1988as,Andreas:2012mt}. Additional bounds come from searches for $\pi^0 \to \gamma Z'$ decays at NA62~\cite{NA62:2019meo}, and $e N \to e N Z'$ at NA64~\cite{NA64:2016oww,NA64:2017vtt}. For compilations of bounds from fixed-target experiments on $Z'$ in the MeV range, see e.g. Refs.~\cite{Essig:2013lka,Tsai:2019mtm,Bauer:2018onh}. 

{\it Neutrino trident searches} --- These are searches for the production of a $\mu^+ \mu^-$ pair from the scattering of a $\nu_\mu$ off the Coulomb field of a nucleus, known as neutrino trident production~\cite{Altmannshofer:2014pba}. While the strongest bounds are currently obtained from CCFR~\cite{CCFR:1991lpl} and CHARM-II~\cite{CHARM-II:1990dvf}, the DUNE Near Detector (ND) is expected to collect a large sample of trident events and to improve significantly over these~\cite{Ballett:2018uuc,Altmannshofer:2019zhy}. 

{\it Precision measurements of $\nu-e$ scattering} --- A light $Z'$ coupled to $L_e$ could be probed using $\nu-e$ scattering measurements, for which the cross section can be computed with a very good level of precision. Very competitive limits are obtained from low-energy solar neutrinos in Borexino~\cite{Harnik:2012ni}, or from reactor neutrinos at the GEMMA~\cite{Beda:2010hk} and TEXONO~\cite{TEXONO:2009knm} experiments (for recent analyses see e.g. Refs.~\cite{Lindner:2018kjo,Bilmis:2015lja}). Finally, LSND and CHARM-II data on $\nu-e$ scattering would also be sensitive to light mediators, but the constraints obtained in this case are less stringent (see e.g. Ref.~\cite{Bilmis:2015lja}). Future measurements of $\nu-e$ scattering at DUNE~\cite{Ballett:2019xoj,Chakraborty:2021apc} may also be sensitive to this scenario. 

{\it Colliders} --- For light $Z'$ masses, the most relevant constraints come mainly from: (1) $Z'$ production in $e+e-$ collisions in Babar, leading to both visible~\cite{Lees:2014xha} and invisible~\cite{Lees:2017lec} final states; (2) LHCb searches for $Z'$ bosons decaying into di-muon final states (both prompt and displaced vertex decay searches have been performed~\cite{LHCb:2019vmc}).

{\it Cosmology} --- Due to the limited lifetime of the neutron, Big Bang Nucleosynthesis (BBN) is greatly sensitive to the Hubble expansion rate and, therefore, to the number of relativistic degrees of freedom  $N_{eff}$~\cite{Fields:2019pfx,Kahniashvili:2021gym,Coffey:2020oir}. This severely constrains the possibility of having new light particles in thermal equilibrium with neutrinos at temperatures of a few MeV. On the other hand,  Large Scale Structure  (LSS) data  sets bounds on the couplings to light particles with masses at or below the eV-scale since these may hinder the free streaming of the neutrinos at matter-radiation equality Era, such as $\nu_i \to \nu_j \phi$~\cite{Basboll:2008fx,Escudero:2019gfk,Barenboim:2020vrr}. Cosmological bounds can be evaded in two very different range of mediator masses. First, in the range above $> 5~\mathrm{MeV}$ the contribution to the neutrino energy density from the $Z'$ decay into neutrinos is sufficiently suppressed by the Boltzmann factor. Second, for $M_{Z'} \ll \mathcal{O}(\mathrm{eV})$ and a very weak coupling ($g' \lesssim 10^{-10}$), the production rate of $Z'$ in the early Universe will be sufficiently small, efficiently suppressing the contribution to $N_{eff}$. 

\begin{figure}
 \centering
 	\includegraphics[scale=0.6]{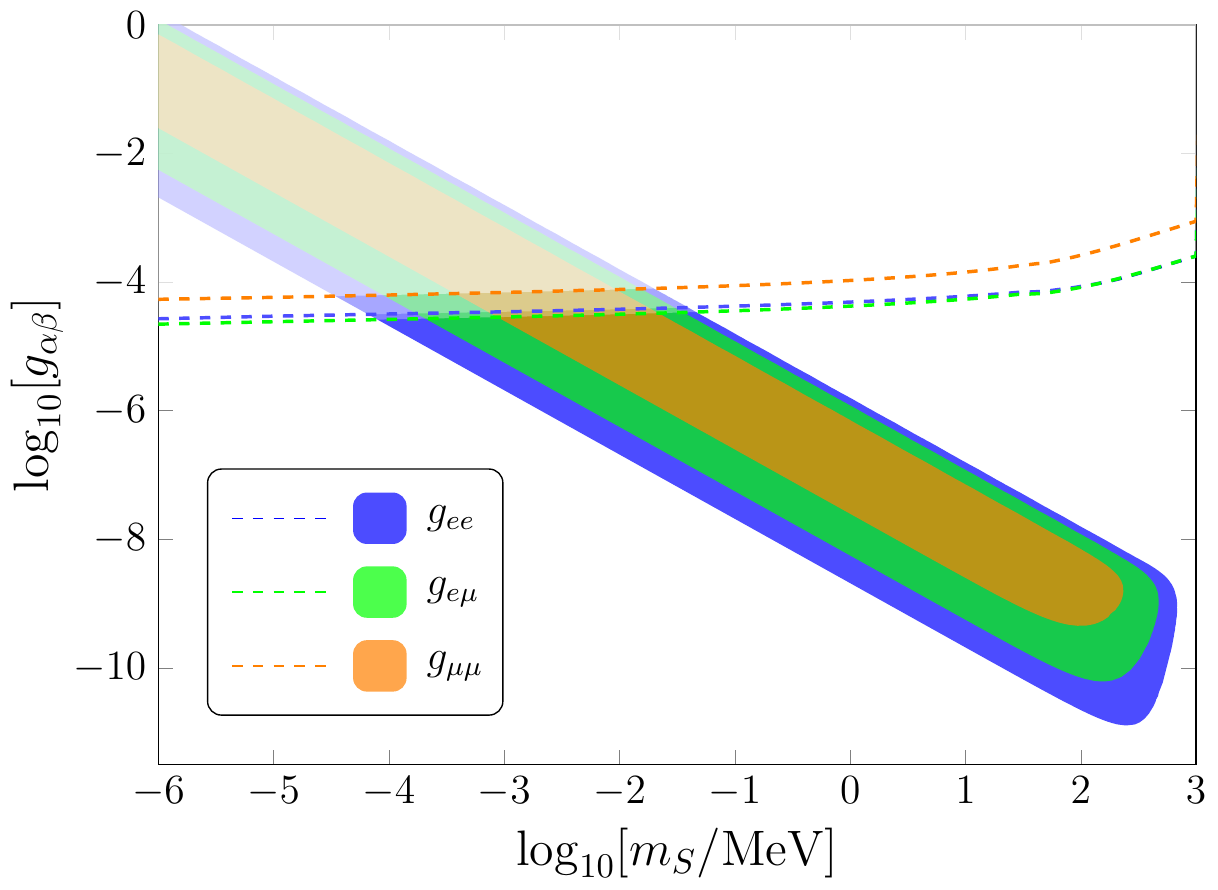}
 \caption{ \textbf{The parameter range excluded by supernova cooling consideration for a scalar, $S$, coupled to neutrinos.} In the region above the dashed lines, the new particle gets trapped inside the supernova and the obtained bounds do not apply. Figure taken from Ref. \cite{Heurtier:2016otg}. \label{Fig:supernova}}
\end{figure}

{\it Supernovae} --- New light particles with mass below $<{\rm few}\times 10$ MeV can be abundantly produced in core-collapse supernovae, which would lead to an additional cooling mechanism, see e.g. Refs.~\cite{Dent:2012mx,Chang:2016ntp,Heurtier:2016otg}. Figure~\ref{Fig:supernova}, shows the derived constraints on the coupling of a scalar mediator from supernova cooling considerations. Note that, when the coupling to the light scalar is too large, the new particles  become trapped inside the supernova core and  cannot freely transfer energy outside, so the bound vanishes. The bounds are valid as long as the new particles do not decay back to neutrinos outside the collapsing star. 
Additional effects caused by new light particles on supernovae include:
(1) If the mean free path of produced light particles inside the supernova core is comparable to the core size ({\it i.e.,} $\sim 1$~km$-$15~km), it can alter supernova core evolution. 
(2) If the decay length of new particles is much larger than the core size but smaller than the size of the progenitor star, it can warm up the outer layers, affecting shock revival and leading to observable effects.  
(3) The new couplings provide new scattering channels for neutrinos,  prolonging the neutrino diffuse time.
(4) New couplings can alter the flavor composition as well as the spectrum of emitted neutrinos which can be tested by future supernova observations. 
In the future, the observation of the neutrino flux from a nearby supernova by Hyper-Kamiokande~\cite{Hyper-Kamiokande:2016srs}, DUNE~\cite{DUNE:2020ypp} and JUNO~\cite{JUNO:2015zny} would greatly improve these constraints. 

{\it Star cooling} --- By requiring that the energy losses induced by the new interaction in stars  do not exceed those from standard neutrino emission, strong bounds can be set on its coupling strength (see e.g. Ref.~\cite{Dreiner:2013tja} for an analysis using white dwarfs). This can be used to set rather strong constraints on light $Z'$ models, see e.g. Refs.~\cite{Harnik:2012ni, Wise:2018rnb,Coloma:2020gfv,Jaeckel:2010ni}. Note, however, that these bounds typically rely on the assumption that the new mediator interacts with electrons.

{\it Fifth-force searches and equivalence principle tests} --- For ultra-light mediators ($< 10^{-13}~\mathrm{eV}$) the force induced by the new interaction becomes long-range. Fifth-force searches try to observe a deviation from the standard Newton (Coulomb) potential $\propto 1/r$ for gravitational (electromagnetic) interactions~\cite{Salumbides:2013dua,Adelberger:2009zz,Salumbides:2013aga,Schlamminger:2007ht,Lamoreaux:2012ka}. Equivalence principle tests, on the other hand, try to search for differences in the potential felt by different materials~\cite{Schlamminger:2007ht}. Bounds from both types of constraints have been recasted to flavored $U(1)'$ models in Refs.~\cite{Wise:2018rnb, Coloma:2020gfv,Harnik:2012ni}).

{\it Secret neutrino interactions} --- In general, a new light mediator may induce new neutrino-self interactions, or {\it secret neutrino interactions}, which may manifest in scenarios where neutrinos encounter high neutrino column densities.  Secret neutrino interactions are motivated as solutions to important open issues; {\it e.g.}, the origin of neutrino mass~\cite{Chikashige:1980ui, Gelmini:1980re, Georgi:1981pg, Gelmini:1982rr, Nussinov:1982wu, Blum:2014ewa}, tensions in cosmology~\cite{vandenAarssen:2012vpm, Cherry:2014xra, Barenboim:2019tux, Escudero:2019gvw}, the muon anomalous moment~\cite{Araki:2014ona, Araki:2015mya}, and the LSND anomaly~\cite{Jones:2019tow}. Secret neutrino interactions can occur when ambient neutrino densities are high, {\it i.e.}, in the early Universe~\cite{Cyr-Racine:2013jua, Ahlgren:2013wba, Archidiacono:2013dua, Forastieri:2015paa, Oldengott:2017fhy, Huang:2017egl, Escudero:2019gvw, Blinov:2019gcj}, in supernova cores~\cite{Kolb:1987qy, Shalgar:2019rqe}, or when astrophysical neutrinos propagate through the relic neutrino background over cosmological-scale distances~\cite{Kolb:1987qy, Farzan:2014gza, Ioka:2014kca, Ibe:2014pja, Ng:2014pca, Kamada:2015era, DiFranzo:2015qea, Kelly:2018tyg, Murase:2019xqi, Bustamante:2020mep, Creque-Sarbinowski:2020qhz, Esteban:2021tub}.  For details see the dedicated Snowmass White Paper, Ref.~\cite{SelfInter-Snowmass}.

\begin{figure}
 \centering
 	\includegraphics[width=0.8\textwidth]{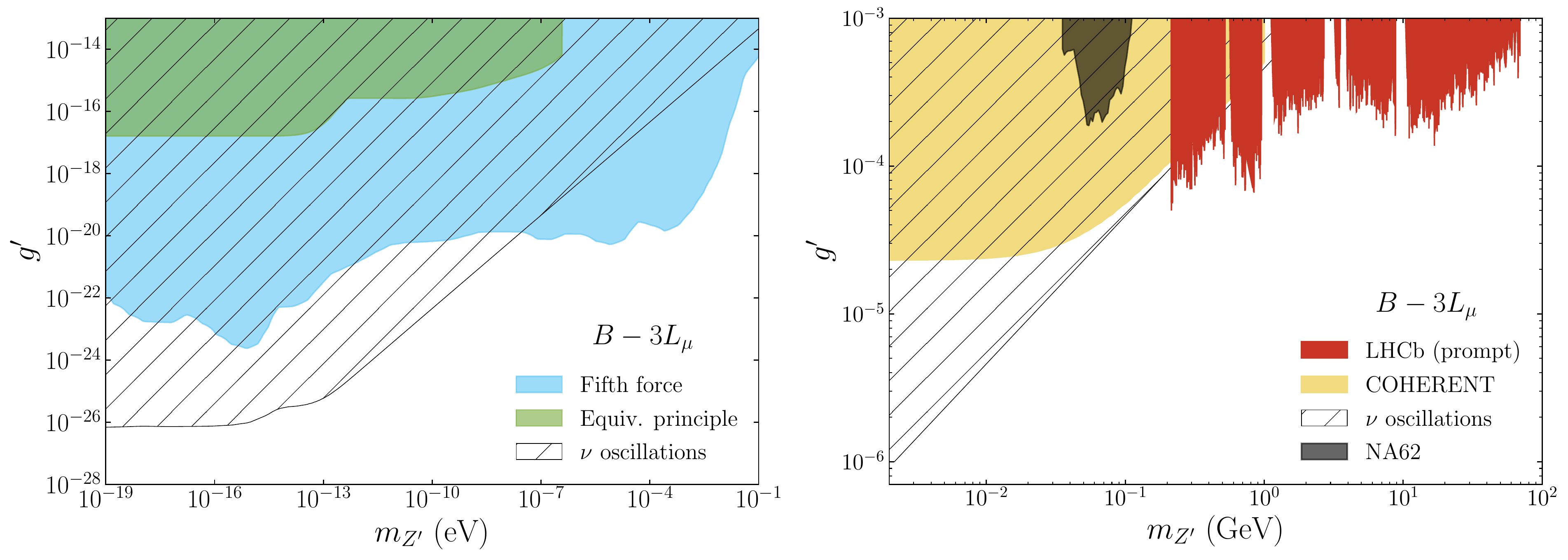}
 \caption{\textbf{Regions of parameter space ruled out from oscillation data, for a light $Z'$ arising from a model with gauged $U(1)'$ symmetry.} The left panel shows the comparison for ultra-light mediators in the long-range domain, while the right panel shows the comparison in the NSI limit (see text for details). Figure taken from Ref. \cite{Coloma:2020gfv}. Results are shown for $U(1)_{B-3L_\mu}$ as an example (see Refs.~\cite{Coloma:2020gfv, Ballett:2019xoj,Harnik:2012ni,Wise:2018rnb} for similar comparisons with other $U(1)'$ charges). \label{Fig:Zprime-bounds}}
\end{figure}

From the discussion above it becomes clear that models which are not coupled directly to $L_e$ will be constrained by fewer laboratory experiments, as in this case BaBar, and most bounds from $\nu-e$ scattering and fixed-target experiments do not apply. However, strong constraints are still obtained from COHERENT data on CE$\nu$NS as well as from neutrino oscillation data, as long as the interaction takes place with quarks. A comparison between the bounds derived from oscillations~\cite{Coloma:2020gfv,Esteban:2018ppq} to those from other experiments is shown in Fig.~\ref{Fig:Zprime-bounds}, using $U(1)_{B - 3L_\mu}$ as an example (see Refs.~\cite{Coloma:2020gfv, Ballett:2019xoj,Harnik:2012ni,Wise:2018rnb,Bustamante:2018mzu} for similar comparisons using other $U(1)'$ charges). The left panel shows the case for ultra-light mediators in the long-range domain, while the right panel shows the comparison in the NSI limit (for $m_{Z'} > 5~\mathrm{MeV}$, where cosmological bounds on $N_{eff}$ do not apply). As can be seen, oscillation data set very powerful constraints on the interaction in the both cases and considerably improves over other bounds (sometimes by several orders of magnitude).  

\subsubsection{Effective operators involving extra neutrino states}
\label{subsec:inter-newops}

So far before this section we have discussed new interactions involving SM neutrinos. 
However, we know that the SM needs to be extended with new particles to give neutrinos a mass. In this context, it is worth considering the addition of heavy sterile neutrino states to the SM particle content, since these are one of the simplest SM extensions that can serve for this purpose. In this case new operators may be generated, involving the sterile neutrino and the SM neutrino fields. In this section we will briefly discuss the phenomenological consequences of operators of dimension $d=5,6$ in this context. 

At $d=5$, the introduction of a right-handed heavy neutral lepton (HNL) allows to write a dipole moment interaction of the form  
\begin{equation}
\mu_\nu \bar{\nu_L}\sigma^{\mu\nu} N_R F_{\mu\nu}, \label{eq:diNnu} 
\end{equation}
where $\mu_\nu$ is the transition magnetic moment, $F_{\mu\nu}$ is the electromagnetic field strength tensor, and $\sigma^{\mu\nu} \equiv \frac{i}{2}\left[ \gamma^\mu, \gamma^\nu\right]$. A dipole interaction of this form leads to a multitude of phenomenological consequences in direct dark matter searches, neutrino experiments and neutrino scattering data, as well as in astrophysics and cosmology (see e.g. Ref.~\cite{Brdar:2020quo} for a recent compilation of bounds). Although such operator may be probed at neutrino oscillation experiments~\cite{Plestid:2020vqf,Coloma:2017ppo,Atkinson:2021rnp,Schwetz:2020xra} it does not have an impact on neutrino flavor, which is the scope of the present document. For details on the phenomenological consequences of this operator, see the Snowmass White Paper on Heavy Neutral Leptons, Ref.~\cite{HNL-Snowmass}.

At $d=6$, the same operators as in Sec.~\ref{subsec:inter-NSI} may be generated, replacing a light neutrino by a heavy neutrino field. For example,
\begin{equation} 
(\bar{N}\gamma^\mu \nu )(\bar{f} \gamma_\mu P f) \quad \mathrm{or} \quad (\bar{N}\gamma^\mu N )(\bar{f} \gamma_\mu P f) \, , \label{eq:VecNnuqq} 
\end{equation} 
in the case of a vector mediator. The first operator in Eq.~\eqref{eq:VecNnuqq} is generated, for example, from the mixing between the light and heavy neutrinos once the $Z$ boson is integrated out. In this case, the HNL may be produced in meson decays at fixed-target or at neutrino-beam experiments. Once the HNL has been produced, it may enter a near detector and decay through the same effective interaction back into SM particles, leaving a visible signal (see e.g. Refs.~\cite{Berryman:2019dme,Ballett:2019bgd,Coloma:2020lgy,Krasnov:2019kdc} for recent studies on the production and decay of the HNL at DUNE, or  Refs.~\cite{Jodlowski:2020vhr,Jho:2020jfz,Bakhti:2020szu,Ansarifard:2021elw} for FASER$\nu$). The second operator in Eq.~\eqref{eq:VecNnuqq}, on the other hand, leads to interactions between the HNL and the SM fermions. As an example, operators of this form were obtained in Ref.~\cite{Pospelov:2011ha} in the context of a seesaw model with an extra $U(1)_B$ symmetry. In this case, the new operators generate a new MSW potential that the new (not-so) sterile neutrinos experience while propagating in matter~\cite{Kopp:2014fha}. This scenario could also lead to novel signals in direct dark matter and solar neutrino experiments~\cite{Pospelov:2011ha,Harnik:2012ni,Pospelov:2012gm}, thanks to their non-vanishing scattering rate with nuclei.

\subsubsection{Neutrino interactions with dark matter}
\label{subsec:inter-DMbg}

Small but nonzero neutrino mass and the nature of Dark Matter (DM) are two most pressing questions before particle physicists. A coupling between neutrinos and DM would link these two mysteries to each other. In this section, we discuss the observable consequences from such couplings at neutrino oscillation experiments, which strongly depends on the mass of the DM: 

{\it PeV mass range} --- Decay of PeV dark matter into neutrino pairs has been suggested as the source of PeV neutrino events observed by IceCube. There are however alternative sources such as AGN \cite{IceCube:2021imv}, Tidal Disruption Events \cite{Winter:2021lyo,Winter:2020ptf,Biehl:2017hnb,Lunardini:2016xwi} and galactic PeVatrons \cite{Carpet-3Group:2021ygp}. Further data in future will help to reconstruct the energy spectrum and flavor ratio of events more accurately, providing a discriminant between the sources. Moreover, multimessenger methods (seeking temporal and directional  correlation with the photons of different wavelength from the same source) will be very helpful to test DM as the source of  ultra high energy neutrinos.

{\it GeV-TeV mass range} --- DM in the GeV-TeV mass range are usually WIMP candidates. The SK collaboration has performed a dedicated analysis for the WIMP induced DM annihilation in the galactic center in the mass $1-10^4$~GeV~\cite{Frankiewicz:2015zma}. Also, Ref.~\cite{Blennow:2019fhy} has studied the neutrino portal to DM in this mass range with the possibility that DM couples to the SM particles with either scalar or vector couplings.  

{\it MeV mass range} --- DM in the mass range of few MeV to tens of MeV coupled to neutrinos is well motivated DM candidate as it can be related to the neutrino mass generation mechanism~\cite{Boehm:2006mi}. For such light DM candidate the number density of DM ($n_{DM}=\rho_{DM}/m_{DM}$) will be sizable. The annihilation of galactic dark matter into neutrino pairs can lead to a  monochromatic flux of neutrinos with energy equal to the DM mass to be detected by future experiments such as Hyper-Kamiokande \cite{Olivares-DelCampo:2018pdl}, as well as the proposed liquid scintilator detector LENA~\cite{Autiero:2007zj}. As shown in \cite{Farzan:2014gza}, the scattering of relic supernova neutrinos off the background DM can lead to a dip in the energy spectrum of the diffuse supernova neutrino backgrounds which can be discerned by hyper-Kamiokande.  Finally, the interaction of neutrinos with light dark matter in the early universe can impact  the structure formation \cite{Olivares-DelCampo:2017feq}. Also, by using the Borexino data one can put an upper bound on the monochromatic antineutrino flux for DM mass in the range of 2-17 MeV~\cite{Olivares-DelCampo:2017feq}.

{\it keV mass range} --- Sterile neutrino ($\nu_s$) of a few keV mass has been widely advocated as DM candidate~\cite{Dodelson:1993je}. Its characteristic observational signature is the monochromatic $\gamma$ line from $\nu_s \to \gamma \nu_a$ which can be detected by $X$-ray observatories focused on the galactic center(s) and galaxy clusters. See e.g. Ref.~\cite{Boyarsky:2018tvu} for a recent review. 

{\it Ultralight Dark Matter} --- Thermally produced dark matter cannot be lighter than a few keV. However, if the production is not thermal the lower bound on the DM mass can be dramatically relaxed. In particular, bosonic DM can be as light as $10^{-21}-10^{-20}$ eV. For such light dark matter mass the de Broglie wavelength  exceeds the average distance between two close-by DM particles $(\rho_{DM}/m_{DM})^{-1/3}$, and the DM should be described by a classic field which oscillates in time. For example, a real non-relativistic scalar DM can be described as $\phi(x)=\sqrt{2\rho_{DM}(x)}/{m_{DM}} \cos[m_\phi(t - \vec{v}\cdot\vec{x})]$, where $\rho_{DM}$ is the DM density at spacetime coordinate $x=(t,\vec{x})$, and $\vec{v}$ is the virialized DM velocity. Coupling of neutrinos to the ultralight DM background opens up the possibility of very exciting observable effects. As shown in	Refs.~\cite{Berlin:2016woy,Krnjaic:2017zlz,Brdar:2017kbt,Dev:2020kgz,Losada:2021bxx}, a Yukawa interaction between real scalar DM with neutrinos induces a time varying effective mass in the dark matter background which can lead to time modulation of the parameters measured in neutrino experiments, averaged distortions of the neutrino oscillation probabilities, or even time-dependent modifications to the matter potential felt as neutrinos propagate. If $\phi$ is a complex scalar, we can define a current for it. The current-current interaction between $\phi$ and neutrinos also induces an effective mass  for neutrinos but without time modulation.  Such  a current-current interaction can come from gauging lepton flavor symmetries, $L_\alpha-L_\beta$ and  assigning lepton flavor to $\phi$ \cite{Farzan:2018pnk}. The Lorenz structure of the effective mass will be of form $\nu^\dagger \nu$ so like the MSW effects, it can be dominant only for very high energy neutrinos. The flavor ratio extracted from the  IceCube data already sets a bound on the current-current coupling but the predicted non-trivial flavor ratio can be tested by combining information from various extremely-high-energy neutrino telescopes (sec.~\ref{sec:ehe-neutrinos}).

Finally, neutrino interaction with DM is also advocated as a solution to cosmological tensions but recent studies show that it can only ease the so-called $\sigma_8$ tension but it cannot solve  the $H_0$ tension \cite{Mosbech:2020ahp}.

\subsection{Neutrino decay}

\textbf{Main authors:} M.~Bustamante, P.~B.~Denton, A.~M.~Gago
\vspace{2mm}

As neutrinos have mass, they have radiative decay channels, e.g., $\nu_j\to\nu_i+\gamma$ \cite{Petcov:1976ff,Marciano:1977wx}, but their expected lifetimes are far too long to be tested~\cite{Pal1982, Nieves:1982bq}.
However, if neutrinos couple to a new light or massless particle, they may decay into it, and the neutrino decay rate would be enhanced and could be probed in a variety of environments.
A relatively minimal model for neutrino decay involves a light or massless Majoron, which is a spin-0 gauge singlet with non-zero lepton number, possibly related to neutrino mass generation~\cite{Chikashige:1980ui,Gelmini:1980re,Schechter:1981cv,Acker:1991ej}.
Additional models include mirror models \cite{Maalampi:1988vs}, SUSY models \cite{Gabbiani:1990uc,Enqvist:1992ef,Aboubrahim:2013gfa}, left-right symmetric models \cite{Kim:2011ye}, neutrino masses generated by a topological formulation of gravitational anomaly \cite{Dvali:2016uhn}, unparticles \cite{Georgi:2007ek,Zhou:2007zq}, and others.

The decay rate of $\nu_i$, with mass $m_i$ and lifetime $\tau_i$ depends on the dampening factor
\begin{equation}
 \exp\left(-\frac{t}{\gamma_i \tau_i}\right)
 =
 \exp\left(-\frac{L}{E} \frac{m}{\tau}\right) \;,
\end{equation}
were $t$ is the time elapsed since the production of $\nu_i$, $L \simeq t$ is the distance traveled, $\gamma_i \equiv E/m_i$ is the Lorentz boost, and $E$ is the neutrino energy.
Neutrino decay constraints are often expressed in terms of the combination $\tau_i/m_i$, since the absolute neutrino mass scale is unknown, but the neutrino energy is; thus, the $m_i$ factor accounts for the Lorentz boost.

Phenomenologically, neutrino decay is often classified into two main categories: invisible and visible decay\footnote{Additional scenarios involving a heavier decaying sterile neutrino invoked in the context of MiniBooNE are covered in a dedicated Snowmass White Paper, Ref.~\cite{LightSteriles-Snowmass}.}.
Invisible decay is where the decay products are undetected because either they are sterile neutrinos or they are too low of energy to be detected in a given experiment.
Visible decay involves the detection of the regenerated lower-energy neutrinos.

Neutrino decay, invisible and visible, has been probed in a wide range of experiments.
In addition, the atmospheric mass ordering often has a significant impact on neutrino decay phenomenology.
Simply put, in the normal ordering neutrino decay tends to lead to a deficit in muon and tau neutrinos and a possible increase in electron neutrinos for visible decay while in the inverted ordering the opposite is expected. Invisible neutrino decay can potentially enhance Earth matter effects of core-collapse supernova neutrinos besides the similarity of average energies and fluxes between electron and other flavors of neutrinos~\cite{Delgado:2021vha}. Finally, we note that if the decay basis is different from the flavor or mass bases then additional care is needed and neutrino decay may appear as unitarity violation \cite{Berryman:2014yoa}.

\subsubsection{Theoretical Formalism - Invisible and Visible Decay }
The phenomenology of neutrino decay is related to the underlying physics model governing the decay and depends on whether the field is a scalar or pseudoscalar, whether it carries lepton number, and whether the neutrinos are Dirac or Majorana, making the phenomenology very rich.

Following from \cite{deGouvea:2019goq}, we first discuss the Dirac case.
If $\phi_n$ is a scalar field with lepton number $n$, then the Lagrangian is modified by the terms,
\begin{align}
\mathcal L
&\supset
\frac{\tilde g_{ij}}{\Lambda}(L_iH)\nu_j^c\phi_0+\frac{y_{ij}}{2}\nu_i^c\nu_j^c\phi_2+\frac{\tilde h_{ij}}{2\Lambda^2}(L_iH)(L_jH)\phi_2^*+\text{h.c.}
\nonumber \\
&=
g_{ij}\nu_i\nu_j^c\phi_0+\frac{y_{ij}}{2}\nu_i^c\nu_j^c\phi_2+\frac{h_{ij}}{2}\nu_i\nu_j\phi_2^*+\text{h.c.}\,,
\end{align}
where $\Lambda$ is the effective energy scale of the dimension-five and -six operators, $g_{ij}=\tilde g_{ij}v/\Lambda$, $h_{ij}=\tilde h_{ij}v^2/\Lambda^2$, $v$ is the Higgs vacuum expectation value, $y_{ij}=y_{ji}$ and $h_{ij}=h_{ji}$.
One can additionally have pseudoscalar couplings with an $i\gamma_5$ term included.
The decay width for mass state $i$ is \cite{deGouvea:2019goq,Coloma:2017zpg,Abdullahi:2020rge,Kim:1990km},
\begin{equation}
\Gamma=\frac{m_i^2}{16\pi E_i}\sum_{j;m_j<m_i}|g_{ij}|^2\frac{f(x_{ij})}{x_{ij}}\,,
\end{equation}
where $x_{ij}=m_i/m_j$ and $f(x)=x/2+2+2\log(x)/x-2/x^2-1/2x^3$.
The lifetime is related to the width by $\tau_i=m_i/E_i\Gamma$.
Additional expressions exist for the pseudoscalar case.
The outgoing neutrinos are emitted with a particular spectrum depending on the nature of the decay and the decay product (helicity-flipping or helicity-conserving).
For the lepton number-conserving scalar with $g_{ij}$, some of the daughter neutrinos will be right-handed and thus invisible, but none will be neutrinos. 
For the lepton number-violating scalar, the $y_{ij}$ decays will all be invisible and the $h_{ij}$ decays will all be visible, but anti-neutrinos.

If neutrinos are Majorana a similar procedure follows, with
\begin{equation}
\mathcal L\supset\frac{\tilde f_{ij}}{2\Lambda^2}(L_iH)(L_jH)\phi+\text{h.c.}
=\frac{f_{ij}}{2}\nu_i\nu_j\phi+\text{h.c.}\,,
\end{equation}
where $f_{ij}=\tilde f_{ij}v^2/\Lambda^2$ and $f_{ij}=f_{ji}$ and a neutrino will decay to both neutrinos and anti-neutrinos, both visible.

There are other ways to generate neutrino decay from more exotic scenarios, however.
E.g., in Ref.~\cite{Dvali:2016uhn}, neutrino masses are generated by a gravitational condensate which leads to an enhanced decay over the existing expected decay.

\subsubsection{Current Bounds - Invisible and Visible Decay}

\begin{figure}
\centering
\includegraphics[width=\textwidth]{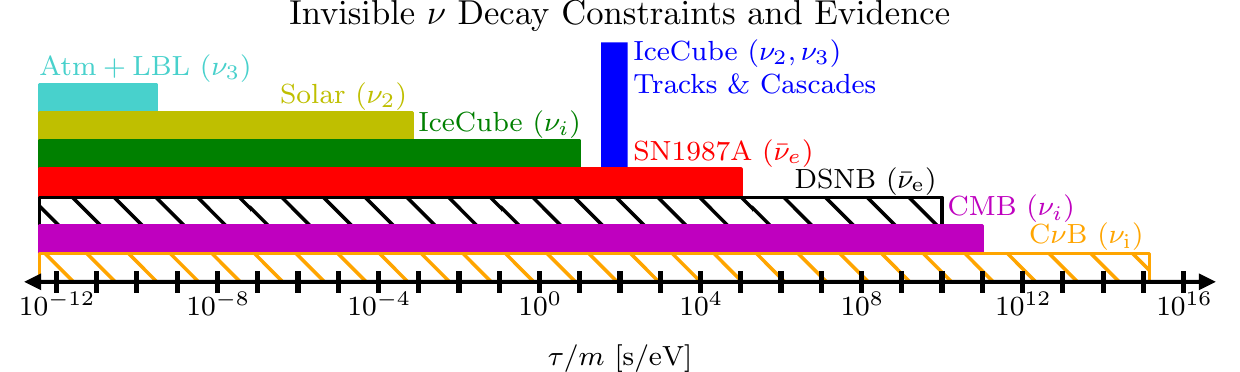}
\caption{\textbf{Constraints on invisible neutrino decay from several experiments} \cite{Abrahao:2015rba,SNO:2018pvg,Bustamante:2016ciw,Hirata:1987hu,Escudero:2019gfk,Long:2014zva}. The blue region represents a hint for neutrino decay \cite{Denton:2018aml,Abdullahi:2020rge} in IceCube data. Dashed regions represent anticipated sensitivities from future measurements. Figure adapted from \cite{Abdullahi:2020rge}.}
\label{fig:neutrino decay schematic}
\end{figure}

The bounds for invisible decay that can be imposed by the neutrinos observed from SN1987A~\cite{Kamiokande-II:1987idp} are rather loose, given that only $\nu_e$ were detected, and that there are significant uncertainties in the emission of neutrinos. In Section~\ref{sec:decay_perspectives}, we discuss the expected sensitivity to the neutrino lifetime-mass ratio from the observation of a future Galactic core-collapse supernova.

Using supernova relic neutrino observations, and considering visible decay plus normal hierarchy, an upper bound for the neutrino lifetime-mass ratio of $ \tau_2/m_2, \tau_3/m_3 \lesssim 10^{10}$ s/eV can be extracted~\cite{Ando:2003ie}. However, as noticed in Ref.~\cite{Ando:2003ie}, the robustness of these upper bounds should be treated with caution, given the large uncertainties embodied in the predictions of supernova relic neutrino flux. Further discussion on the theoretical treatment of the supernova relic neutrino flux, including the 
three-neutrino flavor transitions and decays can be found in Ref.~\cite{Fogli:2004gy}.

Additionally, measurements of the CMB constrain neutrino decay, in a somewhat model-dependent fashion, to be $\tau_i/m_i\gtrsim10^{11}$ s/eV \cite{Hannestad:2005ex,Escudero:2019gfk,Chacko:2020hmh,Escudero:2020ped}, although these bounds may be relaxed by an additional 3--4 orders of magnitude in a separate analysis \cite{Barenboim:2020vrr}.

It is known that neutrino decay can produce rather different astrophysical neutrino flavor ratios compared to those obtained when only the standard oscillation mechanism is considered~\cite{Beacom:2002vi,Beacom:2003zg}. Therefore, using the IceCube flavor ratios a lower bound  $(\tau_2/m_2,\tau_3/m_3) \gtrsim 10$ s/eV is imposed, considering normal hierarchy~\cite{Baerwald:2012kc, Bustamante:2016ciw}. These bounds are tied to the uncertainties in the prediction and measurement of the high-energy neutrino flavor composition. 

Figure \ref{fig:neutrino decay schematic} summarizes the invisible neutrino decay picture and table~\ref{tab:boundsNudec} shows  the current experimental lower bounds on the neutrino lifetime-to-mass ratio $\tau_3/m_3$ using 
data from atmospheric experiments, long-baseline experiments, and constraints on the solar anti-neutrino flux. These bounds correspond to normal ordering, i.e., $\nu_3$ is the decaying neutrino. Given that the experimental power for constraining $\tau/m$ is proportional to $L/E$, it is not surprising to see that the best limit in Table~\ref{tab:boundsNudec}, for invisible decay, involves atmospheric neutrino data: $\tau_3/m_3 \geq 2.9 \times 10^{-10}$ s/eV. It is also clear from
this table that, for the same data set (e.g., MINOS+T2K), the limits improve by one order of magnitude from invisible to visible decay. This is due to the fact that visible decay produces lower-energy active neutrinos and allows chirality-changing transitions ($\nu_i \rightarrow \bar{\nu}_j$). Within this context, it is remarkable the very stringent limit of
  $\tau_3/m_3 \gtrsim 10^{-5}$ s/eV obtained from anti-neutrino data, which represents a rather novel approach for getting it. This bound was first estimated in Ref.~\cite{Funcke:2019grs}. However, in Tab.~\ref{tab:boundsNudec} we quote the values obtained from the dedicated analysis in Ref.~\cite{Picoreti:2021yct}, which provide a more accurate result. 

\begin{center}
\begin{table}[h]
\centering
\begin{tabular}{ | C{2.5cm} ||  C{5.5cm}  C{3.0cm}  C{1.5cm} |}
 \hline
  Decay mode & Experiments & $\tau_3/m_3$(s/eV)& {Reference } \\  
 \hline
 \hline
\multirow{4}{*}{Invisible} & SK I + SK II+ K2K+ MINOS & $2.9 \times 10^{-10}$
 &  \cite{Gonzalez-Garcia:2008mgl} \\ 
 & MINOS+ T2K  & $ 2.8 \times 10^{-12}$ &  \cite{Gomes:2014yua}  \\
 & OPERA & $ 1.3 \times 10^{-13}$ &   \cite{Pagliaroli:2016zab}\\
& NOVA+T2K & $ 2.3 \times 10^{-12}$  &\cite{Choubey:2018cfz} \\
\hline
\multirow{2}{*}{Visible} 
& MINOS+T2K & $ 1.0 \times 10^{-11}$ & \cite{Gago:2017zzy}  \\
& Borexino+KamLAND & $1.0(7.0) \times 10^{-5\dagger}$ & \cite{Picoreti:2021yct}\\  
\hline
\hline
\end{tabular}
\\
\footnotesize{$^\dagger$ These two bounds corresponds to $\nu_3 \rightarrow \bar{\nu}_2 + X$ and $\nu_3 \rightarrow \bar{\nu}_1 + X$ }
\caption{\textbf{Current experimental bounds on $\tau_3/m_3$ at 90 \% C.L..} Values are provided for visible/invisible decay modes and obtained from atmospheric,
long-baseline neutrino data, and from KamLAND and Borexino anti-neutrino data, as indicated.}
\label{tab:boundsNudec}
\end{table}
\end{center}
On the other hand, when $\nu_2$ decays into invisible daughters, we find the lower bound $\tau_2/m_2 \geq 7.2 \times 10^{-4}$ s/eV at 95.45 \% C.L. in~\cite{Berryman:2014qha}, a similar value appears in~\cite{Picoreti:2015ika}, but given at 99 \% C.L., see also the similar constraint from SNO \cite{SNO:2018pvg}. Both limits are extracted using solar and/or reactor neutrino data. In~\cite{Berryman:2014qha} it is also found the lower bound 
$\tau_1/m_1 \geq 4.2 \times 10^{-3}$ s/eV, assuming $\nu_1$ as the decaying mass neutrino eigenstate. 

A number of low- to modest-significance anomalies exist which have been interpreted in the context of neutrino decay.
Starting with the hints at the shortest lifetimes, they are:
\begin{enumerate}[parsep=2pt,topsep=0pt]
\item Comparing different long-baseline accelerator experiments with sensitivity leads to a few very-low-significance hints around $\tau_3/m_3\sim10^{-12}$ s/eV:
  \begin{itemize}
    \item An analysis of MINOS and T2K data in the invisible decay scenario finds $\tau_3/m_3\simeq1.6\times10^{-12}$ s/eV at $\sqrt{\chi^2_{\text{osc}}-\chi^2_{\text{osc}+ \text{decay}}}=\sqrt{\Delta\chi^2}=2.3$~\cite{Gomes:2014yua}.
    \item An analysis of OPERA $\nu_\tau$ appearance data in the invisible decay scenario combined with existing measurements of from SK, MINOS, and T2K data finds $\tau_3/m_3\simeq2.6\times10^{-13}$ s/eV at $\sqrt{\Delta\chi^2}=1$ \cite{Pagliaroli:2016zab}.
    \item An analysis of NOvA and T2K appearance and disappearance data in the invisible decay scenario finds $\tau_3/m_3\simeq5.0\times10^{-12}$ s/eV at $\sqrt{\Delta\chi^2}=1.2$ \cite{Choubey:2018cfz}.
  \end{itemize}
\item Comparing the high-energy astrophysical track and cascade spectra from IceCube leads to a modest significance hint around $\tau_{2,3}/m_{2,3}\sim10^2$ s/eV. The track spectra at IceCube is somewhat harder than the cascade spectra \cite{IceCube:2016umi,IceCube:2018pgc} which has been interpreted in the context of invisible \cite{Denton:2018aml} and visible \cite{Abdullahi:2020rge} decay suggesting $\tau_3/m_3\simeq\tau_2/m_2\simeq85$ s/eV at $\sqrt{\Delta\chi^2}=3.4$.
\item Comparing Planck's measurements of the CMB in TT, TE, and EE correlations, lensing measurements, and BAO measurements leads to low significance hints for invisible decay around $\tau/m\sim10^{10}$ s/eV. Cosmological measurements prefer an invisible decay scenario with $\tau\simeq(2-16)\times10^9$~s $\times(m/0.05$~eV$)^3$ for all three states at $\sqrt{\Delta\chi^2}=1.3-2.0$ where the ranges depend on which data sets are included; any combination of data lead to some hint for neutrino decay \cite{Escudero:2019gfk}.
\end{enumerate}

We note that these hints are largely incompatible with each other.
That is, if, e.g., the cosmological hint is verified then the accelerator and astrophysical hints would be ruled out due to the large hierarchy in lifetimes considered.
This does not apply, however, if the model leading to neutrino decay does not apply in certain environments (e.g., early Universe, extragalactic space, etc.), in which case the shorter lifetime scenarios would become viable again.
In addition, the data here has only been examined in the invisible case except for the astrophysical data; the scenario may be different for the visible case.

\subsubsection{Future Perspectives}
\label{sec:decay_perspectives}

Table~\ref{tab:futureboundsNudec} shows the expected sensitivities for $\tau_3/m_3$, at 90 \% C.L. and for visible and invisible decay, that will be achieved in in-development or projected atmospheric, reactor, and 
long-baseline neutrino experiments. In this table, the most restrictive sensitivity limit for invisible decay is $\tau/m \sim 10^{-10}$ s/eV that would be reached in future atmospheric neutrino experiments. This expected bound is similar to the one obtained with atmospheric neutrino data (see Tab.~\ref{tab:boundsNudec}), while the sensitivity for invisible decay placed by DUNE will be improved in one order of magnitude with respect to the best one placed using long-baseline neutrino data. The latter observation will also be valid in the case of visible decay. 

\begin{center}
\begin{table}[ht]
\centering

\begin{tabular}{ | C{2.5cm} ||  C{5.0cm}  C{3.0cm}  C{2.0cm} |}
 \hline
  Decay mode & Experiments & $\tau_3/m_3$(s/eV) & Reference \\  
 \hline
 \hline
\multirow{5}{4em}{Invisible} & JUNO or RENO-50 & $ 4.7 \times 10^{-11}$ & \cite{Abrahao:2015rba}\\
& DUNE & $4.5(5.1)\times10^{-11}$$^*$  & \cite{Choubey:2017dyu,Ghoshal:2020hyo} \\
& MOMENT & $2.8\times10^{-11}$ &\cite{Tang:2018rer}\\
& ESSnuSB &  $ 4.9 (4.2) \times 10^{-11}$$^{**}$ &\cite{Choubey:2020dhw}\\
& INO-ICAL & $ 1.5 \times 10^{-10}$ & \cite{Choubey:2017eyg}  \\
& KM3NeT-ORCA & $2.5 \times 10^{-10}$ & \cite{deSalas:2018kri}  \\
\hline
\multirow{2}{4em}{Visible}& JUNO & 
 $1.0 \times 10^{-10}$ & \cite{Porto-Silva:2020gma}\\  
  & DUNE & 
 $ 2.6(3.3) \times 10^{-10}$ $^{\ddag}$
 & \cite{Coloma:2017zpg,Ascencio-Sosa:2018lbk} \\  
 \hline
\end{tabular}\\
\footnotesize{These sensitivities corresponds to: $^*$ CC events and CC+ NC events, $^{**}$ two different baselines, and 
$^\ddag$ slightly different theoretical treatments.}\\
\caption{\textbf{Expected sensitivities to $\tau_3/m_3$ at 90 \% C.L..} Values are provided for visible/invisible decay and calculated for atmospheric, reactor and 
long-baseline neutrino experiments, as indicated.}
\label{tab:futureboundsNudec}
\end{table}
\end{center}

As follows, we discuss the expected sensitivity potential for neutrino decay of astrophysical neutrinos from different sources, and that will be detected at future facilities:
\begin{itemize}
          \item  {\bf Galactic supernova neutrinos}: Within the context of visible decay it has been shown in Ref.~\cite{deGouvea:2019goq} that 
        DUNE will be able to test $\tau_i/m_i \lesssim 10^{6}$ s/eV for neutrinos emitted from a core-collapse supernova located at a distance of 10~kpc. Meanwhile, Hyper-Kamiokande will probe $\tau_i/m_i \lesssim 10^{7}$ s/eV. Both 
        sensitivities are valid for normal ($\nu_3 \rightarrow \nu_1$) and inverted ($\nu_2 \rightarrow \nu_3$) hierarchy (i.e. $i=3,2$).  
        \item {\bf High-energy astrophysical neutrinos}:  IceCube, combining the current flavor ratios and the high-energy shower rate, will be able to set a limit of $(\tau_1/m_1,\tau_2/m_2) \gtrsim 10$ s/eV in IceCube~\cite{Bustamante:2016ciw}, for inverted hierarchy. On the other hand, the hint of neutrino decay that we already mentioned, which is supported by the fact that the cascade spectrum is softer than the track spectrum,  will be probed in IceCube through the neutrino energy spectrum per flavor plus information about tau neutrinos~\cite{Denton:2018aml,Abdullahi:2020rge}. It is interesting to note that IceCube-Gen2 will have the ability to discriminate between neutrino decay and standard neutrino mixing/other non-standard ones.~\cite{Rasmussen:2017ert}. 
        \item {\bf Cosmic neutrino background}: Through the direct detection of relic neutrinos it will be possible to reach sensitivities for neutrino lifetime-to-mass ratio such as: 
      $\tau_i/m_i \sim 4.36 \times 10^{18}$s/eV for neutrino masses of \cal{O(0.1 eV)}, invisible decays and normal hierarchy ~\cite{Long:2014zva}. 
    \end{itemize}

\subsection{Tests of fundamental physics principles }
\label{sec:CPT}

\textbf{Main authors:} G. Barenboim, T. Katori, R. Lehnert
\vspace{2mm}

Poincar\'e invariance and quantum mechanics 
belong to the most foundational principles of physics.
Their synthesis is intimately associated with local Lorentz-invariant quantum field theory~\cite{Weinberg:1995mt},
which provides the framework for our best description of nongravitational physics,
the Standard Model (SM). 
Moreover, 
versions of both of these principles 
form the springboard for most theoretical explorations 
of physics beyond the Standard Model (BSM).
To identify examples of this fact 
in the neutrino context,
one has to look no further than this document:
the above sections discuss 
additional neutrino states,
nonstandard interactions,
and neutrino decay, 
which can all be formulated 
within local Lorentz-invariant quantum field theory.

The experimental nature of physics 
combined with the singular significance of spacetime symmetries and quantum mechanics 
provide ample impetus 
for the continued scrutiny of these principles. 
At the same time, 
certain BSM physics ideas based on these principles, 
and often also involving aspects of gravity, 
open exciting prospects for small, effective modifications 
to these two cornerstones at currently attainable energies.
Examples include spontaneous CPT and Lorentz breaking in string theory,
through noncommutative field theory, 
and through cosmologically varying scalars~\cite{Kostelecky:1988zi,
Kostelecky:1991ak,
Mocioiu:2000ip,
Carroll:2001ws,
Carlson:2001sw,
Anisimov:2001zc,
Alfaro:1999wd,
Kostelecky:2002ca,
Arkani-Hamed:2003pdi,
Jackiw:2003pm}.
Likewise, 
small, effective departures from quantum-mechanical coherence 
have been predicted in the context of black-hole physics~\cite{Hawking:1982dj,Ellis:1996bz}, string theory~\cite{Ellis:1992eh}, and D-brane foam~\cite{Ellis:1997jw,Benatti:1998vu}, etc.
Various formalisms for the theoretical description of these effects in presently accessible physical systems
are now available and widely used~\cite{Colladay:1996iz,Colladay:1998fq,Barenboim:2001ac,Kostelecky:2003fs,Ellis:1983jz,Lindblad:1975ef}.
In light of these latter developments,
investigations of CPT and Lorentz symmetry as well as quantum mechanics
are not only necessary,
but also acquire an element of urgency. 

The remainder of this subsection 
provides some key ideas and possibilities for testing such fundamental physics principles 
showcasing the bright future for such endeavors.

\subsubsection{CPT violation}

CPT symmetry, the combination of Charge Conjugation, Parity and Time reversal, and it guarantees particle and antiparticle have the same properties such as mass, lifetime, and mixing angles. CPT symmetry is \underline{\bf{the}} cornerstone of our model building strategy. We describe the elementary world in terms of local relativistic quantum field theories, and CPT theorem~\cite{Streater:1989vi} is based on only several assumptions from them, including Lorentz invariance, hermiticity of the Hamiltonian and local commutativity. Therefore the repercussions of potential violation of CPT will severely threaten the most extended tool we currently use to describe physics, {\it i.e.} local relativistic quantum fields.
This is precisely the reason why testing the CPT symmetry is so important. If not found, one of the three ingredients named before must be false and all our model building strategy would need to be revisited.

\begin{figure}[htb!]
    \centering
    \includegraphics[width=0.7\textwidth]{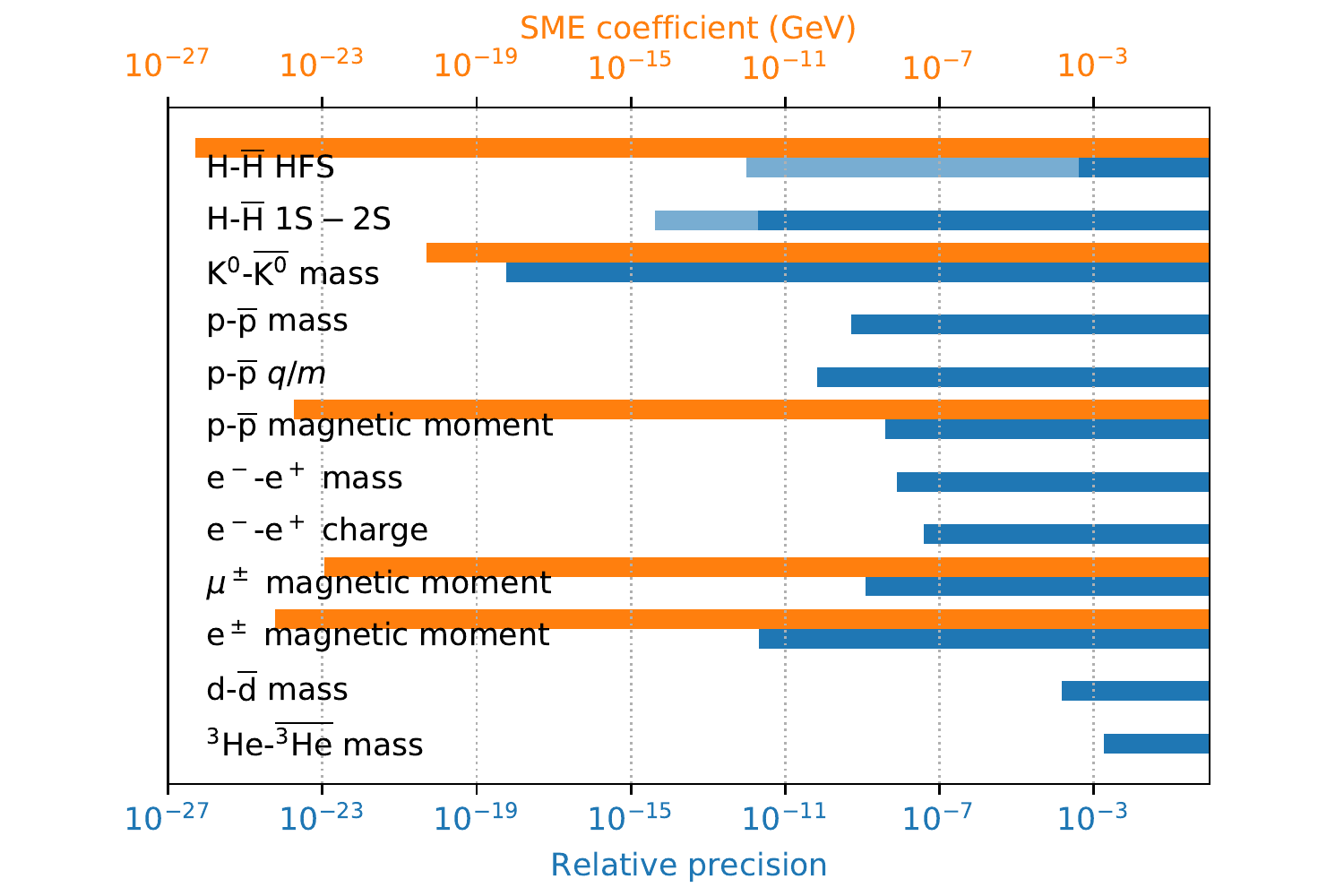}
    \caption{\textbf{Summary of recent bounds on the direct CPT tests}~\cite{Widmann:2021krf}. Blue lines represent relative experimental precision and the corresponding SME coefficients for various quantities, while orange one show the existing experimental values. Although it is not straightforward to map neutrino $\Delta m^2 - \Delta\bar{m}^2$ limits here, it is order 7 higher precision than equivalent $m^2(e^-)-m^2(e^+)$ from this figure.}
    \label{fig:CPT1}
\end{figure}

A summary of the results of the search of CPT violation can be found in  Figure~\ref{fig:CPT1}. 
The most stringent limit is quoted in the Particle Data Group~\cite{ParticleDataGroup:2020ssz} in terms of relative precision, and it comes from the neutral kaon system and seems so robust that leaves little room for imagining a CPT violating world. $|m(K^0) - m(\overline{K}^0)|/{m_K} < 0.6 \times 10^{-18}$.  

This idea, however,  may be deceptive. First, the strength of this limit arises from the choice of scale in the denominator, however, we do not have any model of CPT violation so far and therefore the choice of scale is arbitrary. We could have obtained a equally meaningful bound by choosing the Planck scale. Therefore, before having a full theory of CPT violation to address the question of which is the appropriate scale of the problem, a more meaningful bound would be;  $|m(K^0) - m(\overline{K}^0)| <0.6 \times 10^{-18} \;  m_K \; \simeq 10^{-9} \; \mbox{eV}.$
Second, the kaon is not an elementary particle and its mass is determined by QCD and therefore the most we could say with this bound is that QCD is CPT invariant. To test CPT violation in elementary particles, lepton probes are mandatory. Such a test was carried out with electrons; $|m(e^+) - m(e^-)| <8 \times 10^{-9} \; m_e \simeq  4 \times 10^{-3} \; \mbox{eV}$. 
Third, on top of that being the kaon a boson, the parameter which enters the Lagrangian is its mass squared and the bound should be re-written as $|m^2(K^0) - m^2(\overline{K}^0)| < 0.25~\mbox{eV}^2 $ while that of the electrons becomes
$|m^2(e^+) - m^2(e^-)| <  4 \times 10^{3}~\mbox{eV}^2$.

At this point it becomes evident than neutrinos can improve this bound several orders of magnitude offering the world best bound on CPT invariance. Currently, CPT limits on neutrino mass square differences are $|\Delta m_{21}^2 - \Delta \bar{m}_{21}^2| <  4.7\times 10^{-5}~\mbox{eV}^2$ and $|\Delta m_{31}^2 - \Delta \bar{m}_{31}^2| <  3.7\times 10^{-4}~\mbox{eV}^2$ ~\cite{Barenboim:2017ewj}, and there is a potential to improve it further using cosmological data~\cite{Barenboim:2017vlc}. 

\begin{figure}[t]
    \centering
    \includegraphics[width=0.35\textwidth]{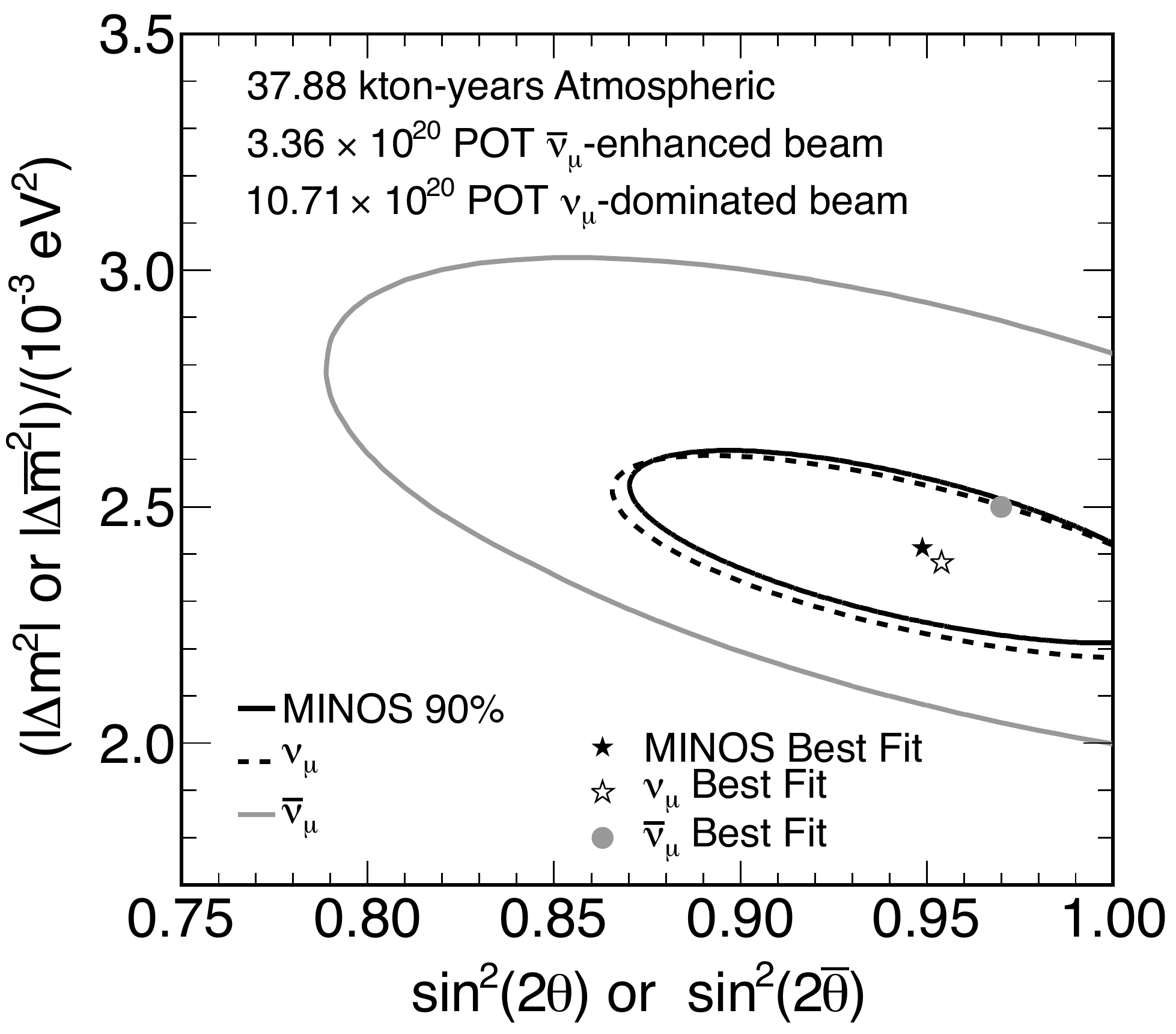}\includegraphics[width=0.55\textwidth]{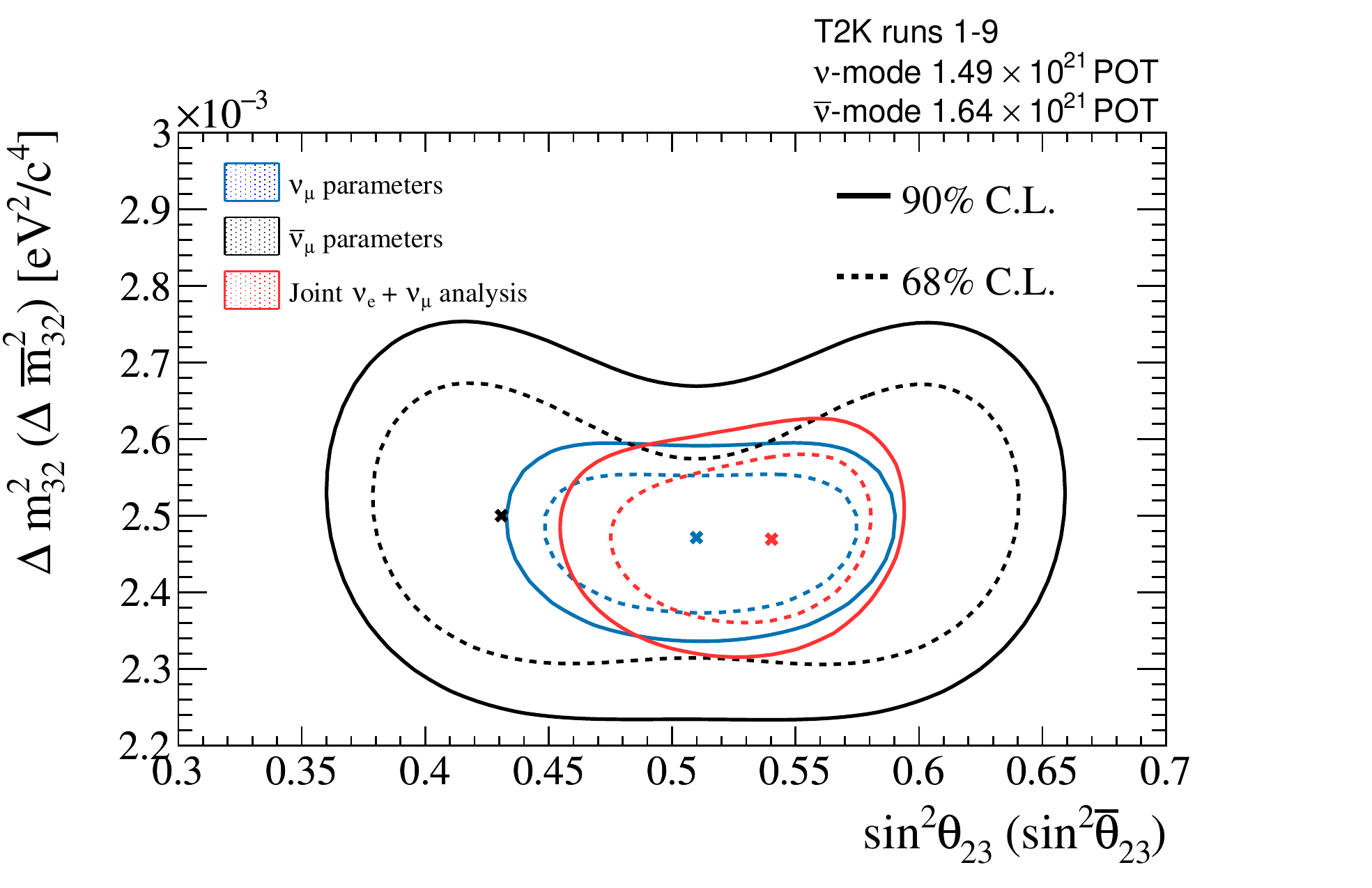}
    \caption{\textbf{Muon neutrino and muon anti-neutrino disappearance  oscillation fits by MINOS (left) and T2K (right)}~\cite{MINOS:2013utc,T2K:2021xwb}. Note that the MINOS far detector is magnetized to  perform charge separations of events, while T2K relies on simulations of neutrino and anti-neutrino dominant beams. One cannot ignore CPT symmetry to measure the CP-violation phase in neutrino oscillation experiments.}
    \label{fig:CPT2}
\end{figure}

Fig.~\ref{fig:CPT2} shows experimental results of neutrino and anti-neutrino oscillation parameter fits. In the long-baseline neutrino oscillation experiments, beams are produced either neutrino dominant or anti-neutrino dominant mode, and this and simulation are used to extrapolate neutrino and anti-neutrino oscillation parameters separately. T2K~\cite{T2K:2021xwb} relies on this to perform neutrino and anti-neutrino parameter extractions. Since they are consistent, the limit of CPT is usually given by the error dominated by statistics. A similar result would be expected from NOvA~\cite{NOvA:2021nfi}. A mild tension between T2K and NOvA oscillation measurements can be eased by nonzero CPT-odd Lorentz violation~\cite{Rahaman:2021leu}, although such possibility is severely constrained by other results~\cite{Kostelecky:2008bfz}.  MINOS~\cite{MINOS:2013utc} had a magnetized far detector and unlikely other experiments it allowed a direct comparison of neutrinos and anti-neutrinos. In future oscillation experiments, statistical errors are expected to be reduced and oscillation parameters will be systematic error dominant. We expect, for example, DUNE can push the CPT test down to $|\Delta m_{31}^2 - \Delta \bar{m}_{31}^2| <  8.1\times 10^{-5}~\mbox{eV}^2$ if the claimed statistics and systematic error are achieved~\cite{Barenboim:2017ewj}. It should also be noted, that no meaningful CP bound can be obtained before CPT symmetry  is tested to the same level~\cite{Tortola:2020ncu}.

On top of that there are plenty of reasons to believe neutrinos might be an ideal probe for CPT violation: quantum gravity is assumed to be non-local, opening the door to a potential CPT violation. Its effects however are expected to be Planck suppressed, {\it i.e.} $\left\langle v\right\rangle^2/M_{\text{P}} $, exactly in the right ballpark for neutrino experiments to see them. Besides, neutrinos enjoy  a unique mass generation mechanism, the see-saw, and therefore their masses are sensitive to new physics and new scales. Scales where non-locality can be expected to show up.  

In summary, if there is a sector where CPT violation can show up, this is in neutrino physics. Amazingly enough, this sector also offers the best bounds. CPT violation in the neutrino sector may also have important implications for the generation of the baryon asymmetry in the Universe: if CPT is violated the baryon asymmetry can be generated in equilibrium (the Sakharov conditions do not need to be fulfilled).

Note as well that  the CPT violating observable chosen is also motivated by the fact that the mass squared is the parameter entering the dispersion relation $E^2 = p^2 + m^2 $ and the natural parameter in relativistic kinematics. Therefore, deviations from this standard dispersion relation are a way to explore CPT violations triggered by Lorentz invariance breaking.

\subsubsection{Lorentz-Invariance Violation}

Lorentz symmetry guarantees 
that the orientation and state of uniform motion of physical systems
do not affect the physical properties of the system.
Small violations of this principle 
are accommodated in various approaches to BSM physics.
For example, a nonzero vacuum expectation value 
of a Lorentz tensor field
selects a preferred spacetime direction 
incompatible with Lorentz invariance.
In the absence 
of a unique and realistic underlying theory,
observable signals of such Lorentz breakdown 
may be described in a model-independent and general way 
via realistic Lagrangian effective field theory (EFT).
We remark in passing 
that Lorentz symmetry is a key ingredient 
of the CPT theorem~\cite{Streater:1989vi},
and it is therefore unsurprising 
that such an EFT also contains corrections to CPT invariance~\cite{Greenberg:2002uu}.
This type of CPT violation maintains $m=\bar{m}$, 
but typically generates different Lorentz-violating dispersion relations 
for particle and antiparticle 
and other CPT-violating effects.

The resulting EFT framework
is known as the Standard-Model Extension (SME)~\cite{Colladay:1996iz,Colladay:1998fq,Kostelecky:2003fs}. 
The SME contains both the SM and General Relativity 
along with all operators for Lorentz violation. 
This provides a powerful, realistic, and calculable framework 
for the identification of Lorentz tests
and the analysis of experimental data. 
Each Lorentz-violating term in the SME Lagrange density
is composed of a Lorentz-breaking operator 
contracted with a coefficient for Lorentz violation
that represents,
e.g., 
the nonzero vacuum expectation value
of a putative Lorentz tensor field;
these coefficients control the size of the violation, 
and each one
governs a physically distinct effect.

\begin{table}[!ht]
	\begin{center}
		{
			\begin{tabular}{|c|c|c|c|c|} 
				\hline
				Experiment     & Source                      & Type     & Limits & Ref. \\\hline\hline 
				LSND           & beam $\bar\nu_e$            & sidereal &$10^{-18}\,$GeV& \cite{LSND:2005oop}\\\hline
				MiniBooNE      & beam $\nu_\mu,\bar\nu_\mu$  & sidereal &$10^{-20}\,$GeV& \cite{MiniBooNE:2011pix}\\\hline
				T2K ND         & beam $\nu_\mu$              & sidereal &$10^{-20}\,$GeV& \cite{T2K:2017ega}\\\hline
				MINOS ND       & beam $\nu_\mu,\bar\nu_\mu$  & sidereal \& CPT&$10^{-20}\,$GeV& \cite{MINOS:2008fnv,MINOS:2012ozn}\\\hline
				Double Chooz   & reactor $\nu_e$             & sidereal &$10^{-20}\,$GeV&\cite{DoubleChooz:2012eiq}\\\hline
				Daya Bay       & reactor $\nu_e$             & sidereal \& spectrum &$10^{-20}\,$GeV& \cite{DayaBay:2018fsh}\\\hline
				MINOS FD       & beam $\nu_\mu,\bar\nu_\mu$  & sidereal &$10^{-23}\,$GeV& \cite{MINOS:2010kat,Rebel:2013vc}\\\hline
				Super-K        & atmospheric $\nu_e,\bar\nu_e,\nu_\mu,\bar\nu_\mu$&sidereal&$10^{-24}\,$GeV& \cite{Metz:2016swz}\\\hline
				AMANDA \& IceCube & atmospheric $\nu_\mu,\bar\nu_\mu$&sidereal \& spectrum&$10^{-24}\,$GeV& \cite{IceCube:2009ckd,IceCube:2010fyu,IceCube:2017qyp}\\\hline
				SNO            & solar $\nu_e$               & seasonal &$10^{-21}\,$GeV& \cite{SNO:2018mge}\\\hline
				IceCube        & astrophys. $\nu_e,\bar\nu_e,\nu_\mu,\bar\nu_\mu,\nu_\tau,\bar\nu_\tau$& flavor ratio &$10^{-27}\,$GeV& \cite{IceCube:2021tdn}\\\hline
			\end{tabular}
		}
		\caption{\textbf{Limits on SME coefficients obtained from neutrino experiments.} 
		Four types of tests are performed, 
		including sidereal-variation searches in the oscillation signals (sidereal), 
		spectrum distortions due to Lorentz-violating neutrino oscillations or mixings (spectrum), 
		seasonal variations of Lorentz-violating solar mixings (seasonal), 
		and searches for anomalous astrophysical neutrino flavor ratios (flavor ratio). 
		Quoted limits are the characteristic orders of $d=3$ CPT-odd SME best limits, 
		and details are described in the references. 
		Note that only MINOS near detector (ND) performed direct CPT tests of SME coefficients. 
		Note also that IceCube astrophysical neutrino flavor-ratio limits 
		depend on the production model. 
			\label{tab:LV}}
	\end{center}
\end{table}

All SME coefficients governing the propagation and flavor oscillations of neutrinos 
have been systematically classified and enumerated 
for operators of arbitrary mass dimension~\cite{Kostelecky:2003cr,Kostelecky:2011gq}. 
The mixing of left-handed neutrinos and their antineutrinos 
resulting from this EFT Lagrangian
is controlled by a 6$\times$6 matrix $H_{\rm LIV}^{AB}$ 
with $A,B=e,\mu,\tau,\bar{e},\bar{\mu},\bar{\tau}$.
The entries of this matrix Hamiltonian 
contain the Lorentz-violating SME coefficients.
The upper (lower) 3$\times$3 block governs the Lorentz-violating mixing of (anti)neutrinos.
As a result of possible CPT violation,
these two blocks are not identical.
Nonzero entries in the two off-diagonal 3$\times$3 pieces of $H_{\rm LIV}^{AB}$ 
are also possible
and result from the presence of Majorana-like contributions 
to the SME Lagrangian.

Several unique effects 
originate from this EFT expression for $H_{\rm LIV}^{AB}$. 
It adds distinct CPT- and Lorentz-breaking contributions
to the $\nu_a\leftrightarrow\nu_b$ 
and to the $\bar{\nu}_a\leftrightarrow\bar{\nu}_b$ mixing probabilities,
where $a,b=e,\mu,\tau$.
A qualitatively new signal is also predicted: 
the mixing $\nu_a\leftrightarrow\bar{\nu}_b$
of neutrinos with antineutrinos, 
described by the off-diagonal 3$\times$3 pieces of $H_{\rm LIV}^{AB}$. 
CPT- and Lorentz-violating contributions 
to oscillation-free neutrino propagation are also present.
Each of the three kinds of mixing effects 
contains both CPT-even and CPT-odd parts. 
An additional feature is associated with novel energy dependences 
in the mixing and propagation
with EFT operators of different mass dimension 
leading to different energy dependencies.

This large variety of interesting LV and CPTV effects 
implies that 
a correspondingly broad range of experimental parameters, 
such as neutrino-beam flavor composition, 
length,
direction, 
and energy, 
as well as detector set-up and capabilities,
all provide different SME sensitivities.
A comprehensive experimental search for CPT and Lorentz violation in neutrinos 
must therefore involve an array of different, complementary neutrino experiments,
each one providing access to a particular region of SME parameter space.

This assessment is borne out 
in past searches for CPT- and Lorentz-violating SME effects
including at
Daya Bay~\cite{Adey:2018qsd},
Double Chooz~\cite{DoubleChooz:2012eiq,Diaz:2013iba}, 
EXO-200~\cite{Albert:2016hbz}, 
IceCube~\cite{IceCube:2010fyu,IceCube:2017qyp},
LSND~\cite{LSND:2005oop}, 
MiniBooNE~\cite{MiniBooNE:2011pix,Katori:2012pe}, 
MINOS~\cite{Adamson:2008aa,Adamson:2010rn,MINOS:2012ozn,Rebel:2013vc}, 
SuperKamiokande~\cite{Messier:2004cu,Abe:2014wla}, 
and T2K~\cite{Abe:2017eot}. 
A few sample measurements 
and their overall sensitivity reach 
are listed in Table~\ref{tab:LV}.
For a subset of SME coefficients, 
sensitivities have been attained at or surpassing a level 
at which Planck-suppressed effects might be anticipated to manifest.
Nonetheless, much of the available SME parameter space
in the neutrino sector remains unexplored.

\begin{figure}[t]
	\centering
	\includegraphics[width=0.70\textwidth]{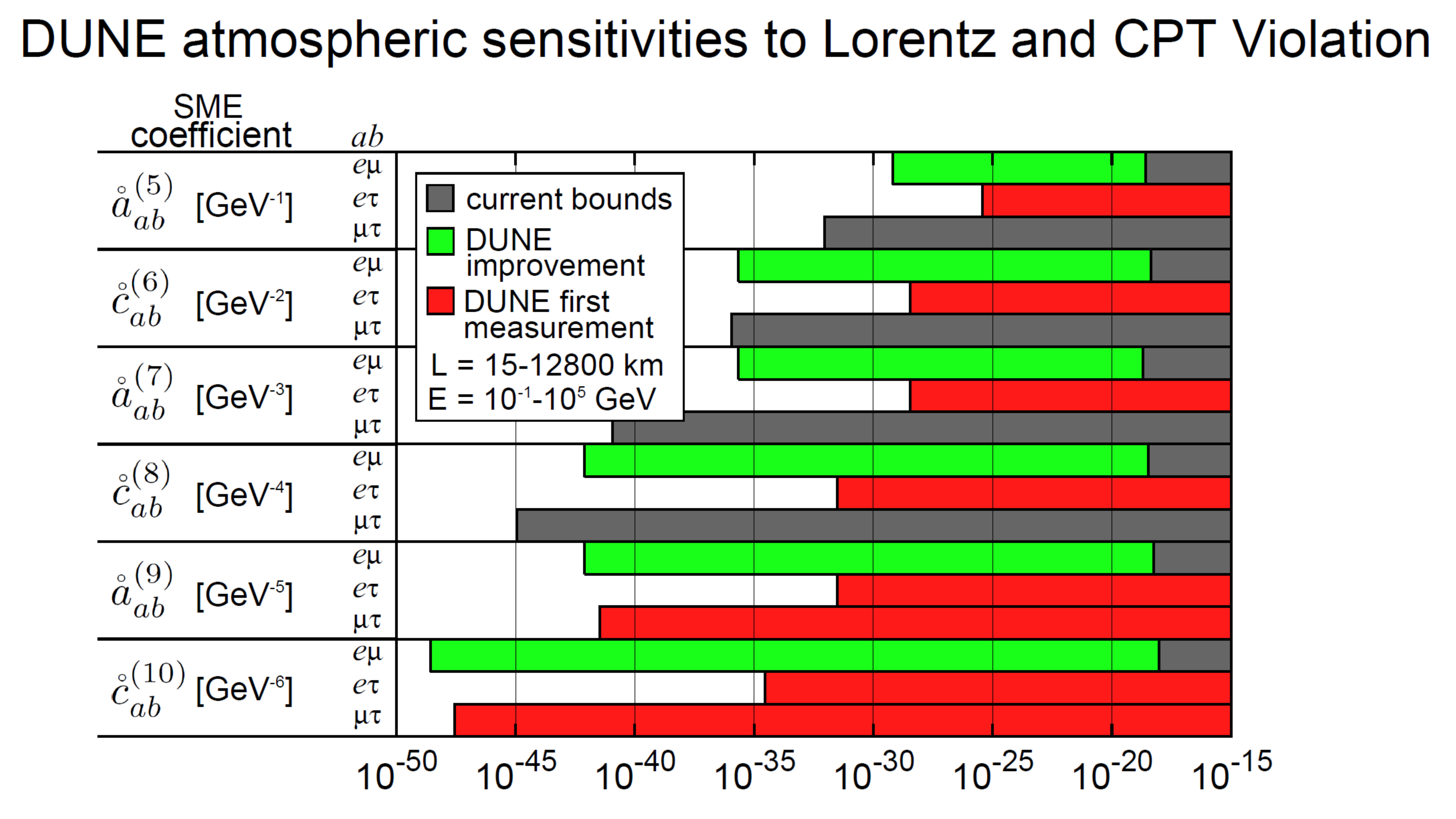}
	\caption{\textbf{DUNE atmospheric sensitivities to CPT- and Lorentz-violating nonminimal isotropic SME coefficients. }
		Studies of DUNE atmospheric neutrinos 
		can obtain first measurements (red) on some nonminimal isotropic coefficients 
		and improved results (green) on others. 
		Gray bars display existing limits from the IceCube atmospheric neutrino data~\cite{IceCube:2017qyp}. 
		Figure adopted from the DUNE Technical Design Report~\cite{DUNE:2020ypp}.}
	\label{fig:LIV1}
\end{figure}

This situation, 
coupled with the expected abundance of incoming neutrino-flavor data in the coming decade,
foreshadows a surge in future CPT and Lorentz tests 
based on neutrino-oscillation measurements. 
For example, 
DUNE
is poised to conduct various first-ever searches for SME coefficients 
and to improve existing SME measurements 
by up to six orders of magnitude
utilizing 
beam, atmospheric, and supernova neutrinos. 
A representative set of projected DUNE sensitivities 
for atmospheric neutrino-flavor investigations 
can be found in the DUNE Technical Design Report~\cite{DUNE:2020ypp} 
and is reproduced here in Fig.~\ref{fig:LIV1}.
High-energy astrophysical neutrinos observed by IceCube is similarly positioned 
for key advances in CPT and Lorentz tests~\cite{Barenboim:2003jm, Bustamante:2010nq, Borriello:2013ala,
Diaz:2013wia,
Stecker:2014xja, Tomar:2015fha, Wang:2016lne, Wei:2016ygk, Lai:2017bbl,  Ellis:2018ogq, Laha:2018hsh}.
Its detection capabilities of highest-energy atmospheric and astrophysical neutrinos 
translates directly to unmatched sensitivities 
to CPT- and Lorentz-violating operators of higher mass dimensions $d$
because their relative effects grow 
with correspondingly higher powers of the neutrino energy.
Although dependent on the neutrino production model,
recent IceCube limits on $d=6$ SME coefficients~\cite{IceCube:2021tdn}, 
for instance, 
are of order $10^{-42}$~GeV$^{-2}$.
This already surpasses na\"ive Planck sensitivity of $E_{\rm Pl}^{-2}\sim 10^{-38}$~GeV$^{-2}$ 
and is expected to be improved at 
future high-statistics neutrino telescopes, 
such as IceCube-Gen2~\cite{IceCube-Gen2:2020qha}.
In general, 
the multitude of such future studies 
will sharpen our understanding of these spacetime symmetries, 
and with their projected experimental reach, 
discovery potential for CPT and Lorentz breakdown exists.

\subsubsection{Quantum Decoherence}
Quantum mechanics states that isolated pure states can never evolve into mixed states.
An examination of a black hole's ultimate fate, on the other hand, may lead us to consider the potential of such development.
Quantum fluctuations in the gravitational field due to micro black holes may lead to the loss of quantum coherence on a microscopic level~\cite{Hawking:1975vcx}. Such quantum foam is inspired by quantum gravity theories including string theory~\cite{Ellis:1992eh,Ellis:1996bz,Ellis:1997jw,Benatti:1998vu}. Quantum theory should therefore be changed in some way if pure states may indeed evolve into incoherent mixtures. 
Fortunately it was shown that decoherence effects may be effectively implemented by introducing a phenomenological modification in the Liouville equation for the density operator of a quantum system~\cite{Ellis:1983jz}. 
Quantum dynamics semigroups and the master equation formalism may both be used to explain the temporal evolution of an open system~\cite{Lindblad:1975ef,Ohlsson:2020gxx} .
The former is a highly generic treatment for systems with non-reversible time evolution, relying on only a few assumptions: probability conservation, rising entropy with time, and full positivity. 

The decoherence term causes the damping of coherence between two mass states $i$ and $j$ ($\sim e^{-\gamma_{ij}\cdot L}$). Since the effect grows with the baseline $L$, long-baseline neutrino oscillation experiments offer unique searches of quantum decoherence in macroscopic coherent systems. Super-K oscillatory shape measurement rejected quantum decoherence as the primary source of neutrino flavor conversion~\cite{Lisi:2000zt,Super-Kamiokande:2004orf}, thus,  decoherence has been studied as a potential sub-dominant phenomenon. Limits on quantum decoherence parameters are studied from available neutrino data~\cite{Liu:1997km,Chang:1998ea,Lisi:2000zt,Benatti:2000ph,BalieiroGomes:2018gtd}. A class of models with nonzero decoherence is especially interesting because that could be used to explain neutrino data anomalies~\cite{Gago:2000qc,Barenboim:2004wu,Farzan:2008zv,Coelho:2017zes},  
or be a source of CPT violation \cite{Carrasco:2018sca}, or for certain models, it would give us a hint of the existence of non-zero CP-Majorana phase \cite{Carrasco-Martinez:2020mlg}. Neutrino coherent oscillation data are also used to test alternative quantum mechanics models~\cite{Formaggio:2016cuh,Smaldone:2021mii}.

\begin{figure}[htb!]
    \centering
    \includegraphics[width=0.45\textwidth]{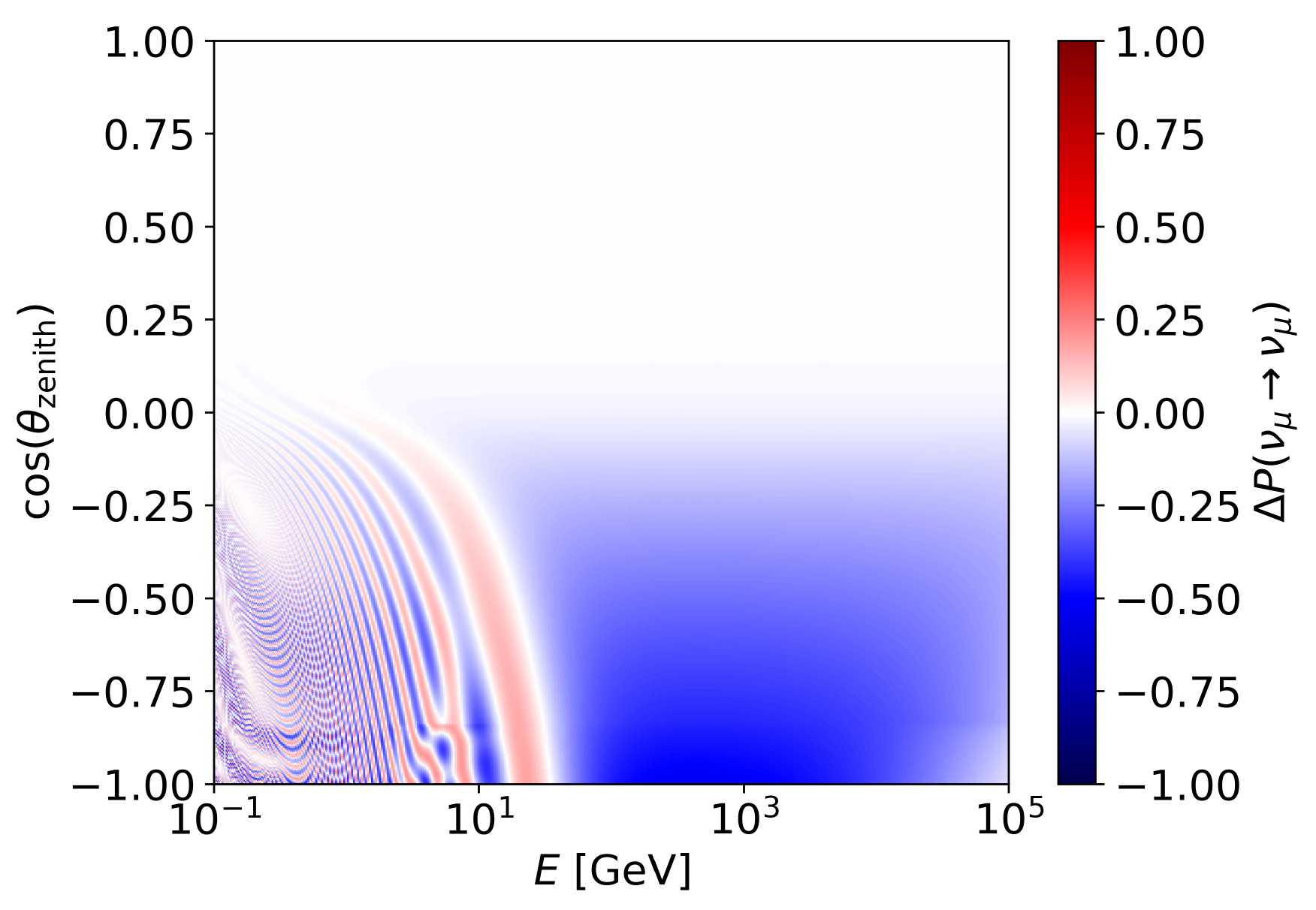}\includegraphics[width=0.45\textwidth]{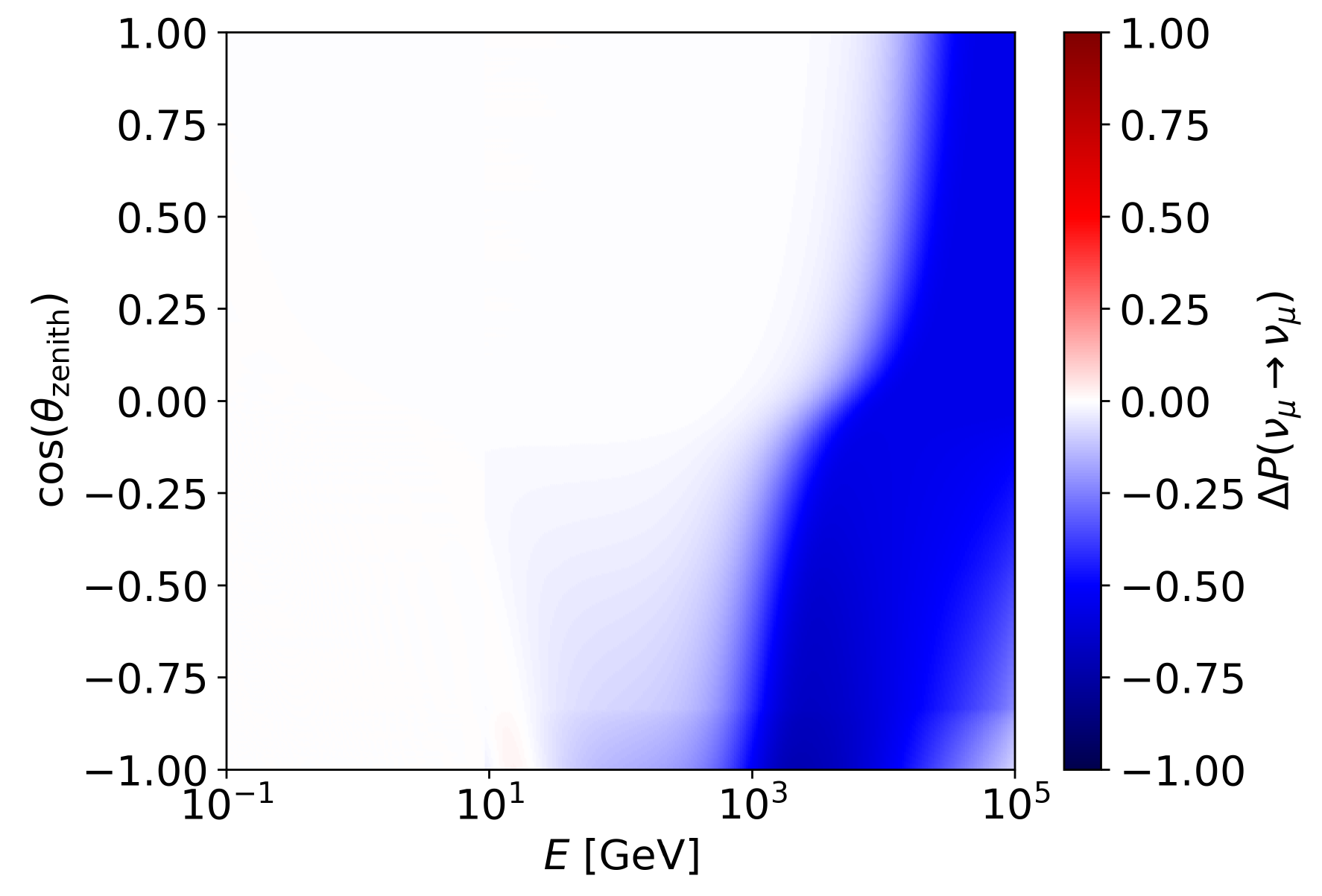}
    \caption{\textbf{Atmospheric $\nu_{\mu}+\bar\nu_{\mu}$ disappearance probability difference between the standard case and decoherence cases}. Differences are shown for $n=0$ (left) and $n=2$ (right). Here, the x-axis is energy, and the y-axis is cosine of zenith angle which is related to the propagation distance. Here, the coherent length is set $\sim 2R_{\oplus}$ and the natural scale is chosen to be 1 TeV, corresponding to $\gamma^0$ sensitivity of $1.3\times 10^{-23}$~GeV ($n=0$) and $1.3\times 10^{-29}$~GeV ($n=2$). Figures are taken from~\cite{Stuttard:2020qfv}.}
    \label{fig:decoherence}
\end{figure}

The decoherence parameter $\gamma_{ij}$ can be a function of energy $E$, $\gamma_{ij}\equiv\gamma_{ij}^0(E/GeV)^{n}$ where $n$ is normally treated as an integer. An experiment is sensitive to the decoherence when $\gamma_{ij}\cdot L\sim 1$ so the sensitivity of an experiment with the decoherence parameter $\gamma_{ij}^0$ can be written in the following way~\cite{Coloma:2018idr}.
\begin{equation}
\gamma_{ij}^0\sim 1.7\times 10^{-19} (L/km)^{-1}(E/GeV)^{-n}~GeV~.
\end{equation}
Although $n=-1, -2,\ldots$ is studied~\cite{BalieiroGomes:2016ykp}, $n>0$ is predicted by some string-theory inspired  models~\cite{Ellis:1996bz,Ellis:1997jw,Benatti:1998vu} and $n=0, 1, 2,\ldots$ are more studied. This makes high-energy neutrinos with long-baseline experiments have advantages to explore the decoherence effect.  This is part of a motivation for the DUNE to use a high-energy mode flux to explore new physics~\cite{Masud:2017bcf,BalieiroGomes:2018gtd,Carpio:2018gum}. Atmospheric neutrinos~\cite{Esmaili:2014ota,Coloma:2018idr,Stuttard:2020qfv} and astrophysical neutrinos~\cite{Hellmann:2021jyz,Stuttard:2021uyw,Ahluwalia:2001xc, Hooper:2004xr, Anchordoqui:2005gj,Klapdor-Kleingrothaus:2000kdx} have even higher energy and longer baseline which can explore further parameter spaces. Fig.~\ref{fig:decoherence} shows oscillation probability differences between the standard oscillation and decoherence cases. For $n=0$, the effect is accessible from neutrinos around $<20$~GeV. For $n=2$, interesting region are above TeV, and neutrino telescopes such IceCube, KM3NeT, and GVD have more chance to explore this case.

\section{Experimental overview and prospects}
\label{sec:exp}

A multitude of experiments are planned to detect neutrinos and set constraints on BSM physics. Here we divide them according to the energies of the neutrinos observed in each case, since this determines the experimental techniques and challenges that have to be met.
\subsection{Low energies: neutrino experiments below the GeV\label{sec:le}}

\textbf{Main authors:} I. Esteban and A. M. Suliga
\vspace{2mm}

Historically, sub-GeV neutrinos have been a key to precisely explore neutrino properties. And, as we demonstrate in this section, they will keep being vital. There are two major reasons for this. On the one hand, low-energy neutrino interactions are relatively well understood~\cite{Erler:2013xha,Tomalak:2019ibg,LlewellynSmith:1971uhs,Vogel:1999zy,Strumia:2003zx,Gonzalez-Garcia:2002bkq}. On the other hand, many natural and artificial nuclear processes produce a large flux of low-energy neutrinos. 

\begin{figure}[hbtp]
    \centering
    \includegraphics[width=0.65\textwidth]{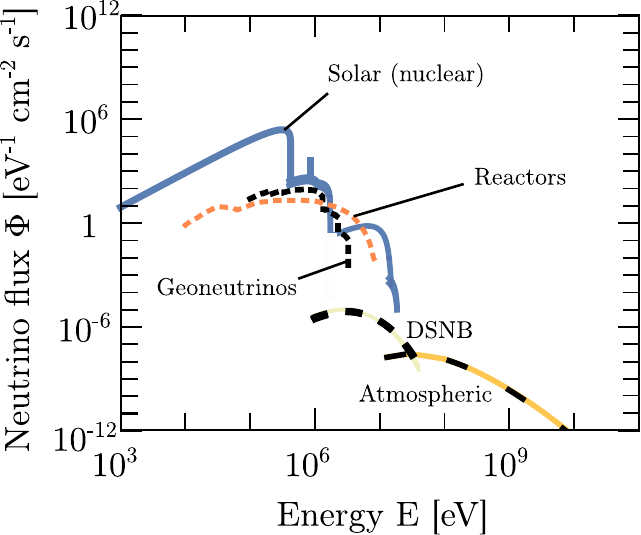}
    \caption{\textbf{Neutrino fluxes at Earth, showing that low-energy neutrinos are dominant.} We do not show the flux from a galactic supernova, that would peak at neutrino energies $\sim 10 \, \mathrm{MeV}$ and outshine the solar neutrino flux by several orders of magnitude~\cite{Mirizzi:2015eza,Scholberg:2012id}. Figure adapted from Ref.~\cite{Vitagliano:2019yzm}}
    \label{fig:subgev_flux}
\end{figure}

\Cref{fig:subgev_flux} illustrates the abundance of sub-GeV neutrinos. We show there the average all-flavor neutrino flux at Earth at energies between 1 keV and 100 GeV. Solid lines show neutrino fluxes, and dashed lines antineutrino fluxes. The reactor flux assumes a thermal power of 140~MW and a detector at 81~m. Notice that the vertical axis encompasses 24 orders of magnitude: neutrino fluxes are much greater at low energies than at high energies.

\subsubsection{Solar and reactor neutrinos}

The sub-GeV neutrino flux at Earth is dominated by solar and reactor neutrinos (c.f. \cref{fig:subgev_flux}). They also come ``for free'', i.e., they are generated by sources not specifically designed for neutrino detection, but we can take advantage of them to carry out powerful neutrino experiments. Furthermore, the fluxes are quite well-understood~\cite{Bahcall:2004pz,Huber:2011wv,Mueller:2011nm,DayaBay:2015lja,DayaBay:2018heb}, which allows for precise measurements of neutrino properties.

But, most importantly, solar and reactor neutrino experiments are \emph{complementary}. They probe similar physics, despite having very different production mechanisms, traversing different baselines with different matter densities, and being detected with different techniques. 
This complementarity extends to standard 3-neutrino oscillations~\cite{Esteban:2020cvm, deSalas:2020pgw,Capozzi:2018dat,Capozzi:2021fjo}, but also to new physics searches: many scenarios introduce different distortions in reactor and solar neutrino oscillations~\cite{Gonzalez-Garcia:2013usa,Esteban:2018ppq,Dentler:2018sju}. 

\begin{figure}[hbtp]
    \centering
    \includegraphics[width=0.65\textwidth]{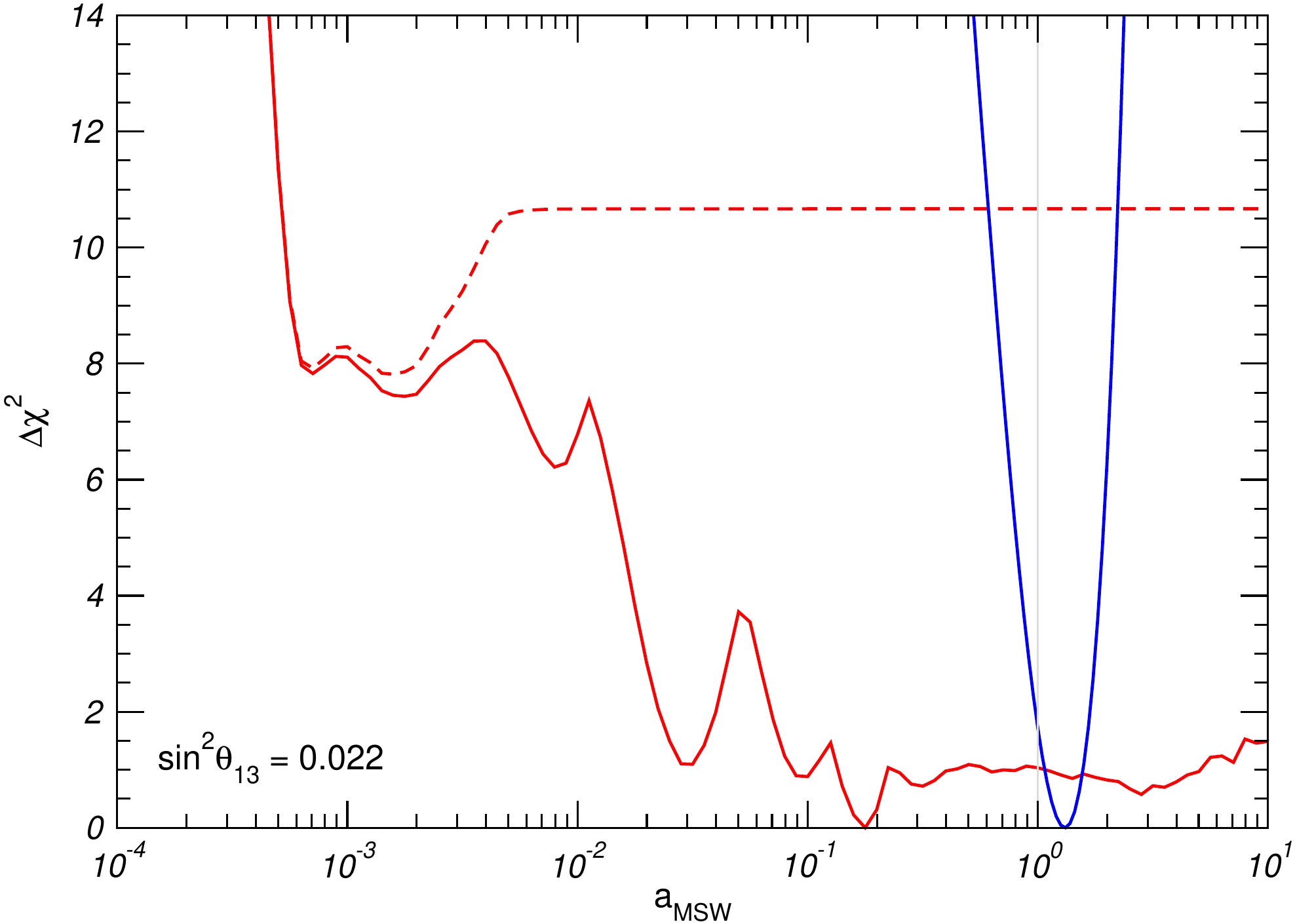}
    \caption{\textbf{Complementarity between solar and reactor neutrinos in measuring neutrino matter effects.} We show $\Delta \chi^2$ for solar (red) and KamLAND reactor (blue) data as a function of an overall rescaling of the neutrino matter potential by a factor $\mathrm{a}_\mathrm{MSW}$ (in the SM, $\mathrm{a}_\mathrm{MSW} = 1$). The red dashed line is obtained neglecting Earth matter effects. Figure adapted from Ref.~\cite{Maltoni:2015kca}.}
    \label{fig:solar_reac}
\end{figure}

\Cref{fig:solar_reac} illustrates the complementarity between solar and reactor neutrino experiments to determine new physics, in this case as an overall rescaling of the neutrino matter potential by a factor $\mathrm{a}_\mathrm{MSW}$. As we see, solar neutrino data allows for several quasidegenerate values of $\mathrm{a}_\mathrm{MSW}$, out of which reactor neutrino data picks out $\mathrm{a}_\mathrm{MSW} \simeq 1$.

This research field has been extremely successful in the last decades, and it is still ongoing. As an example, the Borexino collaboration has recently detected neutrinos from the CNO reaction chain in the Sun~\cite{BOREXINO:2020aww}. The future is even brighter: the JUNO and JUNO-TAO experiment aims to detect reactor neutrinos with unprecedented accuracy~\cite{JUNO:2015zny,JUNO:2020ijm}. 
Meantime, other multi-baseline reactor neutrino experiments, such as PROSPECT~\cite{PROSPECT:2018dtt}, STEREO~\cite{STEREO:2018rfh}, and DANSS~\cite{DANSS:2018fnn} can help to disentangle different facets of possibly complex BSM effects.
And, in the solar sector, next generation experiments, such as DUNE~\cite{Capozzi:2018dat}, Hyper-Kamiokande~\cite{Hyper-Kamiokande:2018ofw}, JUNO~\cite{JUNO:2015zny}, Super-Kamiokande-Gd~\cite{Super-Kamiokande:2021the},  SNO+~\cite{SNO:2015wyx}, and THEIA~\cite{Theia:2019non}, will collect more events with higher precision. Together, they will inaugurate the precision era of solar and reactor neutrino detection. From the theory and phenomenology side, it is essential that we scrutinize the opportunities that this program will bring about. 

\subsubsection{Exploring new avenues\label{sec:lowEnew}}
Within the last decade, solar and reactor neutrinos detected with dedicated kiloton experiments have been the paradigm of sub-GeV neutrino physics. But this does not mean that they are the only option. There exist very promising experiments that could test any possible signal in a complementary way, or provide hints for the larger-scale experiments to scrutinize~\cite{Harnik:2012ni,Plestid:2020ssy,Plestid:2020vqf}. These are mostly experiments with a different main purpose, which illustrates the trend in particle physics of employing the same experiment to look for different physics.

\paragraph{Coherent Neutrino-Nucleus Elastic Scattering}

First theorized in 1974~\cite{Freedman:1973yd}, in 2017 the COHERENT collaboration detected for the first time Coherent Neutrino Nucleus Elastic Scattering (CE$\nu$NS) using neutrinos from a neutron spallation source~\cite{COHERENT:2017ipa}. In this process, a neutrino interacts coherently with an entire atomic nucleus, boosting the cross-section by about 2 orders of magnitude. Although this would allow to detect neutrinos with smaller detectors (and, furthermore, the cross-section is quite well understood), the process is very challenging to detect as the only signature is a nucleus with a $\sim\mathrm{keV}$ recoil energy. However, recent developments in detector technologies motivated by dark matter searches have finally made detecting this process feasible.

The first detection has triggered a very intense phenomenological activity. A non-exhaustive list of BSM
topics covered includes bounds on NSI~\cite{Coloma:2017ncl,Dent:2016wcr,Liao:2017uzy,Dent:2017mpr,Farzan:2018gtr,Abdullah:2018ykz,Coloma:2019mbs,AristizabalSierra:2018eqm,Shoemaker:2017lzs,Giunti:2019xpr,Denton:2018xmq,Coloma:2019mbs,Flores:2020lji} (see also Sec.~\ref{sec:inter}),
constraints on neutrino electromagnetic properties
\cite{
Papoulias:2017qdn, Billard:2018jnl,Cadeddu:2018dux,Miranda:2019wdy,Papoulias:2019txv}, sterile neutrino
searches~\cite{Blanco:2019vyp,Kosmas:2017zbh}, or searches for new weakly-interacting
particles from a hidden sector~\cite{Brdar:2018qqj,Ge:2017mcq,Dutta:2020vop,Dutta:2019nbn}.  On the other hand,
standard physics studies include new constraints on the weak mixing
angle \cite{Canas:2018rng,Cadeddu:2018izq,Huang:2019ene} at very low momentum transfer, as well as
studies of the nuclear structure factors of the target nuclei
~\cite{Cadeddu:2017etk,Ciuffoli:2018qem,Cadeddu:2018izq,Cadeddu:2019eta,Papoulias:2019lfi,Khan:2019cvi,Huang:2019ene,Canas:2019fjw,Cadeddu:2020lky,Miranda:2020tif,Coloma:2020nhf}. This illustrates that these experiments can cross-check possible anomalies or explore new avenues. Furthermore, there is a large variety of proposed future CE$\nu$NS experiments, with both neutron spallation sources (COHERENT~\cite{COHERENT:2019kwz}, 
Coherent CAPTAIN-Mills~\cite{CCM:2021leg,Shoemaker:2021hvm}, ESS-based experiment~\cite{Baxter:2019mcx}) 
and reactor sources 
(CONNIE~\cite{CONNIE:2019swq}, 
CONUS~\cite{CONUS:2020skt}
NEON~\cite{Lee:2021wsh}, 
NUCLEUS~\cite{NUCLEUS:2019igx}
MINER~\cite{MINER:2016igy}
RED-100~\cite{Akimov:2017hee},
RICOCHET~\cite{Ricochet:2021rjo},  
TEXONO~\cite{Kerman:2016jqp}). 
Thus, we should understand the complementarity between different techniques and which approach is best at solving each problem. For additional details on the current and future prospects for CE$\nu$NS measurements, please see the dedicated Snowmass White Paper~\cite{CEvNS-Snowmass}.

\paragraph{Dark matter detectors}

Direct detection dark matter experiments are approaching the multi-ton scale. At this level, they should start being sensitive to solar neutrinos 
(DARWIN~\cite{DARWIN:2016hyl, DARWIN:2020bnc}, 
DarkSide-20k~\cite{Franco:2015pha,DarkSide-20k:2017zyg}). It is therefore very appealing to explore the capabilities of these experiments that, although not mainly built to explore neutrino physics, will provide useful data. Furthermore, they will be uniquely sensitive to low-energy pp neutrinos.

\begin{figure}[hbtp]
    \centering
    \includegraphics[width=\textwidth]{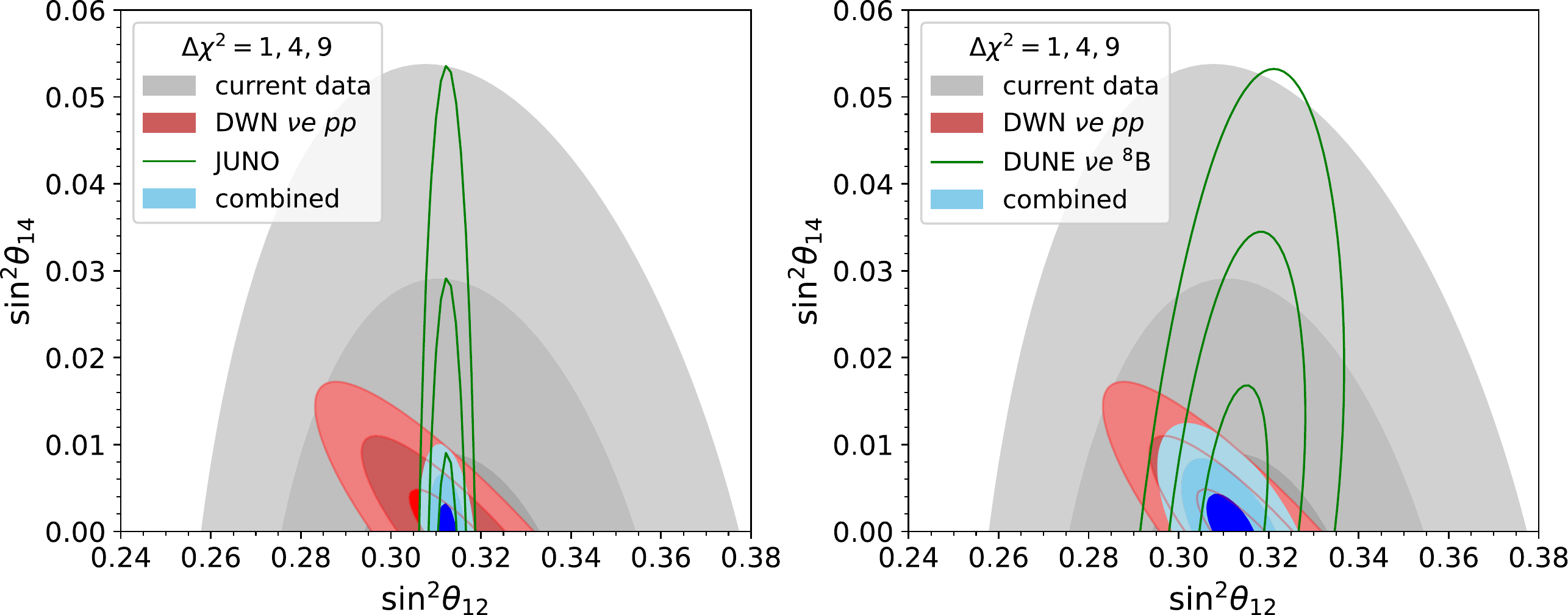}
    \caption{\textbf{Complementarity between different probes of low-energy neutrinos in exploring sterile neutrinos.} Both panels show the sensitivity on the standard mixing angle $\theta_{12}$ and the sterile mixing $\theta_{14}$. As we see, different experiments probe different directions in parameter space. The best results are thus obtained combining them. Figure from Ref.~\cite{Goldhagen:2021kxe}.}
    \label{fig:compl_sterile}
\end{figure}

\Cref{fig:compl_sterile} illustrates the complementarity between reactor neutrino experiments and solar neutrino data from DUNE and the DARWIN dark matter experiment (labelled as DWN). The figure shows the combined sensitivity on $\theta_{12}$ and the mixing angle $\theta_{14}$ parametrizing the mixing with putative sterile neutrinos. 

Exploring sterile neutrinos in $\nu_e$ detection has become more pressing now that the Gallium anomaly has reached the $> 5 \sigma$ level~\cite{Barinov:2021asz,Berryman:2021yan}. Solar neutrino data exclude an sterile neutrino interpretation of the anomaly with $\sim 3 \sigma$, and provide the only constraint in a large sterile neutrino mass range. It is therefore very appealing that a combination of different experimental approaches (\emph{a priori} with different systematic uncertainties) will significantly increase the current sensitivity; and it could be interesting to explore if the complementarity also extends to alternative models that may explain the Gallium anomaly.

\paragraph{Beta-decay spectral measurements} Motivated by sensitivity to the absolute neutrino-mass scale, experiments like KATRIN~\cite{KATRIN:2021uub}, Project 8~\cite{Project8:2017nal}, ECHo~\cite{Gastaldo:2017edk} and HOLMES~\cite{Alpert:2014lfa} aim to measure $\beta$ and electron-capture spectra with unprecedented accuracy near the spectral endpoint. Due to their unprecedented precision, they offer complementary probes of many scenarios discussed in Sec.~II~\cite{Arcadi:2018xdd,Diaz:2013saa, Lehnert:2021tbv,Steinbrink:2017uhw,KATRIN:2022ith}.

\subsubsection{Supernova neutrinos}

The detection more than 30 years ago of neutrinos from SN1987A~\cite{Kamiokande-II:1987idp,Bionta:1987qt,Alekseev:1988gp} was a breakthrough in neutrino physics~\cite{Ando:2003is,Balantekin:2007xq,deGouvea:2012hg,Raffelt:2011nc,Heurtier:2016otg,Kachelriess:2000qc,Farzan:2002wx} and astrophysics~\cite{Janka:2006fh,Horiuchi:2017qja,Loredo:2001rx}. 25 $\bar{\nu}_e$ events detected Kamiokande, IMB, and Baksan were enough to deserve a Nobel prize, open a new research field and trigger phenomenology activity that continues today. As the galactic core collapse supernova rate is estimated to be around 3 per century~\cite{Diehl:2006cf,Adams:2013ana}, we must make sure that we do not miss the next event and that we take the most out of the data~\cite{Lund:2010kh,Scholberg:2012id,Tamborra:2013laa,Tamborra:2014hga,Mirizzi:2015eza,Nakamura:2016kkl,Horiuchi:2018gkc,Horiuchi:2018ofe,Nakazato:2020ogl}.

The findings of numerous studies indicate that BSM physics may considerably alter the evolution of core-collapse supernovae. In turn, these supernovae should serve as compelling testing facilities to discover or constrain BSM physics. This includes but is not limited to sterile neutrinos~\cite{Nunokawa:1997ct, Hidaka:2006sg, Hidaka:2007se, Fuller:2008erj, Tamborra:2011is, Raffelt:2011nc, Warren:2014qza, Arguelles:2016uwb, Rembiasz:2018lok, Xiong:2019nvw, Suliga:2019bsq, Syvolap:2019dat, Mastrototaro:2019vug, Tang:2020pkp, Suliga:2020vpz}, non-standard mediators coupling to neutrinos~\cite{Kolb:1987qy, Heurtier:2016otg, Shalgar:2019rqe, Reddy:2021rln}, or also charged fermions~\cite{Chang:2016ntp, Croon:2020lrf,Caputo:2021rux} and quarks~\cite{Farzan:2018gtr, Suliga:2020jfa, Cerdeno:2021cdz, Huang:2021enl}, where any flavor-changing scenarios play a role as well. But core-collapse supernovae are complex phenomena that incorporate symmetries and interplays between particle, nuclear physics as well as hydrodynamics~\cite{Burrows:2020qrp}. Due to this, to make sure we get reliable limits, modeling of the source evolution has to be performed meticulously, taking into account the arising feedback from invoked BSM physics, as that might drastically modify the obtained limits, e.g.,~\cite{Suliga:2020vpz}.

Luckily, for the next Milky Way core collapse supernova we will have much better detectors~\cite{SNEWS:2020tbu} that will also complement advances in astrophysics. Furthermore, we should be able to detect all neutrino flavors: Super-Kamiokande~\cite{Super-Kamiokande:2007zsl} (or its successor, Hyper-Kamiokande~\cite{Hyper-Kamiokande:2018ofw,Hyper-Kamiokande:2021frf}) will lead the statistics by detecting $\bar{\nu}_e$, DUNE will detect $\nu_e$~\cite{DUNE:2020ypp}, and JUNO~\cite{JUNO:2015zny} together with large-scale direct dark matter and CE$\nu$NS detectors will observe other flavors. By looking at the excess over large background rates, IceCube and KM3NeT should also be able to detect supernova neutrinos~\cite{Kopke:2017req,ColomerMolla:2020qds}. It is thus interesting to explore the complementarities that detecting different flavors will bring about, and whether uncertainties related to collective neutrino flavor effects can be overcome (see, e.g., Refs.~\cite{Dighe:1999bi,Dighe:2003be,Ando:2004qe,deGouvea:2019goq}). In addition, this multi-channel observation will serve well in probing and constraining non-standard physics in the neutrino sector, described in the previous paragraph.

Finally, even if we are unlucky and a galactic supernova does not take place within the near future, supernovae have been exploding throughout the entire history of the Universe, generating a Diffuse Supernova Neutrino Background (DSNB)~\cite{seidov,Krauss:1983zn,Wilson:1986ha}.
Its detection is also valuable by itself regardless of the occurrence of the next galactic supernova, as it is essential to observe neutrinos from multiple supernova events, which is by definition the DSNB, to understand the whole core-collapse supernova population~\cite{Lunardini:2009ya, Lunardini:2012ne, Moller:2018kpn, Kresse:2020nto, Horiuchi:2020jnc}. It also opens up a window to probe non-standard interactions occurring between DSNB and cosmic neutrino background~\cite{Jeong:2018yts} or dark matter~\cite{Farzan:2014gza}. 
Although backgrounds  affecting the detection of this flux are large, the enrichment of Super-Kamiokande with gadolinium should dramatically reduce them and make detection of $\bar{\nu}_e$ from the DSNB an actual possibility~\cite{Beacom:2003nk,Kaplinghat:1999xi,Ando:2002ky,Fukugita:2002qw,Keil:2002in,Lunardini:2010ab,Beacom:2010kk,Super-Kamiokande:2021the}. In the future, Hyper-Kamiokande~\cite{Hyper-Kamiokande:2018ofw}, DUNE~\cite{DUNE:2020ypp} and JUNO~\cite{JUNO:2015zny} could detect both $\bar{\nu}_e$ and $\nu_e$; and together with dark matter detectors put an upper bound on contributions from other flavors~\cite{Tabrizi:2020vmo,Suliga:2021hek}. Both the generic DSNB detection as well as its flavor content would bring about interesting physics studies~\cite{Ando:2003ie,Fogli:2004gy,DeGouvea:2020ang,Tabrizi:2020vmo}.
\subsection{Medium energies: neutrinos with energies between the GeV and TeV scales\label{sec:me}}

\textbf{Main authors:} D. V. Forero and M. Ross-Lonergan
\vspace{2mm}

As we increase in neutrino energy, we enter the realm of man-made neutrino sources from accelerators, and atmospheric neutrino measurements. In this section we focus on the energy range between the GeV and TeV scales, where oscillation effects are observable, leaving for Sec.~\ref{sec:he} the discussion of ultra-high-energy neutrinos and neutrino telescopes. 

\subsubsection{Past and Contemporary Neutrino Beam Experiments}
\label{sec:me-beams1}

At neutrino beams, high-energy proton beams are collided with a stationary target to produce charged mesons (predominantly pions and kaons). These are then focused inside a magnetic horn where their subsequent decays produce a relatively collimated and intense beam of neutrinos. By choosing the polarity of the magnetic focusing horns either positive ($\pi^+$,$K^+$) or negative $(\pi^-$, $K^-$) mesons can be selectively chosen, leading to a predominately neutrino ($\nu_\mu$) or anti-neutrino ($\overline{\nu}_\mu$) beam respectively. The fact that the initial proton beam, the magnetic horn and target are all carefully designed and controlled means that the resulting spectrum of neutrino species and kinematics can be well understood in comparison to many natural sources of neutrinos. This is further bolstered by the use of a near-detector, located close to the neutrino beam source, that measures the expected spectrum of neutrino well before any oscillations are expected to take place under the assumption of three neutrinos. Precise measurement of the neutrino interactions at the near detector allows the spectrum at the far detector be estimated to a much higher precision, leading to higher sensitivity to oscillatory effects that take place in the interim. This section summarizes past and contemporary neutrino beam experiments which could be sensitive to BSM effects on neutrino flavor, while a review of future proposals is provided in Sec.~\ref{sec:me-beams2}. 

\paragraph{MINOS/MINOS+}

MINOS \cite{MINOS:1995txm} was an on-axis, long-baseline experiment with a near detector located at Fermilab, and a far detector at the Soudan Underground Laboratory in northern Minnesota. It ran originally from 2005 to 2012, and for a further 3 years as MINOS+. The neutrino energy of the NuMI beam is configurable, with the majority of the original MINOS data being taken in ``low-energy'' mode with a peak neutrino energy of $\approx 3$ GeV, and MINOS+ taking data in the ``medium-energy'' configuration whose peak neutrino energy is $\approx6$ GeV. Both the far and near detectors were functionally identical magnetized steel plastic scintilator detectors \cite{MINOS:2008hdf}. The near detector had a mass of 980~ton,  located 1km from the target, with the 5.4~kt far detector located 735km from the target. The primary goal of MINOS/MINOS+ was the measurement of the atmospheric oscillation parameters though both $\nu_\mu$ disappearance and $\nu_e$ neutrino appearance channels \cite{MINOS:2008kxu,MINOS:2020llm}. 

Alongside the standard three-neutrino paradigm, MINOS/MINOS+ had a significant program of BSM searches that highlighted the power of long baseline experiments in probing exotic models. These results include searches for Lorentz invariance and CPT violation  using both the far detector \cite{MINOS:2010kat} and near detector \cite{MINOS:2012ozn}, neutrino decoherence \cite{MINOS:2008kxu}, non-standard neutrino interactions \cite{MINOS:2013hmj,MINOS:2016sbv},  light sterile neutrinos searches \cite{MINOS:2017cae} and large-extra dimensions \cite{MINOS:2016vvv}. Alongside their own light sterile neutrino searches, MINOS+ also published a combined study with the Daya-Bay collaboration and Bugey-3 experiment \cite{MINOS:2020iqj}. 

\paragraph{NOvA}

The NuMI Off-Axis $\nu_e$ Appearance (NOvA) Experiment is a long-baseline neutrino experiment looking for oscillations in neutrinos originating in the ``Neutrinos at the Main Injector'' (NuMI) beam \cite{Adamson:2015dkw} at Fermilab in Illinois. The experiment consists of two detectors, one 300~ton near detector located at Fermilab and a second 14~kt far detector located 810 km away on the surface in Ash River, Minnesota. The location of the far detector is $0.8^\circ$ off the primary NuMI neutrino beam axis, leading to a predicted neutrino spectrum peaked strongly at 2 GeV energies. 
There are several planned searches for BSM physics at NO$\nu$A, with a mature analysis on NSI focusing on the $\epsilon_{\mu \tau}$ element by looking for deviations in the CC $\nu_\mu$ spectrum, when compared to three $\nu$ oscillations. The combination of this with a CC $\nu_e$ disappearance study can be sensitive to $\epsilon_\mu$ and $\epsilon_\tau$ individually. NO$\nu$A has also published searches for 3+1 sterile neutrino oscillations in neutral current interactions both for neutrino \cite{NOvA:2017geg} and anti-neutrino \cite{NOvA:2021smv} running mode, where no evidence for oscillations was observed in either channel. 

\paragraph{T2K}

The Tokai-to-Kamioka (T2K) experiment is a 275km long-baseline experiment, using neutrinos from the J-PARC in Tokai, operating since 2009. The primary near detector is located 280m downstream, ND280, and samples the neutrino flux before oscillations. Alongside ND280 there are additional detectors such as INGRID, WAGASCI+BabyMIND used to constrain the flux and better understand interaction cross-sections. The Far detector consists of the 50kt water Cerenkov Super-Kamiokande detector, located $2.5^\circ$ off the neutrino beam axis. This results in the neutrino beam forming a narrow peak at 650MeV, at the expected maximum oscillation point assuming three neutrinos. 

In the past T2K has probed Lorentz and CPT violation \cite{T2K:2017ega} by searching for sidereal modulations in the observed neutrino interaction rate, with plans to improve with larger data sets in the future. T2K has also published bounds on light sterile neutrinos \cite{T2K:2019efw} using a combination of five charged current and three neutral current samples, observing no evidence for new oscillation frequencies. Future searches will be further improved by a series of planned improvements to both the beam and near detector complex. By studying the charged pion yields in a T2K-replica target measured by the NA61/SHINE experiment \cite{NA61SHINE:2016nlf} the beam flux prediction can be constrained and has can reduce the flux uncertainty from about 10\% to around 5\% near the flux peak. This is expected to be further improved as more data is obtained by NA61/SHINE experiment on the replica target.  In addition, a magnet power supply upgrade as well as RF upgrades mean the neutrino beam itself is expected to increase in power from 500 kW to 1MW  by Japanese Fiscal Year 2025. 

\subsubsection{Future and Potential Neutrino Beam Experiments}
\label{sec:me-beams2}

\begin{figure}[htb!]
    \centering
    \includegraphics[width=0.65\textwidth]{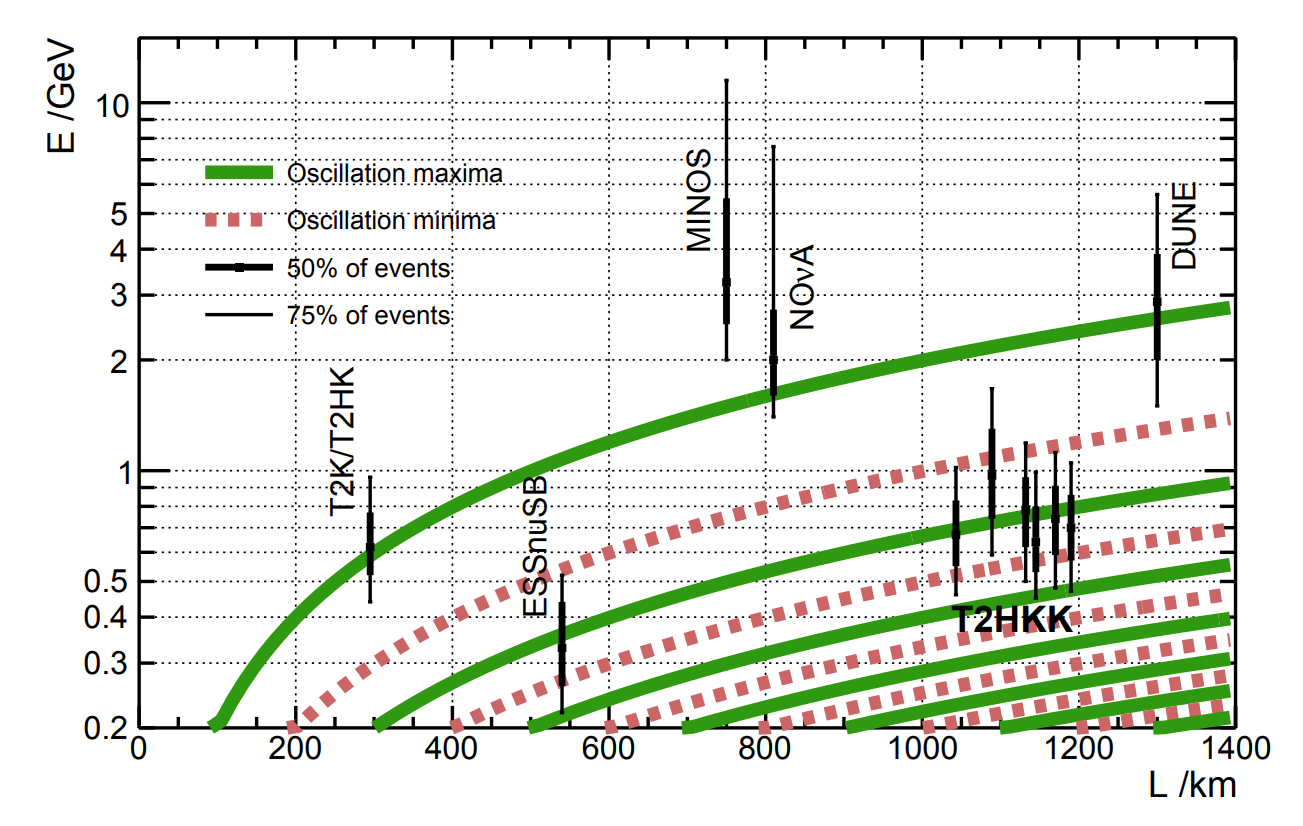}
    \caption{\textbf{Comparison of energy and baselines of various accelerator driven long-baseline experiments.} The various lines for T2HKK represent different possible baselines, all focusing on the second oscillation maxima. Figure from \cite{Hyper-Kamiokande:2016srs}}
    \label{fig:comp_acc_baselines}
\end{figure}

The overall energy and baseline ranges for both contemporary and future planned long-baseline experiments is highlighted in Fig.~\ref{fig:comp_acc_baselines}. All contemporary experiments are situated at or near the first oscillation maxima assuming the standard three neutrino paradigm. DUNE and T2HK also are sitting directly on the first oscillation maxima, but T2HKK (detector situated in Korea) and ESS$\nu$SB target the second oscillation maxima. By sitting at the second oscillation maximum in terms of $L/E_\nu$, the CP violating term in $\nu_\mu$ to $\nu_e$ oscillations is approximately three times larger than at the first oscillation maxima.

\paragraph{T2HK/T2HKK}

The T2HK\cite{Hyper-KamiokandeProto-:2015xww} (Tokai to Hyper-K) experiment continues the usage of the of the T2K beam of neutrinos from the J-PARC in Tokai, upgraded to 2.6 times the power  (1.3~MW), but with the far detector being the next generation underground water Cherenkov detector Hyper-Kamiokande located near Kamioka town. Hyper-Kamiokande build on the extensive experience and knowledge of water Cherekov reconstruction developed during the successful running of the Super-Kamiokande experiment. For more details on the Hyper-Kamiokande detector, see the following section on atmospheric neutrinos. The T2HKK (''Tokai to Hyper-K detector in Korea'') \cite{Hyper-Kamiokande:2016srs} is a proposed expansion featuring both a detector in the Kamioka mine and a second Hyper-Kamiokande detector located in Korea. When considering the Korean candidate site, the longer baseline and thus larger matter effect significantly enhances sensitivities to non-standard interactions \cite{Fukasawa:2016lew,Liao:2016orc}.

\paragraph{DUNE}

The Deep Underground Neutrino Experiment (DUNE) is a next-generation, long-baseline neutrino oscillation experiment, designed to be sensitive to $\nu_\mu$ to $\nu_e$ oscillation. The experiment consists of a 1.2~MW, broadband neutrino beam around $1-5$ GeV, with a near detector complex located at Fermilab, and a $>20$~kt liquid argon time-projection chamber far detector located at the 4850~ft level of Sanford Underground Research Facility in Lead, South Dakota. The combination of high powered beam and comprehensive suite of both near and far detectors provides a huge opportunity for a rich and diverse set of BSM physics searches, which is explored extensively in \cite{DUNE:2020fgq} and \cite{DUNE:2020ypp}. 

Due to the particularly long 1300km baseline DUNE is sensitive to non-standard interactions that affect neutrinos propagation thought the earth and can substantially improve bounds on NC matter NSI parameters \cite{Blennow:2016etl,Coloma:2015kiu,deGouvea:2015ndi}. New CC NSI interactions which can produce BSM effects at the source and detector have been shown to not be discoverable at DUNE \cite{Bakhti:2016gic,Blennow:2016etl}. DUNE will also be able to place world leading bounds on Lorentz and CPT violation \cite{Barenboim:2017ewj,Barenboim:2018ctx} in the neutrino sector, improving bounds by several orders of magnitude. Searches for non-unitary at DUNE \cite{Blennow:2016jkn,Escrihuela:2016ube} have been shown to produce bounds comparable with other constraints from present oscillation experiments. Further studies on sensitivities to new physics at DUNE include neutrino decay, see Tab.~\ref{tab:futureboundsNudec}.

\paragraph{ESS$\nu$SB}

The ESS neutrino Super Beam (ESS$\nu$SB) is a proposed neutrino beam utilizing the 5 MW proton linear accelerator of the European Spallation Source (ESS) in Lund, Sweden, that is currently under construction. Although the final site for the associated far detector is not set, current proposed sites are the Garpenberg mine (baseline of 540 km) or the Zinkgruvan mine (baseline of 360km) both located in Sweden. In combination with relatively low energy expected flux peak of $\approx$0.25 GeV \cite{essnusb_flux}, this allows for the probing of the second oscillation maxima of $\nu_\mu \rightarrow \nu_e$ oscillations.  

Studies of the potential of the ESS$\nu$SB have been performed to bound non-unitarity \cite{Chatterjee:2021xyu}, with the best sensitivities to the non-unitary parameters studied ($\alpha_{21}$) are achieved for the shortest 200 km baseline that was studied, with 360km performing somewhat better that 540km although remaining comparable to current bounds. Studies have shown that ESS$\nu$SB would also be sensitive to CPT violation \cite{Majhi:2021api} and to neutrino decay (see Tab.~\ref{tab:futureboundsNudec}). Due to the shorter baselines, the ESS$\nu$SB is less sensitive to matter effects and therefore to NSI parameters.

\paragraph{Precision beams: nuSTORM and Iso-DAR}

While conventional superbeams (neutrino beams from meson decay-at-rest) from accelerators provide well understood fluxes in comparison to many natural sources, there remains still an overall $\mathcal{O}(5\rightarrow10\%)$ uncertainty, growing larger at lower energies. Alternative approaches, for example neutrino factory (neutrino beams from muon decays in a storage ring) and beta beam (neutrino beams from boosted radioactive decays) have known spectrum and flavor structures. These beams can contribute BSM physics search from neutrino flavor beyond what superbeams can.

{\it NuSTORM}~\cite{nuSTORM:2014phr} --- The Neutrinos from Stored Muons (nuSTORM) facility forms the neutrino beam from the decay of stored muons and pions in a low-energy muon decay ring.  By placing appropriate instrumentation in the decay ring, the total integrated neutrino flux can be estimated with a precision of $\approx 1\%$. The initial pion injection will lead to an intense flux of neutrinos from primarily pion decay, with subsequent rotations of the stored muons providing a neutrino beam that can be calculated precisely using the known three-body Michel decay parameters alongside the instrumentation of the muon decay ring.

{\it IsoDAR}~\cite{Alonso:2022mup} --- An accelerator-driven single-isotope antineutrino source (IsoDAR) uses a high-power cyclotron to deliver protons on a beryllium target, and ejected neutrons are captured by a surrounding $^7\mathrm{Li}$ sleeve. Then subsequent decays ($^8\mathrm{Li}\rightarrow ^8\mathrm{Be}+e^-+\bar\nu_e$) produce a pure $\bar\nu_e$ beam with known spectrum. Currently, there is a ongoing effort to build such facility at the Yemilab underground facility, South Korea. Here, 60~MeV cyclotron is proposed to operate from 2027, with 2.3~kt of liquid scintillator counter.

{\it MOMENT}~\cite{Cao:2014bea} --- The MuOn-decay MEdium-baseline NeuTrino beam facility is proposed to generate a very intense beam of neutrinos and antineutrinos (from a proton beam expected to reach 15~MW power at 1.5~GeV energy)  via decay of positive and negative muons to match the second oscillation maximum with the neutrino beam energy peaked at around 200--300~MeV and a baseline of 150~km, although different baselines can be considered~\cite{Vihonen:2022thz}. BSM physics potential studies for non-standard neutrino interactions~\cite{Tang:2017qen} and invisible neutrino decays~\cite{Tang:2018rer} has been performed assuming a large
gadolinium-doped water Cherenkov detector of 500~kt.

\paragraph{Near detectors}

New generation neutrino oscillation facilities feature beams of unprecedented luminosity which, when observed by their near detectors (ND), will provide a very powerful tool to explore extremely small new physics effects as long as they do not require long baselines to develop. These searches can be divided in three main categories:
\begin{itemize}
   \item Searches for the decay products of long-lived new particles produced in the beam. This is the case of new light and weakly interacting gauge bosons, dark fermions, or even heavy neutrinos in the MeV-GeV range as outlined in Section~\ref{sec:steriles}. These scenarios are discussed in more detail in the dedicated Snowmass White Papers in Refs.~\cite{HNL-Snowmass} and~\cite{DarkSectors-Snowmass}.
   \item Precision measurements of neutrino scattering processes. Due to the large cross section uncertainties involved in neutrino-nucleus scattering, better prospects are expected for $\nu-e$ elastic scattering. As discussed in Sec.\ref{subsec:inter-LIGHTnew}, these measurements can be used to constrain models with light mediators, as long as they are coupled to electrons. For example, in flavored $U(1)'$ models such as $L_e-L_\mu$, a new mediator opens up a new t-channel diagram for $\nu-e$ scattering through the $Z^\prime$, which leads to interference with the existing SM processes. This has been studied for the DUNE ND, both in the context of a total rate analysis~\cite{Ballett:2019xoj,Dev:2021xzd} and for a binned energy spectrum \cite{Chakraborty:2021apc} which shows that in a bin-by-bin analysis the effect of destructive interference is reduced compared to the total rate approach. The DUNE ND could provide competitive bounds for mass ranges ($\approx 0.02-0.2$ GeV) of the $L_\mu - L_e$ $Z^\prime$.
   \item Additionally, new physics effects that may lead to neutrino flavour change at very short or even zero baselines are best probed at near detectors. These include the \emph{zero distance effects} discussed for non-unitarity and LED, but also light sterile neutrinos with large enough $\Delta m^2$ so as to be in the averaged-out regime as discussed in Sec.~\ref{sec:steriles} as well as CC NSI affecting neutrino production and detection processes, described in Sec.~\ref{sec:inter}. Indeed, the most stringent constraints in the right column of Tab.~\ref{tab:NUbounds} stem from the non-observation of this \emph{zero-distance effect} and in the case of $\alpha_{\mu \mu}$ the MINOS/MINOS+ ND plays a crucial role, as discussed in~\cite{Forero:2021azc}. Future prospects for the DUNE ND to some of these BSM scenarios have been studied e.g. in Refs.~\cite{Coloma:2021uhq,Giarnetti:2020bmf,Miranda:2018yym}.
   \end{itemize}
Given the very high statistics that can be collected with intense beams at near detectors, the bottleneck for the sensitivity to these searches  is always the level of understanding of the signal sample. That is, systematic uncertainties are critical and must be evaluated and modelled thoroughly in order to derive reliable constraints.

\subsubsection{Neutrinos from colliders: FASER\texorpdfstring{$\nu$} and SND@LHC}
Two  experiments at CERN aim to measure for the first time the intense flux of neutrinos created in $pp$ collisons at the LHC. Constructed using 1000 emulsion layers interleaved with 1-mm tungsten plates, FASER$\nu$\cite{FASER:2019dxq,FASER:2020gpr} sits 480m from the ATLAS interaction point and was installed during the LHC Long Shutdown 2. FASEr$\nu$ aims to measure primarily $\nu_\mu$ interactions but with sensitivity to $\nu_e$ and $\nu_\tau$ interactions also. The energy of the expected $\nu_\mu$ neutrinos peaks $\approx 0.5$ TeV, with $\nu_\tau$ and $\nu_e$ components being slightly higher $\mathcal{O}$(TeV) energies. Its primary goal is the detection of these collider neutrinos across all three flavors and measure their cross sections at $\mathcal{O}$(TeV) energies, higher than any man-made neutrino beam thus far. Beyond this, it has been shown that FASER$\nu$ will provide strong bounds on neutrino non-standard interactions \cite{Falkowski:2021bkq} as described in Sec.~\ref{subsec:inter-eft}. Complementary to FASER$\nu$, the Scattering and Neutrino Detector at the LHC (SND@LHC) \cite{SHiP:2020sos} will search for collider neutrinos off-axis angles relative to the beamline larger than those covered by FASER$\nu$, with the the corresponding neutrinos mostly originating from charm decays with energies between 350 GeV and a few TeV, with sensitivity to all neutrino flavours. The detector will be a small-scale prototype of the scattering and neutrino detector of the SHiP experiment \cite{Anelli:2015pba}.

\subsubsection{Atmospheric neutrinos}
Neutrinos produced by the interaction of incoming cosmic rays with the upper atmosphere of the Earth provide us with an abundant source of neutrinos. These neutrinos, referred to as \emph{atmospheric neutrinos}, are the byproduct of the decay of mesons that are produced by energetic protons colliding with the elements that composes the Earth atmosphere. Not only does Nature provide us with a natural neutrino source, but its flux spans several orders of magnitude in energy, ranging from few hundred MeV to hundreds of TeV. These atmospheric neutrinos were the first natural neutrinos to ever be observed, at experiments in the East Rand Proprietary Mines in South Africa \cite{PhysRevLett.15.429} and the Kolar Gold Fields mine in India \cite{Achar:1965ova}. Atmospheric neutrinos has since been observed in many experiments including MACRO~\cite{MACRO:1998ckv}, Soudan2~\cite{Soudan2:2003qqa}, Super-Kamiokande~\cite{Super-Kamiokande:2002weg} and IceCube-DeepCore~\cite{IceCube:2011ucd}. Historically, the first observation of the neutrino flavor conversion, consistent with the neutrino oscillation phenomenon, was performed in Super-Kamiokande. This observation was made possible thanks to both a wide atmospheric neutrino flux and that neutrinos can travel different distances from the production point to the detector, which ranges from few tens of kilometers to the Earth diameter. 

In the following we describe the main experimental features of current atmospheric neutrino experiments and the BSM studies that have been performed so far to search for any modifications to the expected neutrino flavor profile. This will help to introduce the future experiments and their experimental prospects.

\paragraph{The Super-Kamiokande (SK) experiment}

SK has been observing both solar and atmospheric neutrinos for more than two decades providing valuable information that has converged in our current understanding of the three-flavor neutrino oscillation mechanism.
 
The SK detector, located underground inside a zinc mine in Kamioka-Japan, is a cylindrical 50 kt water Cerenkov neutrino detector with nominal fiducial volume of 22.5 kt.
SK is able to observe atmospheric neutrino events spanning four orders of magnitude. 
Fully-contained events have energies ranging from a few hundred MeV to $\sim 10\,\text{GeV}$. Partially contained events (with vertices inside the fiducial volume, but leptons that leave the inner detector) have energies ranging from a few GeV to tens of GeV. Finally, up-going muon events (those which start in the surrounding rock and pass through both the outer and inner detector volumes) have energies between a few GeVs and TeVs.
The data collected since SK started operating in 1996 (divided into four different running periods) have been used to search for a wide set of BSM effects. In particular, tests of non-standard neutrino interactions (NSI)~\cite{Super-Kamiokande:2011dam}, tests of Lorentz invariance~\cite{Super-Kamiokande:2014exs} and light sterile neutrino searches~\cite{Super-Kamiokande:2014ndf} have been performed. 

In 2018, the Super-Kamiokande experiment entered in a new phase, after the water purification system was upgraded to allow operation while doped with gadolinium sulfate (SK-Gd). This improves the differentiation between neutrinos and antineutrinos, thanks to neutron capture by gadolinium (neutron tagging), and is expected to impact the physics reach of SK program.  Operation with 0.01\% loading started in 2020. In addition, the atmospheric neutrino analysis is expected to be improved by better modeling of neutrino interactions from measurements at the T2K near detector~\cite{T2K:2020jav} and MINERvA~\cite{MINERvA:2021csy}.

\paragraph{The Hyper-Kamiokande (HK) experiment}

Hyper-Kamiokande is the successor to the SK experiment with a 8.4 times larger effective volume. Hyper-Kamiokande is expecting to start operation in 2027. The detector tank will have a total volume of 258~kt. As SK it is separated into an inner and an outer veto detector. The inner region containing 217~kt of water and will have an array of 40,000 high QE Box \& Line (B\&L) PMTs with 50~cm diameter. The new PMT type improves the photon detection efficiency with a better timing resolution, and charge resolution of SK PMTs~\cite{Hyper-Kamiokande:2020aij}.

With a larger detector mass than SK and running for 20 years (high statistics) plus the mentioned detection improvements plus better modeling of neutrino interactions (reduction of systematical uncertainties), precise measurements with atmospheric neutrinos are expected in HK. This can be used to test for BSM scenarios that affect neutrino flavor, similar to what has been done at SK, such as Lorentz invariance violation~\cite{Super-Kamiokande:2014exs}, non-standard neutrino interactions~\cite{Super-Kamiokande:2011dam}, and light sterile neutrino oscillations~\cite{Super-Kamiokande:2014ndf}.

\paragraph{IceCube DeepCore}

The IceCube neutrino detector is a $1\,\mathrm{km}^3$ of ice instrumented with 86 strings with 60 digital optical modules (DOMs) each, located at the South Pole, Antarctica. The basic mechanism is similar to that of Water Cerenkov detectors: the DOMs contain the PMTs that detect the Cerenkov radiation emitted by charged particles (produced in neutrino interactions) traveling faster than the speed of light in the ice. The strings are arranged in a hexagonal grid, buried below the ice surface at depths ranging from 1450~m to 2450~m, and with a typical separation between strings of 125~m (or, 50~m between DOMs). 

At the center of the IceCube detector, DeepCore is comprised of eight additional strings with more efficient PMTs (increasing light collection), and with a typical separation of 17~m between strings (7~m between DOMs). With an increased DOM density, DeepCore is sensitive to lower neutrino energies than Icecube. Moreover, its geometry was optimized to maximize the detection efficiency for neutrino events in the range of 10~GeV to 100~GeV. In this energy range, the high cosmic-ray background is mitigated by using IceCube surrounding strings as an active muon veto~\cite{IceCube:2011ucd}. At DeepCore, muon tracks are reconstructed with an angular resolution of $12^\circ$ at 10~GeV. The muon energy can be reconstructed from the muon track length and adding the energy of the hadronic shower (estimated from the total amount of light collected by the detector) provides a proxy for the neutrino energy~\cite{IceCube:2017zcu}. The median energy resolution is about 30\% at 8~GeV~\cite{IceCube:2017ivd}.

Using three years of atmospheric neutrino data from the DeepCore detector~\cite{IceCube:2014flw} a search for a light sterile neutrino was performed in Ref.~\cite{IceCube:2017ivd}, resulting in limits on the sterile-active mixing matrix elements $|U_{\mu 4}|^2<0.11$ and $|U_{\tau 4}|^2<0.15$ (at 90\% C.L), for a value of the neutrino mass splitting $\Delta m_{41}^2=1\,\text{eV}^2$. Using the same data sample, a limit on the NC NSI parameter $\varepsilon_{\mu \tau}$ was found to be in the range $-0.0067<\varepsilon_{\mu \tau}<0.081$ (at 90\% C.L.)~\cite{IceCube:2017zcu}, performing a similar analysis as Super-Kamiokande~\cite{Super-Kamiokande:2011dam}. Recently, a more general analysis considering all NSI couplings with quarks (including the effect of the NSI phases) was performed in Ref.~\cite{IceCubeCollaboration:2021euf}, using three years of Icecube DeepCore data.

\paragraph{IceCube Upgrade}
An additional array of 7 strings, with an separation between DOMs of 3~m, will be added to DeepCore within the following year. This upgrade will impact important experimental features as the energy and directional resolution, lowering the detector energy threshold while increasing the efficiency below 10~GeV. The detector and ice calibration will also be improved by the deployment of new calibration devises. A significant reduction of systematic uncertainties is expected which, together with the high IceCube statistics, will improve precision on the determination of standard oscillation parameters and at the same time extend the BSM physics sensitivity.

\subsection{High energies: neutrinos above the TeV scale\label{sec:he}}

\textbf{Main authors:} C. Arg\"uelles and T. Katori
\vspace{2mm}

The quest to observe neutrinos from high-energy astrophysical sources, where hadronic activity has been inferred to be present, has led to the development of a category of detectors known as \textit{neutrino telescopes}.
In order to achieve the large effective masses needed to observe these very small fluxes, these detectors use naturally occurring materials --- \textit{e.g.} lakes, oceans, glaciers, mountains, \textit{etc} --- as targets for the neutrino interactions.

Current and proposed neutrino telescopes are listed in Table~\ref{tab:he-experiments}, where they have been organized by the detector technology and their acceptance to different neutrino flavors.
These experiments can be roughly grouped into two categories: those that aim to observe and characterize high-energy neutrinos ($1~{\rm TeV} < E_\nu< 10~{\rm PeV}$) and those that aim to discover extremely-high-energy neutrinos ($E_\nu > 10~{\rm PeV}$).
The high-energy neutrino telescopes (HENT) can measure the high-energy part of the atmospheric neutrino flux and astrophysical neutrino sources. 
HENT will be able to perform new physics searches using both of these sources, which we discuss in Sec.~\ref{sec:he-atmo} and Sec.~\ref{sec:he-astro} respectively.
Extremely high-energy neutrino telescopes (EHENT) aim to observe for the first time neutrinos produced in interactions of cosmic rays with the cosmic microwave background. 
These so-called cosmogenic neutrinos have not yet been observed; however, as we will discuss in Sec.~\ref{sec:ehe-neutrinos} observation of anomalous events on detectors hunting for cosmogenic neutrinos has produced significant discussion.

Neutrino flavor conversions are quantum effects happening on a macroscopic scale.
Neutrino flavor conversion probabilities are the solution of the effective neutrino Hamiltonian, which is given by
\begin{equation}
H^{eff}(\nu_\alpha\rightarrow\nu_\beta)\sim
m^2/2E+V+\mathrm{new~physics~(E^0, E^1, E^2,\ldots)~.}
\end{equation}
Here, the first term is the neutrino mass term which is proportional to $E^{-1}$, $V$ is the matter potential, and the last term represents new physics one can explore from the flavor conversion.
Since new physics terms are in general zero or positive power of energy at larger energies, the mass term is suppressed, potentially making new physics terms become the dominant term.
On top of this, longer baseline oscillation can explore smaller couplings.
The quantum effect, long-baseline, and high-energy combination make HENT and EHENT extremely sensitive to many new physics models. 

The neutrino telescopes shown in Table~\ref{tab:he-experiments} can also be organized by their capacity to identify different neutrino flavors. 
In this respect, neutrino telescopes can be roughly organized as all-flavor detectors or tau-neutrino-specific detectors.
The latter are in a special category since tau-neutrinos  interactions generate secondary neutrinos that carry a significant fraction of the primary energy~\cite{Ritz:1987mh,Halzen:1998be,Iyer:1999wu,Dutta:2000jv,Becattini:2000fj,Beacom:2001xn,Dutta:2002zc,Jones:2003zy,Yoshida:2003js,Bugaev:2003sw,Bigas:2008sw,Alvarez-Muniz:2018owm,Safa:2019ege,Safa:2021ghs,Soto:2021vdc}
The capacity to identify flavor in these experiments will be crucial for the target models of this white paper.
At present multi-flavor neutrino detectors can differentiate well between muon-neutrino charged-current interactions and all other neutrino interactions but have limited capacity to separate between tau neutrinos and electron neutrinos. 
The latter point motivates tau-neutrino-specific experiments that aim to single out that component.
 
\begin{table}[!ht]
\begin{center}
{
\begin{tabular}{ |c|m{10em} |c|c|c|}
 \hline
 \textbf{Energy Range} & \textbf{Experiment} & \textbf{Technology} & \textbf{Detected Flavor} & \textbf{Ref.}
 \\ \hline\hline 
   
    $\lesssim 10^3$ GeV & JUNO & Liquid scintillator & All Flavors &\cite{JUNO:2015zny}\\\hline
    $\lesssim10^3$ GeV & DUNE & LArTPC & All Flavors &\cite{DUNE:2020lwj}\\\hline
    $\lesssim 10^3$ GeV & THEIA& WbLS & All Flavors &\cite{Theia:2019non}\\\hline
    $\lesssim 10^3$ GeV & Super-Kamiokande &Gd-loaded Water C &  All Flavors &\cite{Super-Kamiokande:2002weg} \\\hline
    $\lesssim 10^4$ GeV & Hyper-Kamiokande &Water Cherenkov &  All Flavors &\cite{Hyper-Kamiokande:2018ofw} \\\hline
    $\lesssim 10^5$ GeV & ANTARES & Sea-Water Cherenkov &  $\nu_\mu,\,\bar{\nu}_\mu$ (CC) &\cite{ANTARES:2011hfw}\\\hline\hline
    $\lesssim 10^6$ GeV & IceCube/IceCube-Gen2  & Ice Cherenkov &  All Flavors &\cite{IceCube:2016zyt,IceCube-Gen2:2020qha}\\\hline
    $\lesssim 10^6$ GeV & KM3NeT & Sea-Water Cherenkov & All Flavors &\cite{KM3Net:2016zxf}\\\hline
    $\lesssim 10^6$ GeV & Baikal-GVD & Lake-Water Cherenkov  &  All Flavors &\cite{Baikal-GVD:2021sbo}\\\hline
    $\lesssim 10^6$ GeV & P-ONE & Sea-Water Cherenkov &  All Flavors &\cite{P-ONE:2020ljt}\\\hline\hline
    $1 - 100$ PeV & TAMBO &  Earth-skimming WC&  $\nu_\tau,\,\bar{\nu}_\tau $ (CC) &\cite{Romero-Wolf:2020pzh}\\\hline
    $\gtrsim 1$ PeV & Trinity &  Earth-skimming Image &  $\nu_\tau,\,\bar{\nu}_\tau $ (CC) &\cite{Otte:2019aaf}\\\hline
    $\gtrsim 10$ PeV & RET-N &  Radar echo &  All Flavors &\cite{RadarEchoTelescope:2021rca}\\\hline
    $\gtrsim 10$ PeV & IceCube-Gen2 &  In-ice Radio &  All Flavors &\cite{IceCube-Gen2:2020qha}\\\hline
    $\gtrsim 10$ PeV & ARIANNA-200 & On-ice Radio &  All Flavors &\cite{ARIANNA:2019scz}\\\hline
    $\gtrsim 20$ PeV & POEMMA &  Space Air-shower Image&  $\nu_\tau,\,\bar{\nu}_\tau $ (CC) &\cite{POEMMA:2020ykm}\\\hline
    $\gtrsim 100$ PeV & RNO-G &  In-ice Radio &  All Flavors &\cite{RNO-G:2020rmc}\\\hline
    $\gtrsim 100$ PeV & ANITA/PUEO &  Balloon Radio &  
    All Flavors&\cite{ANITA:2019wyx,PUEO:2020bnn}\\\hline
    $\gtrsim 100$ PeV & Auger/GCOS &  Earth-skimming WC &  $\nu_\tau,\,\bar{\nu}_\tau $ (CC) &\cite{PierreAuger:2019ens,Horandel:2019qwu}\\\hline
    $\gtrsim 100$ PeV & Beacon &  Earth-skimming Radio &  $\nu_\tau,\,\bar{\nu}_\tau $ (CC) &\cite{Wissel:2020sec}\\\hline
    $\gtrsim 100$ PeV & GRAND &  Earth-skimming Radio &  $\nu_\tau,\,\bar{\nu}_\tau $ (CC) &\cite{GRAND:2018iaj}\\\hline
\end{tabular}
}
\caption{
\textbf{Neutrino detectors and neutrino telescopes sorted from smaller to larger energies.}
These are grouped into three categories separated by the double lines: large neutrino detectors (top), high-energy neutrino telescopes (HENT, middle), and extremely high-energy neutrino telescopes (EHENT, bottom). 
The detector technology and the neutrino interaction that they are sensitive to are shown in the right most columns with references. 
The label `All Flavors' implies that they can also detect both charged- and neutral-current interactions.
Table adapted from~\cite{Arguelles:2019ouk}.
}
\label{tab:he-experiments}
\end{center}
\end{table}

\begin{figure}[htb!]
    \centering
    \includegraphics[width=\textwidth]{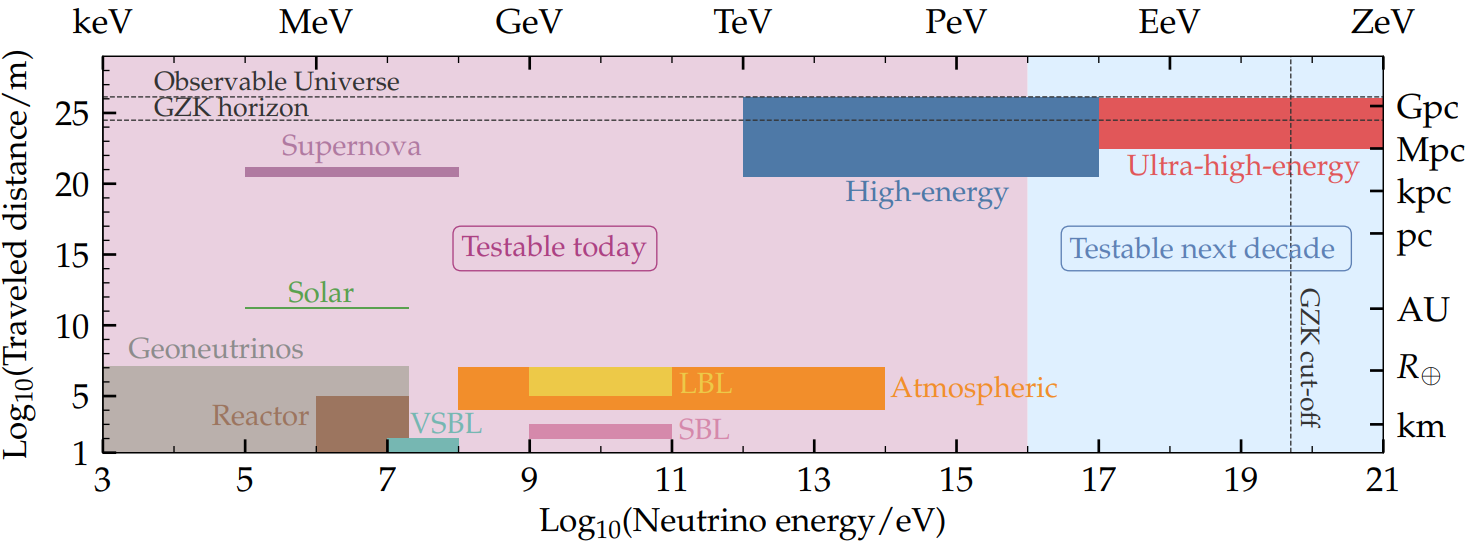}
    \caption{
    \textbf{Energy and distance scales relevant for neutrino telescopes.}
    Three high-energy neutrino fluxes are labeled as atmospheric (orange), high-energy astrophysical (blue), and ultra-high-energy (red). 
    The region explored by the current experiment is shown in pink, while next-generation is in light blue.
    Adapted from~\cite{Arguelles:2019rbn}.
    }
    \label{fig:L_E}
\end{figure}

Figure~\ref{fig:L_E} shows the energy ranges and the relevant neutrino fluxes.
The three fluxes discussed in this section are shown there: atmospheric (orange), high-energy astrophysical (blue), and ultra-high-energy (red). 
Since neutrino flavor changing depends on the distance traversed by the neutrino the distance from the neutrino source to us is shown in the vertical axis. 
These spans distances from the Earth radius to several gigaparsecs in length.
The region probed by experiments today is signaled by a pink background, while the region that is expected to be probed by next-generation detectors is shaded light blue. 

Due to the fact that neutrino telescopes need to instrument large volumes, they are sparsely instrumented. 
This implies that the information we can obtain on each individual event is limited compared to what is available in hodoscopic detectors such as DUNE.
Four observables can be extracted out of the neutrino events observed by these detectors.
These are: direction, morphology, energy, and time of arrival of the event.
Of these four observables, the one more accurately measured is the time of arrival.

The capacity to reconstruct the direction and energy of the events is correlated with the event morphology~\cite{IceCube:2013dkx}. Morphological information is correlated with the neutrino flavor and interaction type. 
Current neutrino telescopes morphological categories are~\cite{IceCube:2015gsk}: tracks, cascades, and double cascades. 
Tracks are produced by charged-current muon-neutrino interaction, though they can also be produced by charged-current tau-neutrino interactions when the tau decays leptonically to a muon.
The cascades are produced in all neutral-current neutrino interactions, as well as charged-current electron neutrino interactions and most of the tau-neutrino interactions.
Double cascades are a smoking gun signature of tau neutrino interactions and have recently been observed by IceCube~\cite{IceCube:2020abv}.
Additionally, the resonant production of $W$-bosons in antineutrino-electron scattering~\cite{Glashow:1960zz,Gauld:2019pgt,Huang:2019hgs} can be identified by selecting events that contain no outgoing track and where the shower has energy close to the expected resonant energy with hadronic activity.
Recently, IceCube has detected the first of these events~\cite{IceCube:2021rpz}, which are interesting as they provide a unique handle on the ratio of neutrinos to antineutrinos.
In general, the ratio of neutrinos to antineutrinos cannot be disentangled on an event-by-event basis, but only statistically at sub-10 PeV energies due to the difference in the cross-section.

\begin{figure}[htb!]
    \centering
    \includegraphics[width=0.6\textwidth]{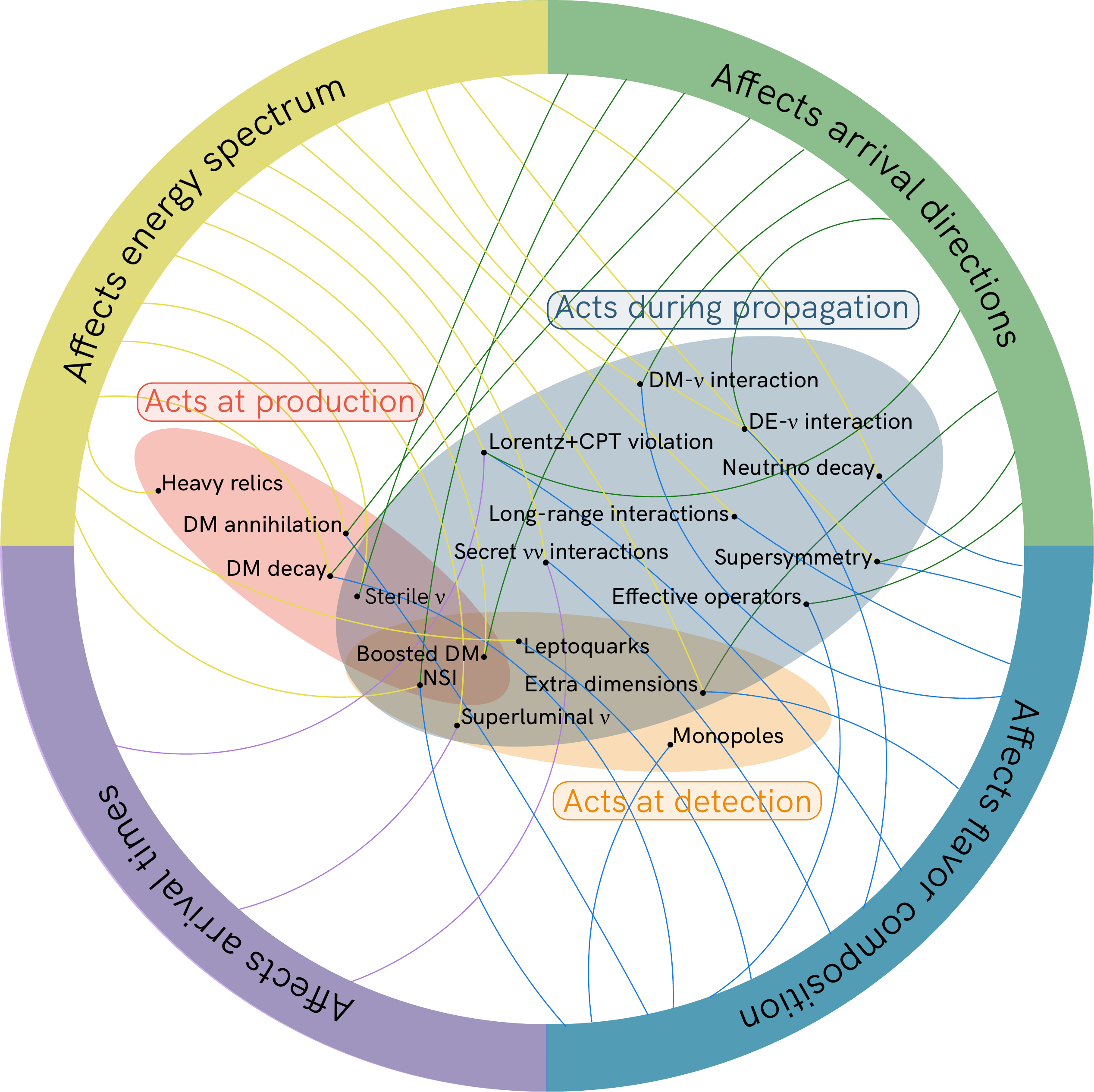}
    \caption{
    \textbf{Models and how they modify high-energy neutrino observables.}
    An incomplete list of models is shown in the center of the figure.
    When a new physics scenario modifies one of the four observables available in neutrino telescopes (flavor, energy, time, and direction) they are linked by a line from the model to the observable.
    The location where the new physics happens (production, propagation, and detection) is illustrated by shaded inner regions.
    Adapted from~\cite{Arguelles:2019rbn}.
    }
    \label{fig:model_classification}
\end{figure}

Figure~\ref{fig:model_classification} shows how these four observables get modified in the presence of new physics. 
New physics flavor scenarios can be present in the production of the neutrino, \textit{e.g.} dark matter decaying into neutrinos~\cite{Arguelles:2019ouk,Feldstein:2013kka, Esmaili:2013gha, Bai:2013nga, Bhattacharya:2014vwa, Zavala:2014dla, Rott:2014kfa, Esmaili:2014rma, Fong:2014bsa, Murase:2015gea, Ahlers:2015moa, Aisati:2015vma, Troitsky:2015cnk, Chianese:2016opp, Chianese:2016kpu, Bhattacharya:2017jaw, Chianese:2017nwe, Aartsen:2018mxl, Sui:2018bbh, Bhattacharya:2019ucd, Arguelles:2019boy, Chianese:2019kyl,Ema:2013nda, Ema:2014ufa, Anchordoqui:2015lqa, Ema:2016zzu}, and in the detection of the neutrinos, \textit{e.g.} secret neutrino interactions that modify the neutrino-nucleon cross section~\cite{Barger:2013pla, Dutta:2015dka, Dey:2015eaa, Mileo:2016zeo, Chauhan:2017ndd, Dey:2017ede, Becirevic:2018uab,Illana:2004qc, Illana:2005pu, Illana:2014bda,Mack:2019bps}. 
The propagation of the neutrinos can be affected by new physics which involves the flavor changing, and this can explore many topics including new neutrino properties (neutrino decay~\cite{Beacom:2002vi, Meloni:2006gv, Maltoni:2008jr, Bhattacharya:2009tx, Bhattacharya:2010xj, Mehta:2011qb, Baerwald:2012kc,Pakvasa:2012db, Pagliaroli:2015rca, Huang:2015flc, Shoemaker:2015qul, Bustamante:2016ciw, Denton:2018aml, Abdullahi:2020rge, Bustamante:2020niz}, sterile neutrinos~\cite{Brdar:2016thq, Rasmussen:2017ert, Arguelles:2019tum,Crocker:2001zs, Keranen:2003xd, Beacom:2003eu, Esmaili:2009fk}, mass-varying neutrinos~\cite{Hung:2003jb}), new neutrino interactions (neutrino -- dark-matter-background scattering~\cite{deSalas:2016svi, Farzan:2018pnk,Weiler2006,Barranco:2010xt, Miranda:2013wla, Reynoso:2016hjr,Arguelles:2017atb, Kelly:2018tyg, Alvey:2019jzx, Choi:2019ixb}, neutrino -- dark-energy scattering~\cite{Ando:2009ts, Klop:2017dim}, neutrino self-interaction~\cite{Keranen:1997gz, Weiler2006, Hooper:2007jr, Lykken:2007kp, Ioka:2014kca, Ng:2014pca, Ibe:2014pja, Blum:2014ewa, Araki:2014ona, Kamada:2015era, DiFranzo:2015qea, Cherry:2016jol, Mohanty:2018cmq, Barenboim:2019tux, Bustamante:2020mep, Creque-Sarbinowski:2020qhz},~non-standard interactions~\cite{Blennow:2009rp, Gonzalez-Garcia:2016gpq}),  long-range force~\cite{Bustamante:2018mzu}, Lorentz violation~\cite{Barenboim:2003jm, Bustamante:2010nq, Borriello:2013ala, Stecker:2014xja, Tomar:2015fha, Wang:2016lne, Wei:2016ygk, Lai:2017bbl, IceCube:2017qyp, Ellis:2018ogq, Laha:2018hsh,IceCube:2021tdn}, quantum decoherence~\cite{Liu:1997km, Ahluwalia:2001xc, Hooper:2004xr, Anchordoqui:2005gj, Stuttard:2020qfv}, extra-dimension~\cite{Aeikens:2014yga}, etc.

\subsubsection{High-energy atmospheric neutrinos\label{sec:he-atmo}}

Atmospheric neutrinos are produced in collisions of primary cosmic rays and the Earth's atmosphere.
These neutrinos have varying propagation distances from their production point, on average above 20~km above the Earth's surface, to their detection.
The incident angle of detection of these neutrinos is then a proxy for the propagation distance, where the maximum propagation distance is an up-going neutrino travelling the Earth's diameter.
The atmospheric neutrino flux has a steeply falling power-law energy distribution, which at low energies follows the cosmic-ray spectrum and at higher energies becomes on spectral index softer due to the interactions of mesons with the atmosphere.
These neutrinos, predominantly produced by kaon decay, have been observed to have energies as large as hundreds TeVs; making them the highest energy neutrinos from terrestrial origin.
Thus atmospheric neutrinos are the only system to study neutrino flavor transitions at baselines up to $\sim 12,800$~km and energies up to $\sim 100$~TeV.
A regime far beyond what accelerator-based neutrino experiments can achieve.
Since neutrino oscillations act as natural interferometers, therefore, atmospheric \textit{neutrino interferometry} is unique to test various exotic flavor mixing due to new physics.  

The energy region of 1~TeV or lower can be measurable by neutrino detectors with tanks. Super-Kamiokande reported 318 up-going showering muon events from 1645 days of data~\cite{Super-Kamiokande:2007uxr}.
These high-energy muons lose energy stochastically most likely originated from muon neutrino over 1~TeV.
Thus, we expect similar or a few times larger sample sizes of such events in neutrino tank detectors such as  JUNO~\cite{JUNO:2015zny}, DUNE~\cite{DUNE:2020lwj}, 
THEIA~\cite{Theia:2019non}, 
Super- and 
Hyper-Kamiokande~\cite{Super-Kamiokande:2002weg,Hyper-Kamiokande:2018ofw}. The advantages of these detectors are their resolution compared with neutrino telescopes without detector tanks~\cite{Schneider:2021wzs}. In this energy region, however, low-energy arrays of neutrino telescopes, such as ANTARES~\cite{ANTARES:2011hfw}, DeepCore~\cite{IceCube:2011ucd}, IceCube-Upgrade~\cite{IceCube-Gen2:2020qha}, and ORCA~\cite{KM3Net:2016zxf} have overwhelmingly larger sample sizes. For example, DeepCore has an effective 10 Mton volume recorded over $300,000$ neutrino events in the eight years run. Neutrino telescopes can use large sample sizes to study the $\nu_\mu$ disappearance and $\nu_\tau$ appearance channels to look for new physics through neutrino oscillations~\cite{IceCube:2017qyp,IceCube:2019dqi,IceCube:2020tka,IceCube:2020phf,IceCube:2022ubv}, and tank detectors use all channels however much smaller sample sizes. Thus, detector tanks designed for accelerator-based neutrino experiments and neutrino telescopes designed for extraterrestrial neutrino detection overlap in this energy region but differences in technologies make them complimentary.

\begin{figure}[htb!]
    \centering
    \includegraphics[width=0.45\textwidth]{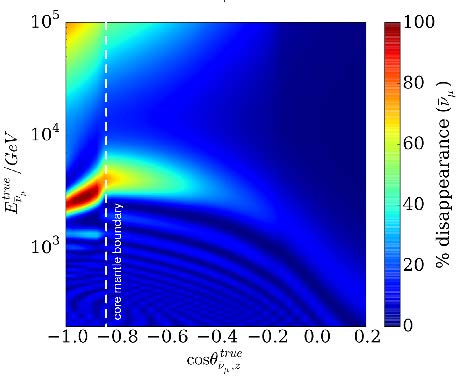}
    \includegraphics[width=0.45\textwidth]{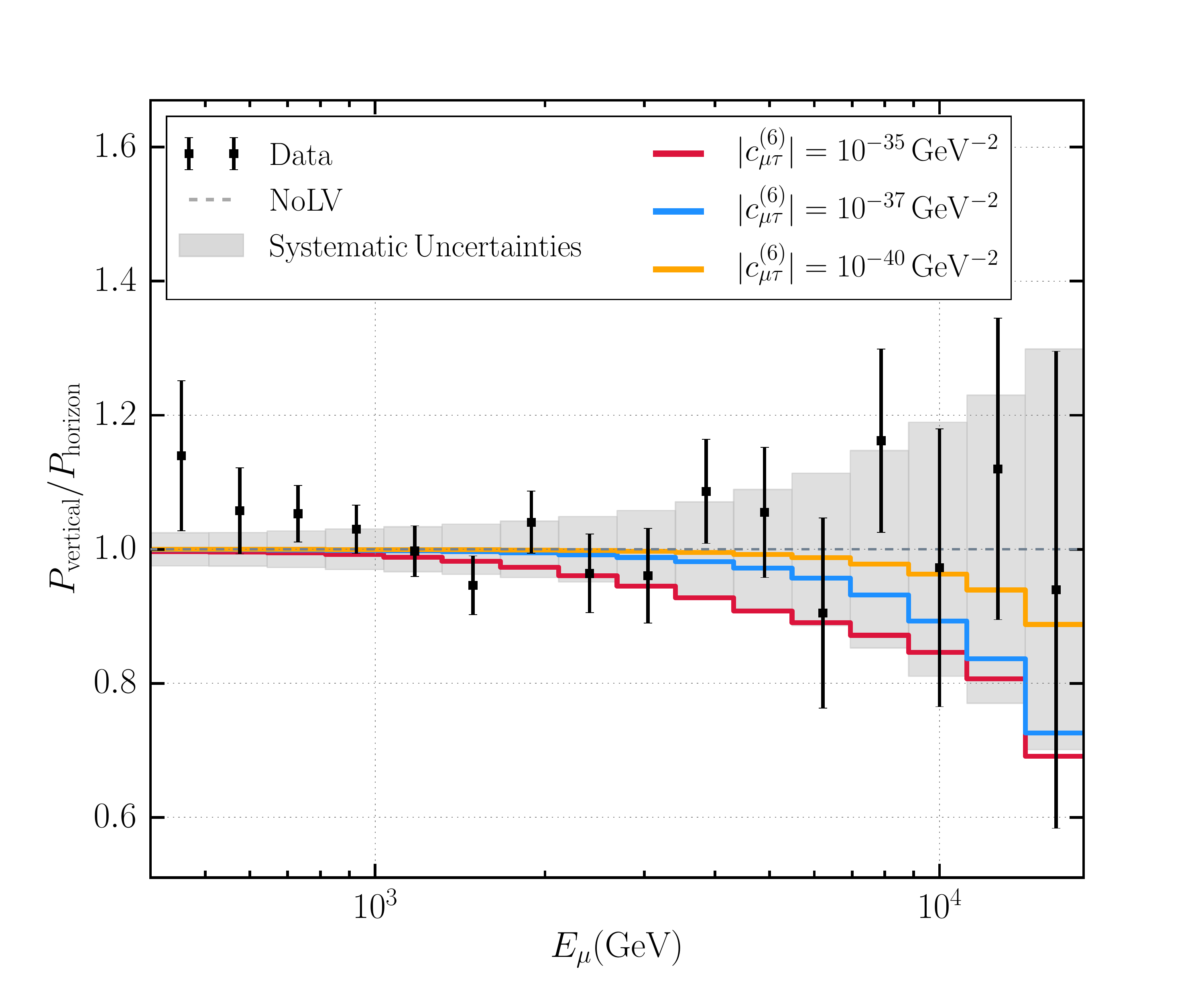}
    \caption{
    \textbf{Signatures of flavor change in TeV neutrinos.}
     Left panel shows the antineutrino survival probability as a function of the neutrino energy and the cosine of the zenith angle.
    Earth-core-through going muon-antineutrinos ($\cos\theta = -1$) experience a significant matter-resonant-enhanced depletion at TeV energies for an eV sterile neutrino with non-zero muon-neutrino mixing. Figure is taken from Ref.~\cite{IceCube:2020tka}.
    Right panel shows the ratio of the muon-neutrino survival probability between a scenario with Lorentz violation and the standard case as a function of the neutrino energy. 
    Different color lines indicate different values of the dimensionful coupling of the effective operator that governs the neutrino interaction with the Lorentz violating field. Figure is taken from~\cite{IceCube:2017qyp}.
    }
    \label{fig:atm_signature}
\end{figure}

Currently, the TeV or larger energy region is only accessible by neutrino telescopes such as IceCube/IceCube-Gen2~\cite{IceCube:2016zyt,IceCube-Gen2:2020qha}, KM3NeT~\cite{KM3Net:2016zxf}, and Baikal-GVD~\cite{Baikal-GVD:2021sbo}.
Over the next decade, these experiments will be joined by next-generation detectors such as P-ONE~\cite{P-ONE:2020ljt} and TAMBO~\cite{Romero-Wolf:2020pzh}. 
This is a unique region to look for exotic effects which disturb standard neutrino oscillation.
Figure~\ref{fig:atm_signature} shows examples from IceCube.
The left figure shows an oscillogram with eV-scale sterile neutrino, where $\nu_\mu$ disappearance peak shows up in the TeV range~\cite{IceCube:2020tka,IceCube:2020phf}.
The right figure shows the double ratio of $\nu_\mu$-disappearance oscillation probability~\cite{IceCube:2017qyp}.
The numerator is the vertical events (long propagation) and the denominator is horizontal events (short propagation and no exotic oscillation).
If the new physics, for example Lorentz violation or quantum decoherence~\cite{Stuttard:2020qfv}, are small effects, we only expect these at high energy where neutrino mass effect is suppressed, then neutrino telescopes with atmospheric neutrinos can be the only place to explore this. 

\subsubsection{High-energy astrophysical neutrinos~\label{sec:he-astro}}

High-energy astrophysical neutrinos were discovered by the IceCube Neutrino Observatory in the South Pole~\cite{Aartsen:2013jdh}.
To date, no other neutrino telescope has measured astrophysical neutrinos, though hints exist in the ANTARES data~\cite{ANTARES:2017srd}.
IceCube has established and characterized the astrophysical neutrino flux using all available morphologies~\cite{Aartsen:2013jdh,Aartsen:2014gkd,Aartsen:2014muf,Aartsen:2015zva,Aartsen:2017mau,IceCube:2018pgc,IceCube:2020acn,IceCube:2020wum,IceCube:2021uhz}. 
First, they were measured using high-energy starting events~\cite{Aartsen:2013jdh,Aartsen:2014gkd}, which are predominantly cascades, and, in their latest analysis, reported that the flux energy distribution is compatible with an unbroken single power-law with a spectral index of 2.87~\cite{IceCube:2020wum}.
Second, evidence was found using northern sky muon-neutrino events, which are morphologically tracks, and, in their latest analysis, this channel reported a spectral index of 2.37~\cite{IceCube:2021uhz}.
Finally, the astrophysical flux was measured by using all-sky cascades and obtained a spectral index of 2.53~\cite{IceCube:2020acn}.
Other notable measurements of the diffuse astrophysical neutrino flux include measurement of it by selecting starting muon-neutrino events~\cite{IceCube:2018pgc}, by selecting medium energy starting events~\cite{Aartsen:2014muf}, by searching for double bang signatures~\cite{IceCube:2020abv}, and finally by searching for signatures for resonant $W$-production in electron antineutrino scattering~\cite{IceCube:2021rpz}. 

Studies of the arrival direction of IceCube's neutrinos indicate that the diffuse astrophysical component is compatible with an isotropic distribution~\cite{IceCube:2021xar}.
Searches for contributions from the galactic center have so far yielded only constraints and the galactic component has been shown to be no larger than approximately 10\% of the diffuse astrophysical flux~\cite{ANTARES:2018nyb}. 
Searches for neutrino sources by either looking for clustering, comparing with catalogues of known sources, and looking for real-time and spatial correlation with gamma-rays have also been performed. 
Evidence has been found for a neutrino source in the direction of the TXS0506+056 blazar by correlating the arrival time of a neutrino alert and a gamma-ray flare~\cite{IceCube:2018dnn}.
To date, the sources of bulk astrophysical neutrino flux are unknown and they are the targets of current and next-generation HENTs. 

\begin{figure}[htb!]
    \centering
    \includegraphics[width=0.45\textwidth]{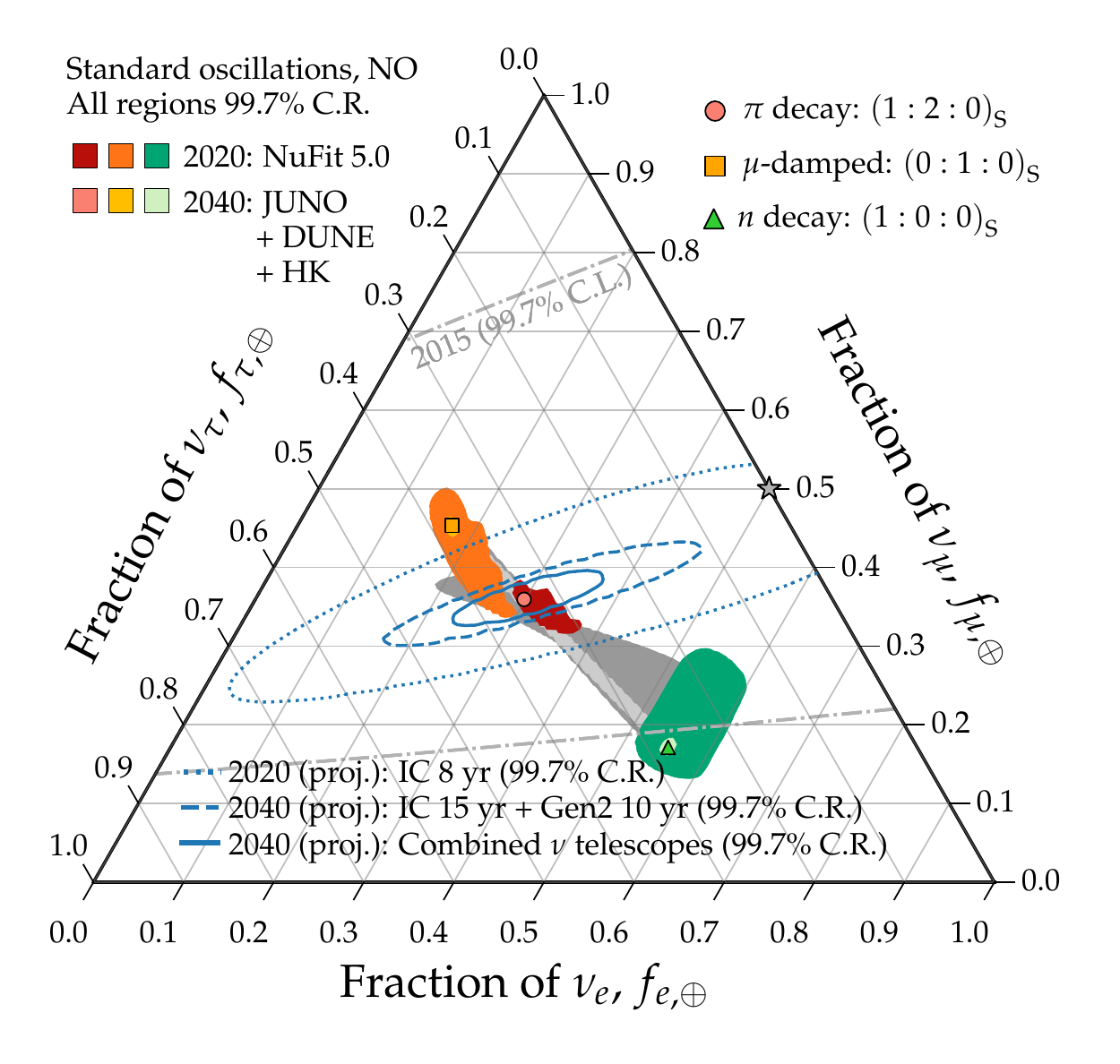}
    \includegraphics[width=0.45\textwidth]{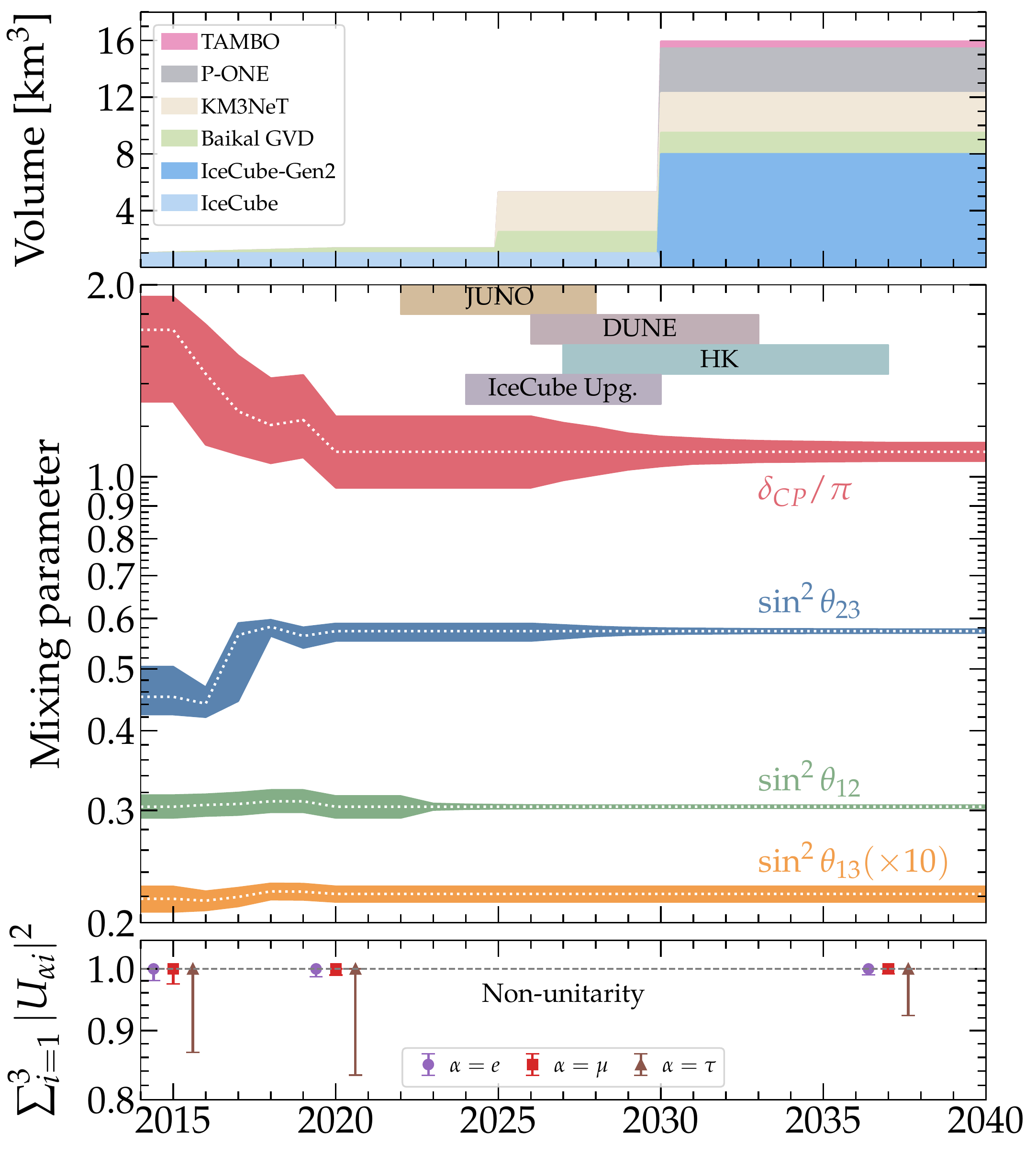}
    \caption{\textbf{Left: Flavor composition of astrophysical neutrinos at Earth now and in the future.}
    Production scenarios are indicated by the different colors.
    The hue of the colors indicated the assumed uncertainties on the oscillation parameters: dark hues use the current uncertainties, while light hues use the expected uncertainties in 2040.
    Current constraint on the flavor composition are shown in gray, while projections under different scenarios are shown in blue with different linestyles.
    \textbf{Right: Evolution of high-energy neutrino telescopes volumes, neutrino mixing angles, and unitarity constraints.}
    Top: effective volume of high-energy neutrino telescopes stacked.
    Middle: Uncertainties in the neutrino oscillation parameters as a function of time. Experiments relevant for neutrino oscillation parameter measurements are also shown with their proposed timelines.
    Bottom: Constraints on the unitarity of the neutrino mixing matrix for different neutrino flavor rows.
    Figures are from Ref.~\cite{Song:2020nfh}.
    }
    \label{fig:he}
\end{figure}

Measurements of the flavor composition of astrophysical neutrinos using different samples are shown in Fig.~\ref{fig:he}, left.
Current measurements are limited by the small sample size and difficulty in separating between different flavors.
The expected flavor composition from standard production scenarios and known neutrino oscillation parameters yields a democratic flavor composition~\cite{Baerwald:2012kc,Mena:2014sja,palomares-Ruiz:2015mka,Arguelles:2015dca,Bustamante:2015waa,Gonzalez-Garcia:2016gpq,Rasmussen:2017ert,Farzan:2018pnk,Ahlers:2018yom,Arguelles:2019rbn,Arguelles:2019tum}.
The current constraints are compatible with this expectation and only at-earth flavor configurations that significantly deviate from the democratic behaviour are ruled out, \textit{e.g.} 100\% electron-neutrinos or 100\% muon neutrinos~\cite{IceCube:2020wum,IceCube:2021tdn}.
Despite this limitation, astrophysical neutrino flavor composition has already been used to put some of the most stringent constraints in dark matter-neutrino interactions~\cite{deSalas:2016svi, Farzan:2018pnk,Weiler2006,Barranco:2010xt, Miranda:2013wla, Reynoso:2016hjr,Arguelles:2017atb, Kelly:2018tyg, Alvey:2019jzx, Choi:2019ixb}, and low-energy manifestations of quantum gravity~\cite{IceCube:2021tdn}, among others~\cite{Pakvasa:2012db, Pagliaroli:2015rca, Huang:2015flc, Shoemaker:2015qul, Bustamante:2016ciw, Denton:2018aml, Arguelles:2019rbn, Abdullahi:2020rge, Bustamante:2020niz,Rasmussen:2017ert,Arguelles:2019rbn}.

Figure~\ref{fig:he} right shows the expected progression of high-energy neutrino telescopes effective volume as a function of time. 
Notably, within this decade we will five-fold increase the neutrino effective volume and in the next decade the aggregated neutrino telescopes effective volume will be more than ten-fold larger.
Not only we expect an increase in the sample size, but also improvements in reconstruction using machine learning are expected to yield improved flavor identification.
This is complemented by the development of tau-neutrino-flavor-specific detectors, \textit{e.g.} TAMBO or Trinity, whose independent measurement would break the degeneracy between electron and tau neutrino flavors. 
Figure~\ref{fig:he} left shows the expected flavor composition measurements at Earth when combining the experiments listed in Fig.~\ref{fig:he} right (IceCube~\cite{IceCube:2016zyt}, IceCube-Gen2~\cite{IceCube-Gen2:2020qha}, Baikal GVD~\cite{Baikal-GVD:2021sbo}, KM3NeT~\cite{KM3Net:2016zxf}, P-ONE~\cite{P-ONE:2020ljt}, TAMBO~\cite{Romero-Wolf:2020pzh}).
In this figure, we can also see the expected regions from standard production scenarios, which are also expected to become smaller as both the sources are identified and the neutrino oscillation parameters uncertainties are reduced.
The effect of the latter is shown in this figure by comparing the size of the same color region with the two different hues, the dark hue is with current uncertainties and the lighter one with expected uncertainties.
As illustrated in this figure by 2040 the combined neutrino telescopes, with morphological mis-identification, is expected to be able to disentangle between the different production scenarios and also constraint new physics that produces anomalous flavor compositions. 
Meantime, as shown in Fig.~\ref{fig:he} right, next-generation long-baseline oscillation experiments (JUNO~\cite{JUNO:2015zny}, IceCube-Upgrade~\cite{IceCube-Gen2:2020qha}, DUNE~\cite{DUNE:2020lwj}, Hyper-Kamiokande~\cite{Hyper-Kamiokande:2018ofw}) will improve the oscillation parameter measurements. This is relevant because BSM effects are often sub-dominant processes of the standard flavor conversions, and the ability to access new physics through astrophysical neutrino flavor is limited by oscillation parameter errors. This is shown in various shaded regions in Fig.~\ref{fig:he} left, these predicted flavor ratios on  the  Earth with different astrophysical neutrino flavor production models. 

In what follows we briefly described the HENT that are expected to contributed to the astrophysical neutrino flavor measurements in the 100 TeV to 10 PeV energy range. 

{\it IceCube}~\cite{IceCube:2016zyt} (build): The IceCube Neutrino Observatory is a gigaton scale ice-Cherenkov detector build in the antartic continent. IceCube is comprised of 86 strings which are arranged on a hexagonal grid instrumenting approximately one cubic kilometer of ice. Each string hosts sixty digital optical modules (DOMs) each of which comprises of one photomultiplier tube with its on data acquisition system. 

{\it KM3NeT}~\cite{KM3Net:2016zxf} (underconstruction): The cubic-kilometer scale water Cherenkov detector is under construction in the Mediterranean sea. KM3NeT has two separate sites optimized for different energy ranges. The low-energy site is called KM3NeT/ORCA and is been deployed south of Toulon, France. The high-energy site is called KM3NeT/ARCA and will be deployed close to the sicilian shore. In its final configuration KM3NeT/ARCA is expected to see 11 muon-neutrinos, 41 electron-neutrinos, and 26 tau-neutrinos when assuming an $E^{-2}$ flux compatible with IceCube's measurements.

{\it Baikal-GVD}~\cite{Baikal-GVD:2021sbo} (underconstruction): The Baikal Gigaton Volume Detector is the next generation detector to be deployed in lake Baikal following the NT-200 detector. The detector has recently reached an effective volume of approximately $0.4 km^3$. The full detector will contain 10,386 opticalmodules arranged in 27 clusters of strings and is expected to have an instrumented volume of $1.5 km^3$. Due to the relatively large spacing between clusters Baikal-GVD is expected to have a larger energy threshold than IceCube.

{\it P-ONE}~\cite{P-ONE:2020ljt} (proposed): The Pacific-Ocean Neutrino Experiment is a proposed water Cherenkov detector to be deployed off the coast of Vancouver island in Canada. The full detector is expected to have 70 strings and be completed by 2030.

{\it TAMBO}~\cite{Romero-Wolf:2020pzh} (proposed): The Tau Air-shower Mountain-Based Observatory is a proposed water-Cherenkov detector to be deployed in the Colca Valley in Peru. The Colca Valley geometry is ideal for the detection of Earth-skimming PeV tau neutrinos. In its final configuration the TAMBO tau-neutrino effective area is expected to be ten times larger than IceCube's at a PeV and thirty times larger at ten PeV. 

{\it IceCube-Gen2 (ice)}~\cite{IceCube-Gen2:2020qha} (proposed): The next-generation ice-Cherenkov detector to be deployed in Antartica is expected to have a detection volume a factor of ten times greater than IceCube. 
\subsubsection{Extremely-high-energy astrophysical neutrinos\label{sec:ehe-neutrinos}}

Extremely-high-energy neutrino telescopes (EHENTs)~\cite{Kotera:2021hbp,Venters:2019xwi,Wissel:2020sec,Wang:2021zkm} aim to discovery neutrinos produced in interactions of cosmic rays with the cosmic microwave background~\cite{1966JETPL...4...78Z,PhysRevLett.16.748}. Radio emission is expected from the neutrino interaction at these extremely high energies.
Radio has the advantage of traveling longer distances than optical light and allows for the detector to be more spaced, increasing the effective area.
Additionally, the polarization of the radio emission can be used to reject downward-going cosmic-ray showers. 
Several experiments have used this technique to search for cosmogenic neutrinos, yielding so far constraints~\cite{IceCube:2018fhm,PierreAuger:2019ens,ANITA:2019wyx,ARA:2015wxq} and anomalous results observed by ANITA~\cite{Gorham:2016zah, Gorham:2018ydl}.
These events cannot be explained by standard neutrino interactions~\cite{Romero-Wolf:2018zxt}, since the required flux overshoots the present constraints and due to the lack of observation of the correlated low-energy neutrinos produced in the neutrino transport through the Earth~\cite{Safa:2019ege,IceCube:2020gbx}.
Several explanations have been put forward for this events, these include dark matter~\cite{Borah:2019ciw,Anchordoqui:2018ucj,Cline:2019snp,Heurtier:2019rkz,Heurtier:2019git}, supersymmetric partners~\cite{Anchordoqui:2018ssd,Collins:2018jpg}, leptoquarks~\cite{Chauhan:2018lnq}, sterile neutrinos~\cite{Cherry:2018rxj,Huang:2018als}, secret interactions~\cite{Esmaili:2019pcy}, axions~\cite{Esteban:2019hcm}, as well as more mundane explanations~\cite{Shoemaker:2019xlt}.

\begin{figure}[htb!]
    \centering
    \includegraphics[width=0.5\textwidth]{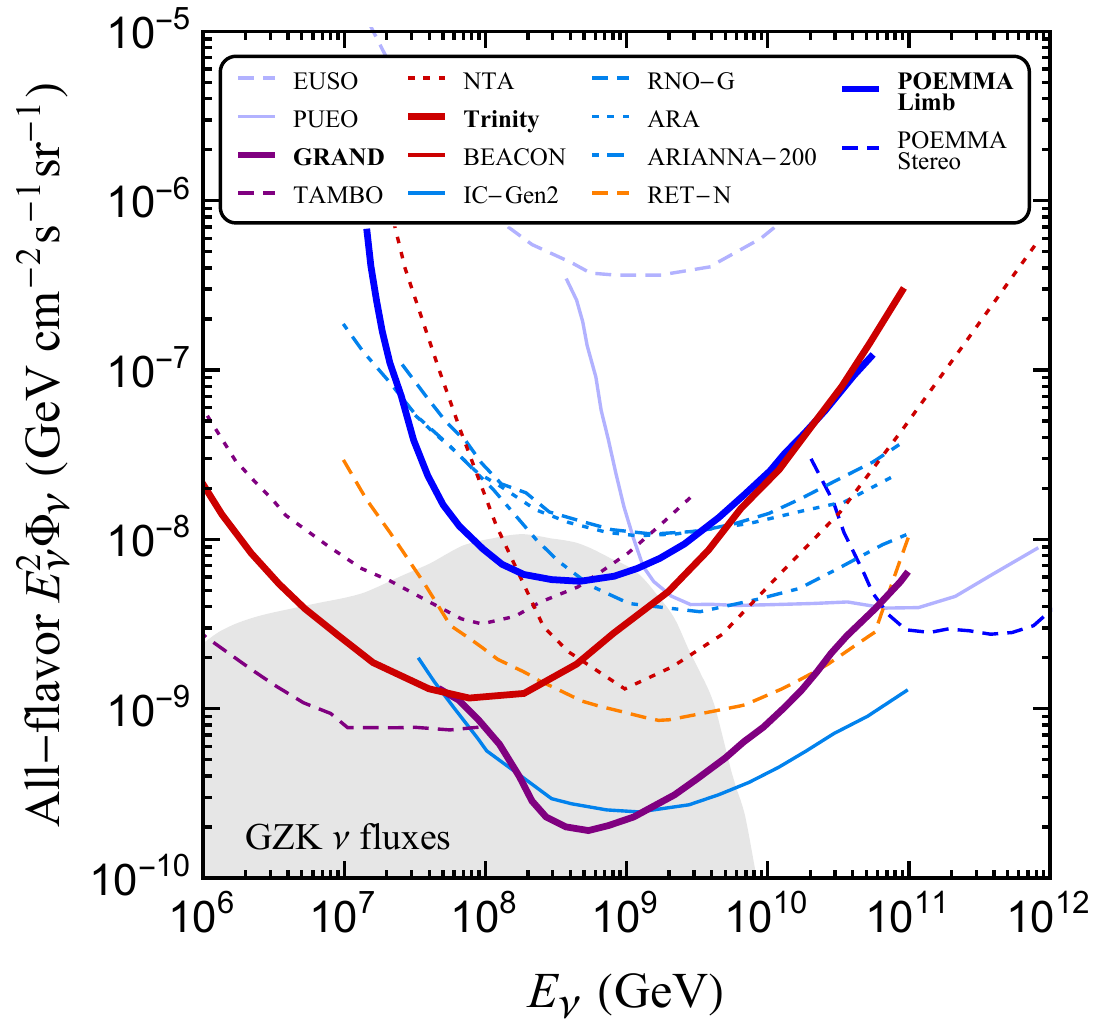}
    \caption{\textbf{Projected sensitivities of future neutrino telescopes to the all-flavor cosmogenic neutrino flux above PeV energies}
    Figure from Ref.~\cite{Huang:2021mki}.
    }
    \label{fig:ehe_sensitivity}
\end{figure}

Figure~\ref{fig:ehe_sensitivity} shows the projected sensitivity of next-generation EHENTs; see Ref.~\cite{Huang:2021mki} for a recent discussion on new physics searches for very high energy tau neutrinos.
Using different detection techniques that vary from earth-skimming neutrinos, Cherenkov detectors, and radio. 
Combining up-going and down-going measurements allow disentangling particle physics (cross-sections) from astrophysics (flux)~\cite{Kusenko:2001gj,Anchordoqui:2001cg}, and constrain both the properties of astrophysical sources and the interactions of neutrinos far above the weak scale~\cite{Anchordoqui:2018qom}.
In what follows, we briefly describe the detectors that are currently probing this energy range and those proposed to follow up:

{\it Auger}~\cite{PierreAuger:2019ens,Horandel:2019qwu} (build): The Pierre Auger Observatory is a hybrid detector consisting of both an array of water Cherenkov surface detectors and atmospheric fluorescence detectors. The main surface array is comprised of 1660 water Cherenkov tanks spread over 3000 squared kilometers and overlooked by 24 air-fluorescence telescopes. The observatory has been in operation since 2008 and so far collected an exposure that exceeds 40,000 ${\rm km^2 sr yr}$. The Global Cosmic Ray Observatory (GCOS) is a conceptual design of similar observatory but covering 40,000  ${\rm km^2}$, roughly 13 times bigger than Auger. 

{\it ANITA/PUEO}~\cite{ANITA:2019wyx,PUEO:2020bnn} (build/under construction): The ANtarctic Impulsive Transient Antenna (ANITA) is an array of radio antennas installed on a helium balloon surveys Antarctica for $\sim30$ days at a time. ANITA has completed four flights and set the strongest constraints on the astrophysical neutrino flux above 
$10^{10}$~GeV, as well as observed anomalous events that are yet unexplained. The follow-up of ANITA is called PUEO, and it has recently been approved for construction. PUEO is expected to have a sensitivity that is an order of magnitude larger than its predecessor below 30 EeV and with lower threshold for tau neutrinos~\cite{PUEO:2020bnn}.

{\it ARIANNA}~\cite{ARIANNA:2019scz} (proposed): Antartic Ross Ice-Shelf ANtenna Neutrino Array is the larger form of ongoing ARIANNA effort. It also aims to detect radio wave from neutrino interactions, however, it utilizes the radio wave reflection by the Ross Ice Shelf, allowing to locate autonomous independent detectors on the ice surface.  

{\it RNO-G}~\cite{RNO-G:2020rmc} (under construction): The Radio Neutrino Observatory in Greenland~\cite{Aguilar:2019jay} is an array of antennas to be deployed in ice designed to measure the neutrino flux above $10^{16}$~eV. The array consists of surface and deep antenna arrays organized in 35 stations. The deep array will survey a larger effective volume, while the surface array will be used to reject incident cosmic rays.

{\it TAMBO}~\cite{Romero-Wolf:2020pzh} (proposed): See description above.

{\it Trinity}~\cite{Otte:2019aaf} (proposed): Trinity is comprised of three stations, each of which deploys an imaging atmospheric Cherenkov telescope on a mountaintop. These are expected to observe the Cherenkov radiation produced by the decay of Earth-skimming tau-neutrinos.

{\it BEACON}~\cite{Wissel:2020sec} (proposed): The Beamforming Elevated Array for COsmic Neutrinos (BEACON) uses a sparse array of clustered antennas to search for the radio signal from upgoing tau neutrinos. The concept uses interferometric triggering on top of a high-elevation mountains to maximize each station's sensitivity. Stations of about ten antennas are separated by 1 kilometer and view a large area of the horizon.

{\it RET-N}~\cite{RadarEchoTelescope:2021rca} (proposed): Radio Echo Telescope-Neutrino observes radio wave reflection by the clouds of ionized media produced by high-energy neutrinos. This is the second stage of RET-CR (Cosmic Ray). 

{\it POEMMA}~\cite{POEMMA:2020ykm} (proposed): POEMMA consists of two identical satellites orbiting the Earth at an altitude of 525 km. These satellites can operate in two different observation modes. First, they can operate in POEMMA-stereo mode, which aims to detect cosmic rays or neutrinos up to 20 EeV by observing the fluorescence emission in up-ward going atmospheric air showers. Second, they can operate in POEMMA-limb mode, where they aim to detect the Cherenkov light produced by an up-ward going tau neutrino decay. These two modes are complementary as they provide all-flavor and single-flavor detection capabilities.

{\it GRAND}~\cite{GRAND:2018iaj} (proposed): GRAND is a radio antenna array expected to be deployed in China over a mountain slope. It aims to detect the radio emission produced by high-energy air-shower showers produced by Earth-skimming tau neutrino interactions. The distance between the antennas is expected to be one kilometer; these are then arranged into twenty GRAND10k clusters. Each GRAND10k cluster is comprised of ten thousand antennas that cover approximately 100 kilometers-squared patch of the mountain slope. 

{\it IceCube-Gen2 (radio)}~\cite{IceCube-Gen2:2020qha} (proposed): Expected to be located in the antarctic glacier near the current IceCube array, the radio component of IceCube-Gen2 is expected to be composed of stations that combine shallow, sub-surface antennas with antennas deployed at depths down to 100~m. The shallower antennas are expected to have improved signal sensitivity and capacity to reject cosmic-ray airshower. These are complemented by the deeper antennas that provide a larger effective volume and sky coverage. Two hundred stations are expected to be deployed in the South Pole over an area of 500 square kilometers.

\section{Summary and outlook}
\label{sec:summary}

In this White Paper we have reviewed the theoretical aspects of Beyond the Standard Model (BSM) scenarios that can impact neutrino flavor measurements, leading to deviations with respect to the three-neutrino standard paradigm in neutrino oscillations, neutrino telescopes and core-collapse supernovae. 
The list of models affecting neutrino flavor patterns is wide and difficult to cover exhaustively in a single document. Therefore, we have defined four general categories accounting for different phenomenological consequences expected in general extensions of the SM. The addition of extra neutrino states (Sec.~\ref{sec:steriles})
is well-motivated from the theoretical point of view: right-handed neutrinos are singlets of the SM gauge group and pose one of the simplest extensions of the SM able to generate neutrino masses and mixings. These may lead to deviations from non-unitarity, to which neutrino oscillation experiments could be sensitive, or could impact the neutrino oscillation pattern through the addition of new oscillation frequencies. This white paper provides an updated summary of current constraints on non-unitarity parameters and LED-induced oscillations. We have also reviewed the phenomenological consequences derived from the plausible existence of new neutrino interactions in Sec.~\ref{sec:inter}. Theory challenges associated to model building which could lead to large effects in the neutrino sector have been reviewed, pointing out possible ways out and model building strategies. We also include a detailed review of constraints applicable to flavored models with light and ultra-light mediators, to which neutrino oscillations in matter are sensitive, as well as a discussion of the phenomenological consequences from dark matter-neutrino interactions. Next, we have reviewed the main consequences of neutrino decay in a variety of setups in Sec.~\ref{sec:decay_perspectives}. Neutrino decay searches offer an indirect way to test the neutrino mass generation mechanism, which is a priori unknown. The phenomenological consequences are diverse, and here we review the main consequences from both \emph{visible} and \emph{invisible} neutrino decay, which impact neutrino flavor measurements in different ways. Tables~\ref{tab:boundsNudec} and~\ref{tab:futureboundsNudec} summarize the current bounds and expected sensitivities for a variety of experiments. Finally, given the singular quantum-mechanical neutrino properties, neutrinos also provide a unique avenue for testing fundamental symmetries as CPT or Lorentz invariance, as well as non-standard quantum decoherence. The phenomenological consequences derived from these scenarios are reviewed in Sec.~\ref{sec:CPT}, which includes a review of current bounds (see, e.g. Tab.~\ref{tab:LV}) as well as a summary of future prospects to improve over these using neutrino oscillation experiments and neutrino telescopes.

Neutrinos from different sources spans several orders of magnitude in neutrino energy (ranging from keVs to EeVs), and therefore to be detected, facilities with different experimental requirements have been (and will be) built. At the same time, BSM effects might be present at neutrino production, propagation, and detection affecting each one of this stages in different ways. In particular, BSM scenarios that impact neutrino flavor leave signals at more than one experiment. 
Fortunately, several future neutrino experiments have been approved or are already being built, and will allow for exciting BSM searches. 
In the future, identifying all neutrino flavors 
is important, which calls for multi-channel observation. 

Understanding complementarity between different techniques and facilities is also relevant. For instance, developments in detector technologies motivated by direct dark matter searches have finally made detecting
coherent neutrino nucleus elastic scattering process feasible. Additionally, direct dark matter searches are starting to be sensitive to solar and atmospheric neutrino scattering and therefore can provide additional constraints on new physics affecting the neutrino floor, as discussed in Sec.~\ref{sec:le}. Section~\ref{sec:me} discussed medium-energy experiments, including neutrino beam experiments and atmospheric neutrino detectors. Neutrino beam experiments can provide beams with relatively well-known spectra, flavor content, and timing information,
which offer multiple advantages with respect to neutrino fluxes from natural sources. 
At these, the use of near detectors is crucial to reduce the level of uncertainties and allow to search for BSM effects using their far detector data.
Past and current experiments have performed BSM searches using this two-detector setup, and future beam-based experiments will build upon past experience and efforts in this regard, leading to state-of-the-art facilities, with higher power, and a reduced level of systematic errors. Atmospheric neutrino experiments are complementary to low-energy and beam-based experiments. The atmospheric neutrino flux spans several orders of magnitude in energy (tens of MeV to hundreds of TeV) and neutrinos can travel different distances from the production point to the detector (tens of km to the Earth diameter). BSM searches benefit from both higher energies and baselines in comparison to long-baseline experiments, and also from the experience gained with multi-purpose detectors used to look for different physics. 
Upgrades of current atmospheric experiments include increasing their volume (to gather more statistics) or improving the detector technology in order to gain energy and angular resolution as well as particle identification, which aids in BSM searches. Finally, neutrino telescopes (described in Sec.~\ref{sec:he}) provide a unique window for high- and extremely high-energy neutrinos. These facilities aim to measure the high-energy part of the atmospheric neutrino flux and astrophysical neutrino sources. 
Extremely high-energy neutrino telescopes aim to detect particles with the highest energies on the Earth surface. 
Given the energy ranges and the very long distances traveled by the neutrinos detected (from the Earth radius to several gigaparsecs), neutrino telescopes provide a unique avenue to probe BSM scenarios.

Clearly, the best way to confront BSM models with data is by exploiting the synergies between different experiments and probing BSM effects across a wide range of energies, and for multiple detector technologies. Thus, next generations experiments will be crucial for discovering or ruling out the new physics scenarios reviewed here, that are of interest of the particle physics community.

\section{Acknowledgements}
\label{sec:acknowledgements}

CAA is supported by the Faculty of Arts and Sciences of Harvard University, and the Alfred P. Sloan Foundation. 
GB acknowledges support from AEI-MICINN PID2020-113334GB-I00 / AEI / 10.13039/501100011033. 
MB is supported by the {\sc Villum Fonden} under project no.~29388. 
PC acknowledges support by Grants CEX2020-001007-S, PID2019-108892RB-I00 and RYC2018-024240-I funded by MCIN/AEI/10.13039/501100011033 and by ``ESF Investing in your future''. 
YF has received  financial support from Saramadan under contract No.~ISEF/M/400279 and No.~ISEF/M/99169. 
TK acknowledges the support from the Science and Technology Council Facilities, UK. 
AMS acknowledges the support of US National Science Foundation (Grant No. PHY-2020275). 
This project has received funding/support from the European Union’s Horizon 2020 research and innovation program under the Marie Sklodowska-Curie grant agreement No 860881-HIDDeN. 






\bibliographystyle{utphys}

\bibliography{common/tdr-citedb}

\end{document}